\documentclass[letterpaper,11pt]{article}
\pdfoutput=1
\usepackage{jheppub}
\usepackage{slashed}
\usepackage{graphicx}
\usepackage{subfigure}
\usepackage{soul}
\usepackage{amsmath}
\usepackage{multirow}
\usepackage{tablefootnote}
\usepackage{hyperref}
\usepackage{epstopdf}
\usepackage{lscape}
\usepackage{url}
\usepackage[dvipsnames]{xcolor}
\usepackage[utf8]{inputenc}
\usepackage{float}
\usepackage{booktabs}
\usepackage{makecell}

\RequirePackage[normalem]{ulem}

\newcommand{\beq}{\begin{equation}}
\newcommand{\eeq}{\end{equation}}
\newcommand{\bea}{\begin{eqnarray}}
\newcommand{\eea}{\end{eqnarray}}
\newcommand{\beqs}{\begin{subequations}}
\newcommand{\eeqs}{\end{subequations}}

\newcommand{\ba}{\begin{array}}
\newcommand{\ea}{\end{array}}

\def\figureautorefname~#1\null{Fig.\,#1\null}
\def\tableautorefname~#1\null{Tab.\,#1\null}

\def\equationautorefname~#1\null{Eq.\,(#1)\null}

\def\m1{M_1}
\def\m2{M_2}
\def\m3{M_3}

\def\ch10{\tilde \chi^0_1}

\def\to{\rightarrow}

\newcommand{\lsim}{\mathrel{\mathop{\kern 0pt \rlap
  {\raise.2ex\hbox{$<$}}}
  \lower.9ex\hbox{\kern-.190em $\sim$}}}
\newcommand{\gsim}{\mathrel{\mathop{\kern 0pt \rlap
  {\raise.2ex\hbox{$>$}}}
  \lower.9ex\hbox{\kern-.190em $\sim$}}}

\definecolor{pink}{RGB}{255,105,180}

\def\cosba{\cos(\beta-\alpha)}

\newcommand{\bpm}{\begin{pmatrix}}
\newcommand{\epm}{\end{pmatrix}}

\newcommand{\tanb}{\tan \beta}

\allowdisplaybreaks[4]

\title{Complete Light Long-Lived Particles searches in Type-I 2HDM}
\author[a]{Xueying Qi}
\author[b]{, Huayang Song}
\author[a,c]{ and Wei Su}

\affiliation[a]{School of Science, Shenzhen Campus of Sun Yat-sen University, No. 66, Gongchang Road, \\ Guangming District, Shenzhen, Guangdong 518107, China }
\affiliation[b]{Particle Theory and Cosmology Group, Center for Theoretical Physics of the Universe, Institute for Basic Science (IBS), Daejeon, 34126, Korea}
\affiliation[c]{Institute of Theoretical Physics, Chinese Academy of Sciences, Beijing 100190, P. R. China}

\emailAdd{qixy33@mail2.sysu.edu.cn}
\emailAdd{huayangs1990@ibs.re.kr}
\emailAdd{suwei26@mail.sysu.edu.cn}

\abstract{
Recently, the study of long-lived particles (LLPs) has attracted increasing attention. In this work, we analyze the full parameter space of the Type-I Two-Higgs-Doublet Model (2HDM) that allows for light long-lived scalar ($H$) and pseudoscalar ($A$) particles. When involving a light beyongd Standard Model (BSM) Higgs, the $\Delta S$ could be the main contribution during the global fit of the oblique parameters, which is different to $\Delta T$ being the main factor for heavy BSM Higgs cases. By imposing theoretical constraints such as vacuum stability and perturbative unitarity, together with current experimental bounds, we summarize a complete region for a potential light $H$ with $\cos(\beta - \alpha) \simeq \frac{1}{\tan \beta}$, light $A$ with $\cos(\beta - \alpha) \simeq \frac{1}{\tan\beta} \frac{2m_H^2 - m_h^2}{m_H^2 - m_h^2}$, and point out the invisible Higgs decay is the most important constraint.  We further identify viable regions for LLPs and propose four benchmark regions that simultaneously accommodate a light long-lived particle and explain the $W$ boson mass anomaly. For these benchmarks, we present the reaches of FASER and FASER~2, where FASER~2 improves the sensitivity by approximately two orders of magnitude compared to FASER.
}


\begin{document}
\maketitle 
\flushbottom
\newpage

\section{Introduction}

Since the discovery of the Higgs boson at the Large Hadron Collider (LHC) in 2012~\cite{ATLAS:2012yve, CMS:2012qbp}, particle physics has entered a new era. The Standard Model (SM) has undergone extensive scrutiny and successfully explained various phenomena in particle physics. However, several experimental observations and theoretical considerations, such as the existence of dark matter, baryon asymmetry of the universe, and strong CP problem suggest the presence of new physics beyond the SM (BSM)~\cite{Crivellin:2023zui, Bertone:2004pz, Cohen:1993nk, Peccei:1977ur, Peccei:1977hh}.

A huge amount of effort has been put in finding any new fundamental particle predicted in various BSM scenarios. However, attempts have been unsuccessful; while hints of BSM physics have been reported in the form of anomalies~\cite{CDF:2022hxs, Muong-2:2021ojo, LHCb:2021trn}, no concrete discovery has been made yet in the TeV range. 
Therefore, the search for so-called long-lived particles (LLPs) has gained considerable attractions in the last few years. The LLPs, once produced at the interaction point, travel a macroscopic distance before decaying, possibly leading to exotic signatures. It can be detected at the conventional detectors such as ATLAS and CMS via searches for displaced signatures~\cite{CMS:2021tkn, CMS:2021kdm, CMS:2021sch, CMS:2022fut, ATLAS:2022gbw, ATLAS:2022pib, CMS:2022qej, ATLAS:2022izj, ATLAS:2023oti, ATLAS:2023zxo, ATLAS:2023cjw, CMS:2023jqi, CMS:2024bvl, CMS:2024trg, CMS:2024ita, ATLAS:2024qoo, ATLAS:2024ocv, CMS:2024nhn, ATLAS:2025fdm, CMS:2025urb, ATLAS:2025pak}, or dedicated detectors such as ANUBIS~\cite{Bauer:2019vqk, Hirsch:2020klk, Dreiner:2020qbi}, CODEX-b~\cite{Gligorov:2017nwh, Dey:2019vyo, CODEX-b:2019jve, Aielli:2022awh}, 
FASER~\cite{Feng:2017uoz, FASER:2018eoc, Kling:2021fwx, FASER:2023tle, FASER:2024bbl}, MATHUSLA~\cite{Alimena:2019zri, Bose:2022obr, Chou:2016lxi, Curtin:2017izq, Evans:2017lvd, Curtin:2018mvb, Curtin:2018ees, Berlin:2018jbm, MATHUSLA:2018bqv, MATHUSLA:2019qpy, Jodlowski:2019ycu, Alidra:2020thg, MATHUSLA:2020uve, MATHUSLA:2022sze}, and ShiP~\cite{Bonivento:2013jag, Alekhin:2015byh, SHiP:2015vad}.
%

Although usually only minimal models are well studied for their LLPs scenario, LLPs naturally appear in numerous well-motivated BSM models. Given that FASER has published its first results~\cite{FASER:2023tle, FASER:2024bbl} with data collected during LHC Run 3, more realistic ultra-violate (UV) models can be tested, where the LLPs have non-minimal interactions that greatly change the expectations for their production (and decay) at the LHC. In previous studies on these UV models, the BSM states are assumed to be heavy, and the low-mass region of the parameter space is considered ruled out because usually the light particles are kinematically accessible in collider experiments. However, the strategies used to detect these light particles are quite different from those used to detect a heavy particle. Therefore, some specific parameter space of the BSM model are still valid for LLPs existing, which should be carefully examined.

In a previous work~\cite{Kling:2022uzy}, we have proposed two benchmark scenarios in Type-I 2HDM, which provide LLPs under all the experimental constraints. They are studied with the FASER detector and constraints on the parameter space in these two scenarios are obtained using FASER 2022 data~\cite{FASER:2024bbl}. In this paper, we will investigate the whole parameter space of the 2HDM to determine the allowed region for a light Higgs from 0.1 GeV to 10 GeV, and explore the potential of being a long-lived scalar. 

Theoretical constraints such as vacuum stability, perturbativity and unitarity restrict the size of the quartic scalar couplings. Therefore the mass splittings among the scalars are severely constrained requiring the presence of a 125 GeV Higgs particle and a sub-GeV particle. We first analyze the allowed parameter space of weakly coupled light scalars $H/A$ in the Type-I 2HDM with theoretical constraints. For experimental constraints, we mainly focus on direct searches for exotic Higgs bosons at colliders and constraints from Higgs invisible decays. Due to the existence of several neutral scalars, we organize the mass hierarchy into four categories: (i) light $H$ with $m_A < m_h/2$; (ii) light $H$ with $m_A > m_h/2$; (iii) light $A$ with $m_H < m_h/2$; and (iv) light $A$ with $m_H > m_h/2$, and determine the allowed parameter space for each case. In scenarios (i) and (iii), we also include constraints from exotic decays $h \to AA/HH \to ffff$ and $4\gamma$~\cite{CMS:2022fyt}. 
We further include constraints from electroweak precision measurements, focusing on the impact of improved precision of the oblique parameters $S$, $T$, and $U$ from 2018 and 2024 $Z$-pole data. Their final allowed regions are similar, while the main contribution from $S$ or $T$ are different. 
The case with both $m_H<5$ GeV and $m_A<5$ GeV is totally excluded. Under these theoretical and experimental constraints, we summarize the full viable parameter space accommodating light long-lived particles in the Type-I 2HDM and define three benchmark scenarios. Furthermore, a light scalar in 2HDM can result in large mass splittings between BSM scalars naturally, which affects the accurate prediction for the W-boson mass. Thus we further analyze the parameter regions that can explain the CDF-II $m_W$ measurement~\cite{CDF:2022hxs} or be consistent with the latest LHC $W$ boson mass value~\cite{CMS:2024lrd, ATLAS:2024erm} at 95\% confidence level (C.L.). We select four benchmark points that both accommodate LLPs and explain the $m_W$ anomaly at CDF-II, and explore the reaches of FASER and FASER~2 for these scenarios. We also consider various constraints on light scalars from beam dump experiments, supernova, and measurements of meson decays.

The rest of this paper is organized as follows. In~\autoref{sec:2hdm}, we briefly review the 2HDM and discuss the advantages of the Type-I 2HDM for accommodating light and long-lived particles. In~\autoref{sec:constraints}, we present the theoretical and experimental constraints and explore the allowed parameter space for light long-lived particles. In~\autoref{sec:parameter}, we summarize the viable parameter regions in the Type-I 2HDM. In~\autoref{sec:mw}, we study to BSM corrections to the $W$ boson mass. In~\autoref{sec:faser}, we show the reaches of FASER and FASER~2 at selected benchmark points. We conclude this work in~\autoref{sec:Conclude}. 

\section{Brief review of 2HDM}
\label{sec:2hdm}
The general 2HDM framework includes two ${\rm SU}(2)_L$ scalar doublets, $\Phi_i$ ($i=1,2$), with hypercharge $Y = +1/2$. They can be parameterized as
\begin{eqnarray}
\Phi_{i}=\begin{pmatrix}
  \phi_i^{+}    \\
  (v_i+\phi^{0}_i+iG_i)/\sqrt{2}
\end{pmatrix}\,,
\end{eqnarray}
\noindent where $v_{1}$ and $v_{2}$ are the vacuum expectation values (vevs) of $\Phi_1$ and $\Phi_2$ after EWSB, satisfying $v^2 \equiv v_1^2 + v_2^2$, with $ v = 246~\mathrm{GeV}$.

\begin{table}[htbp]
\centering
\begin{tabular}{|c|c|c|c|c|}
\hline
2HDM & Type-I & Type-II & Type-L & Type-F \\
\hline
up-type & $\Phi_2$ & $\Phi_2$ & $\Phi_2$ & $\Phi_2$ \\
\hline
$\xi_{huu}$ & $c_{\alpha}/s_{\beta}$ & $c_{\alpha}/s_{\beta}$ & $c_{\alpha}/s_{\beta}$ & $c_{\alpha}/s_{\beta}$ \\
$\xi_{Huu}$ & $s_{\alpha}/s_{\beta}$ & $s_{\alpha}/s_{\beta}$ & $s_{\alpha}/s_{\beta}$ & $s_{\alpha}/s_{\beta}$ \\
$\xi_{Auu}$ & $t_{\beta}^{-1}$ & $t_{\beta}^{-1}$ & $t_{\beta}^{-1}$ & $t_{\beta}^{-1}$ \\
\hline
down-type & $\Phi_2$ & $\Phi_1$ & $\Phi_2$ & $\Phi_1$ \\
\hline
$\xi_{hdd}$ & $c_{\alpha}/s_{\beta}$ & $-s_{\alpha}/c_{\beta}$ & $c_{\alpha}/s_{\beta}$ & $-s_{\alpha}/c_{\beta}$ \\
$\xi_{Hdd}$ & $s_{\alpha}/{s_{\beta}}$ & $c_{\alpha}/c_{\beta}$ & $s_{\alpha}/{s_{\beta}}$ & $c_{\alpha}/c_{\beta}$ \\
$\xi_{Add}$ & $-t_{\beta}^{-1}$ & $t_{\beta}$ & $-t_{\beta}^{-1}$ & $t_{\beta}$ \\
\hline
lepton & $\Phi_2$ & $\Phi_1$ & $\Phi_1$ & $\Phi_2$ \\
\hline
$\xi_{h\ell \ell}$ & $c_{\alpha}/s_{\beta}$ & $-s_{\alpha}/c_{\beta}$ & $-s_{\alpha}/c_{\beta}$ & $c_{\alpha}/s_{\beta}$ \\
$\xi_{H\ell \ell}$ & $s_{\alpha}/s_{\beta}$ & $c_{\alpha}/c_{\beta}$ & $c_{\alpha}/c_{\beta}$ & $s_{\alpha}/s_{\beta}$ \\
$\xi_{A\ell \ell}$ & $-t_{\beta}^{-1}$ & $t_{\beta}$ & $t_{\beta}$ & $-t_{\beta}^{-1}$ \\
\hline
\end{tabular}\\
\caption{Higgs couplings to the SM fermions in the four different types of 2HDM, normalized to the corresponding SM values~\cite{Kling:2020hmi}. }
\label{tab:2hdm_coup}
\end{table}

To avoid tree-level flavor-changing neutral currents (FCNCs), a soft $\mathbb{Z}_2$ symmetry is imposed on the Lagrangian, under which $\Phi_1 \rightarrow \Phi_1$ and $\Phi_2 \rightarrow -\Phi_2$. With this symmetry and assuming CP conservation, the most general Higgs potential can be written as: 
\begin{eqnarray}
\label{eq:L_2HDM}
 V(\Phi_1, \Phi_2) &=& m_{11}^2\Phi_1^\dag \Phi_1 + m_{22}^2\Phi_2^\dag \Phi_2 -m_{12}^2(\Phi_1^\dag \Phi_2+ h.c.) + \frac{\lambda_1}{2}(\Phi_1^\dag \Phi_1)^2 + \frac{\lambda_2}{2}(\Phi_2^\dag \Phi_2)^2  \notag \\
 & &+ \lambda_3(\Phi_1^\dag \Phi_1)(\Phi_2^\dag \Phi_2)+\lambda_4(\Phi_1^\dag \Phi_2)(\Phi_2^\dag \Phi_1)+\frac{\lambda_5}{2}   \Big[ (\Phi_1^\dag \Phi_2)^2 + h.c.\Big]\,,
\end{eqnarray}
\noindent where all parameters are real. After EWSB, the scalar sector of the 2HDM contains five physical eigenstates: a pair of neutral CP-even Higgs bosons $h$ and $H$, a CP-odd Higgs boson $A$, and a pair of charged Higgs bosons $H^{\pm}$. In the alignment limit $\cos(\beta - \alpha) = 0$, the CP-even Higgs boson $h$ has SM-like couplings to fermions and gauge bosons. Therefore we mainly focus on the values of $\cosba$ at the vicinity of the alignment limit. It is convenient to parameterize the potential of the 2HDM using the physical Higgs boson masses $m_h$, $m_A$, $m_H$, and $m_{H^{\pm}}$, the EWSB vev $v$, the CP-even Higgs bosons mixing angle $\alpha$, and the ratio of the Higgs vevs $\tan \beta$ =  $v_{2}$/$v_{1}$. Besides there is an additional soft $\mathbb{Z}_2$ symmetry breaking parameter $m_{12}^2$, usually replaced by $\lambda v^2 \equiv m_H^2 - \frac{m_{12}^2}{\sin\beta \cos\beta}$. 

\begin{figure}[htpb]
  \centering
\includegraphics[width=0.49 \linewidth]{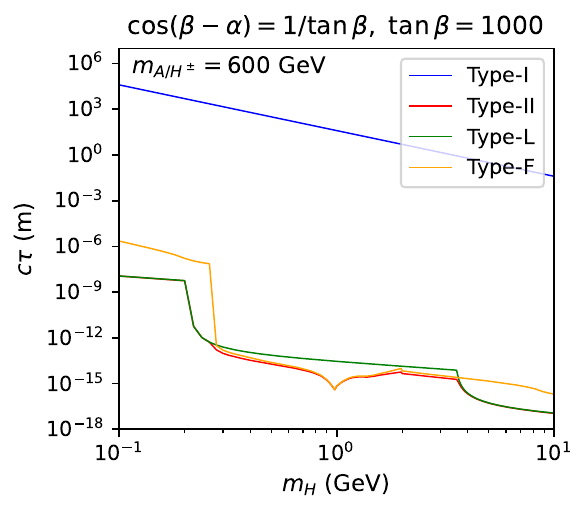}
\includegraphics[width=0.49 \linewidth]{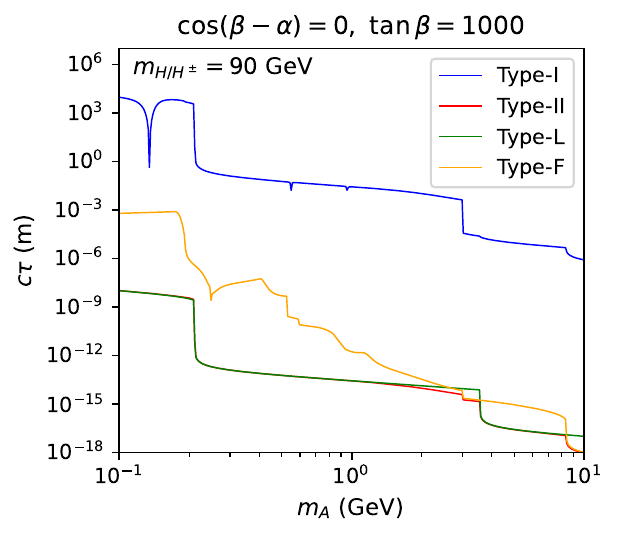}
\caption{Decay length comparison of four types of 2HDMs with benchmarks from Ref.~\cite{Kling:2022uzy}. (Left panel): Decay length $c\tau$ of the light CP-even Higgs $H$ in the Type-I, Type-II, Type-L, and Type-F 2HDMs for $m_{A/H^\pm} = 600~\mathrm{GeV}$ and $\cos(\beta - \alpha) = 1/\tan\beta$. (Right panel): Decay length $ c\tau $ of the light CP-odd Higgs $A$ in the Type-I, Type-II, Type-L, and Type-F 2HDMs for $m_{H/H^\pm} = 90~\mathrm{GeV}$ and $\cos(\beta - \alpha) = 0$. Here, $\lambda v^2 = 0$ and $\tan\beta = 1000$ are fixed~\protect\footnotemark[1].} 
\label{fig:type}
\end{figure}
%
Depending on different assignments of couplings between the doublets $\Phi_1$ and $\Phi_2$ and the SM quarks and leptons, there are four types of 2HDMs: Type-I, Type-II, Type-L, and Type-F. In Table~\ref{tab:2hdm_coup}, We show how the neutral scalars couple to fermions in different types of 2HDMs using the mixing angles $\alpha$ and $\beta$. Among the four types of 2HDMs:
\begin{itemize}
    \item In Type-I, only $\Phi_2$ couples to all fermions. The tree-level couplings of the Higgs bosons to fermions are given by:
    \begin{align}
    \xi_h^f &= \frac{\cos \alpha}{\sin \beta} = \sin(\beta - \alpha) + \cos(\beta - \alpha) \cot \beta, \\
    \xi_H^f &= \frac{\sin \alpha}{\sin \beta} = \cos(\beta - \alpha) - \sin(\beta - \alpha) \cot \beta, \\
    \xi_A^f &=
    \begin{cases}
    \cot \beta, & \text{for } f = u, \\
    -\cot \beta, & \text{for } f = d,\, e.
    \end{cases}
    \end{align}
    \item In Type-II, $\Phi_1$ couples to down-type quarks and charged leptons, while $\Phi_2$ couples to up-type quarks.
    \item In Type-L, $\Phi_2$ couples to all quarks, and $\Phi_1$ couples to charged leptons.
    \item In Type-F, $\Phi_1$ couples to down-type quarks, and $\Phi_2$ couples to up-type quarks and charged leptons.
\end{itemize}
We find that in Type-II, Type-L, and Type-F 2HDMs, there is always at least one scalar whose coupling to fermions is unsuppressed over the entire $\tan\beta$ range. This makes it challenging to achieve weakly coupled long-lived scalars. 

To have a first general idea of 4 types for a light long-lived Higgs, in~\autoref{fig:type} we compare the decay lengths $c\tau$ in the Type-I, Type-II, Type-L, and Type-F 2HDMs, based on two benchmark points for light $H$ (left panel) and light $A$ (right panel) defined by Eqs.(4.17) and (4.18) in Ref.~\cite{Kling:2022uzy} with $\tan\beta=1000$\footnote{Generally $\tanb$ larger than 50 is not allowed for Type-II, Type-L, and Type-F 2HDMs. Here we temporarily relax this requirement to explore the potential LLP candidate.}. It is clear that, in both the light $H$ and light $A$ scenarios, the values of $c\tau$ in the Type-II, Type-L, and Type-F 2HDMs are significantly smaller than in the Type-I case, making them incapable of accommodating long-lived light scalars. 
We illustrate this detailedly in \autoref{sec:type}, where these 3 types fail to accommodate a long-lived light particle. Hence, in the following we only focus on the Type-I 2HDM, where all fermions couple only to the Higgs doublet $\Phi_2$, and the couplings of BSM Higgs could be small enough for a long-lived scalar candidate. We focus on it in the following analysis, get the light Higgs region first and then dig into the light long-lived scalar parameter space.
%
%
\section{Theoretical and Experimental Constraints}
\label{sec:constraints}
In this section, we will figure out the allowed parameter space for the light long-lived particles in the Type-I 2HDM under various constraints. 


\subsection{Theoretical Constraints}
\begin{itemize} 
\item \textbf{Vacuum stability} 
To ensure vacuum stability, the scalar potential must be bounded from below~\cite{Gunion:2002zf}:
\begin{eqnarray}
\label{eq:Vacuum}
\lambda_1 > 0 \ , \ \lambda_2 > 0 \ , \ \lambda_3 > -\sqrt{\lambda_1 \lambda_2} \ , \ \lambda_3 + \lambda_4 - |\lambda_5| > -\sqrt{\lambda_1 \lambda_2} 
\end{eqnarray}
\item \textbf{Perturbativity and unitarity}
For the general perturbativity condition, we must have
\begin{eqnarray}
 |\lambda_i|\lesssim  4\pi(i=1,...,5). 
 \end{eqnarray}
And the tree-level unitarity of the scattering matrix in the scalar sector of the 2HDM imposes the following constraints~\cite{Ginzburg:2005dt}:
\begin{eqnarray}
& &3(\lambda_1 + \lambda_2) \pm \sqrt{9(\lambda_1 - \lambda_2)^2 + 4(2\lambda_3 + \lambda_4)^2 } < 16\pi \ , \ \\
& &(\lambda_1 + \lambda_2) \pm \sqrt{(\lambda_1 - \lambda_2)^2 + 4 \lambda_4^2 }    < 16\pi \ , \  \\
& &(\lambda_1 + \lambda_2) \pm \sqrt{(\lambda_1 - \lambda_2)^2 + 4 \left| \lambda_5^2 \right| }    < 16\pi \ , \ \\
& &\lambda_3 + 2\lambda_4 \pm 3\left| \lambda_5 \right|
  < 8\pi \ , \  \\
& &\lambda_3 \pm \lambda_4  < 8\pi \ , \  \\
& &\lambda_3 \pm \left| \lambda_5 \right|  < 8\pi .
\end{eqnarray}
\noindent 
\end{itemize}

Vacuum stability sets a lower bound on both $\lambda v^2 \equiv m_H^2 - \frac{m_{12}^2}{\sin\beta \cos\beta} \gtrsim 0$ and the mass splitting $m_{H^\pm/A}^2 - m_H^2$~\cite{Kling:2016opi}. Unitarity and perturbativity together impose upper bounds on various parameters, including $m_{H^\pm/A}^2 - m_H^2$, $\lambda v^2$, and $\tan\beta$~\cite{Gu:2017ckc}.
%
We observe that $\tan \beta$ is strongly constrained for large $\lambda v^2$ but unbounded when $\lambda v^2 = 0$. Since the couplings of $H/A$ to fermions are proportional to $1/\tan\beta$ under the alignment limit, and given the requirement for a long-lived light scalar discussed below, a small value of $|\cos(\beta - \alpha)| \sim 0$ is necessary. Therefore, we choose $\lambda v^2 \simeq 0$ to suppress the couplings between the light Higgs bosons and fermions.

\begin{figure}[htbp]
  \centering
\includegraphics[width=0.49 \linewidth]{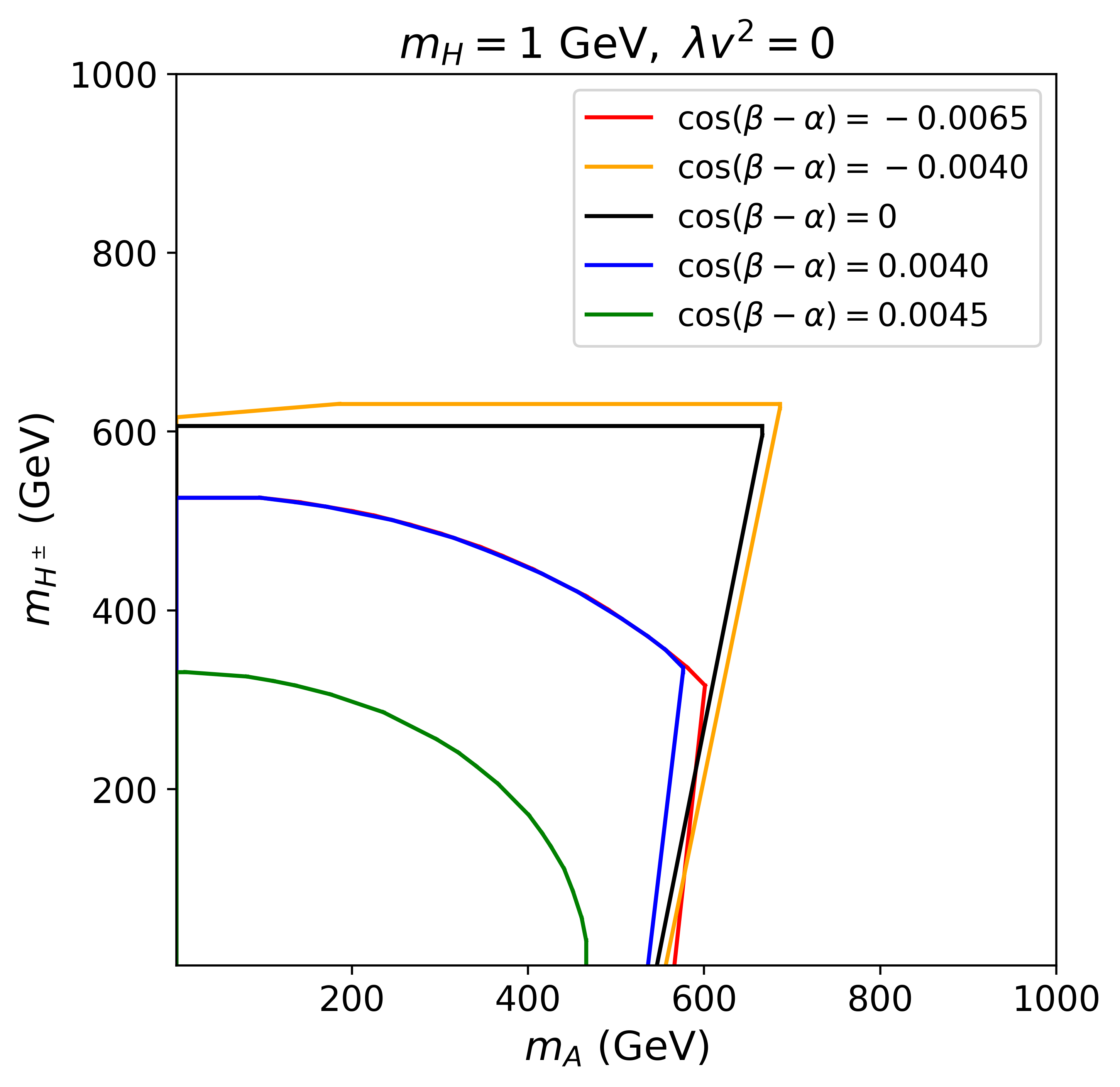}
\includegraphics[width=0.49 \linewidth]{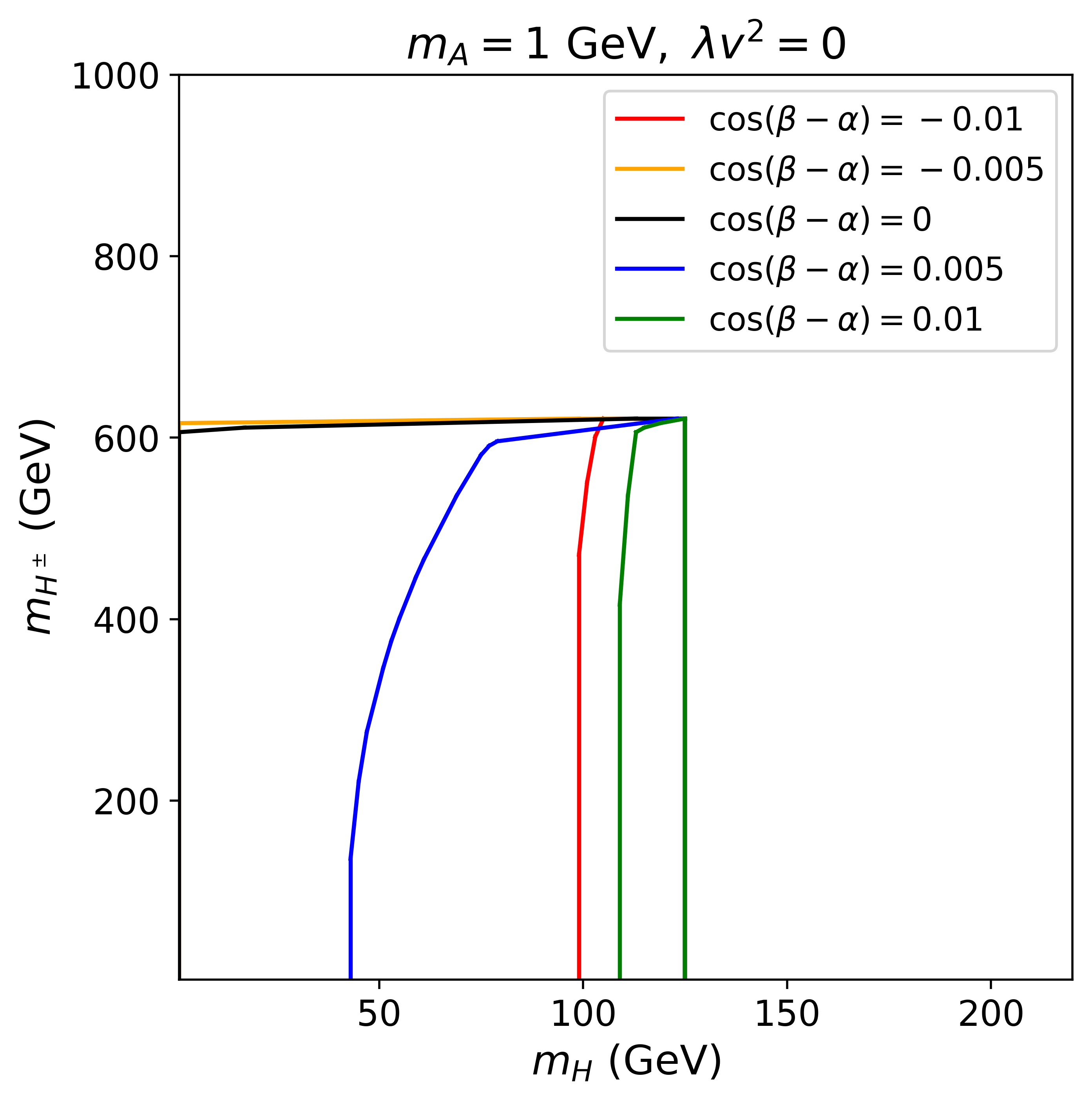}
\caption{Lower left allowed regions in the $m_{A/H}$ vs.\ $m_{H^\pm}$ plane under theoretical constraints, for $m_H = 1~\mathrm{GeV}$ (left) and $m_A = 1~\mathrm{GeV}$ (right), with $\lambda v^2 = 0$ and $\tan\beta = 1000$. The curves of different colors correspond to different values of $\cos(\beta-\alpha)$. 
Due to the varying degrees of constraint imposed by theoretical constraints, when $\cos(\beta-\alpha) \leq -0.007$ or $\cos(\beta-\alpha) \geq 0.005$, no allowed region exists for the light H scenario. Similarly when $|\cos(\beta-\alpha)| > 0.03$ no allowed region exists in the light $A$ scenario.}
  \label{fig:lam}
\end{figure}
For Type-I 2HDM with $\tan\beta > 10$, the viable range is $|\cos(\beta-\alpha)| < 0.35$~\cite{Kling:2020hmi}. To explore the impact of $\cos(\beta - \alpha)$ on scenarios with light non-SM Higgs, in \autoref{fig:lam}, we show the allowed lower left regions in the $m_{A/H}$ vs. $m_{H^\pm}$ plane under theoretical constraints, for light $m_H = 1~\mathrm{GeV}$ in the left panel with $\cos (\beta -\alpha)$ = -0.0065 (red), -0.004 (orange), 0 (black), 0.004 (blue), 0.0045 (green) and $m_A = 1~\mathrm{GeV}$ in the right panel with $\cos (\beta -\alpha)$ = -0.01 (red), -0.005 (orange), 0 (black), 0.005 (blue), 0.01 (green). We fix $\lambda v^2 = 0$ and $\tan\beta = 1000$.
In the left panel of \autoref{fig:lam}, in the alignment limit $\cos(\beta-\alpha)=0$ (black) and for $\cos(\beta-\alpha)=-0.004$ (orange), the allowed regions are maximized, with $m_{A/H^\pm}$ reaching about $600~\mathrm{GeV}$. In particular, for $\cos(\beta-\alpha)=-0.004$ (orange), when $m_A$ increases to about $180~\mathrm{GeV}$, the range of $m_{H^\pm}$ slightly expands, since $\lambda_4 \sim (m_A^2 - m_{H^\pm}^2)/v^2$ and slightly larger $m_{H^\pm}$ can satisfy the perturbativity condition $|\lambda_4| < 4\pi$.
For $\cos(\beta-\alpha)=-0.0065$ (red)  and $0.004$ (blue), the allowed regions shrink significantly. As $m_A$ increases, the viable range of $m_{H^\pm}$ gradually decreases because larger $m_{H^\pm}$ no longer satisfy the unitarity condition. For $\cos(\beta-\alpha)=0.0045$ (green), the allowed region further shrinks for the same reason. When $\cos(\beta-\alpha) \leq -0.007$ or $\cos(\beta-\alpha) \geq 0.005$, no allowed region exists. 

In the right panel of \autoref{fig:lam}, in the alignment limit $\cos(\beta-\alpha)=0$ (black) and for $\cos(\beta-\alpha)=-0.005$ (orange), the allowed region becomes largest, with $m_H \leq 125~\mathrm{GeV}$ and $m_{H^\pm}$ reaching $600~\mathrm{GeV}$. For $\cos(\beta-\alpha)=-0.005$ (orange), the range of $m_H$ is reduced to $43~\mathrm{GeV} \leq m_H \leq 125~\mathrm{GeV}$ due to the unitarity constraint, and the upper limit of $m_{H^\pm}$ increases with $m_H$.
For $\cos(\beta-\alpha)=-0.01$ (red), unitarity further limits $m_H$ to $99~\mathrm{GeV} \leq m_H \leq 125~\mathrm{GeV}$, while for $\cos(\beta-\alpha)=0.01$ (green), the range of $m_H$ is further reduced to $109~\mathrm{GeV} \leq m_H \leq 125~\mathrm{GeV}$; in both cases, $m_{H^\pm}$ can still reach $600~\mathrm{GeV}$. When $|\cos(\beta-\alpha)| > 0.03$, there is no allowed region.

By considering theoretical constraints, the weakly coupled light neutral scalars are allowed in two scenarios with $\cos(\beta - \alpha) \simeq 0$ and $\lambda v^2 = 0$:
\begin{align}
m_H \sim 0 & : \quad m_{A/H^\pm} \lesssim 600 ~\mathrm{GeV} \label{eq:scan_H} \\
m_A \sim 0 & : \quad m_{H^\pm} \lesssim 600 ~\mathrm{GeV}, \quad m_H \lesssim m_h. \label{eq:scan_A}
\end{align}
%
%
%
\subsection{Experimental constraints}
\subsubsection{Direct searches at LEP and LHC}
\label{Sec:LEPLHC}
The Large Electron-Positron Collider (LEP) searches for charged Higgs bosons via $e^+e^- \to H^+H^-$, setting a lower bound of $m_{H^\pm} > 80~\mathrm{GeV}$~\cite{ALEPH:2013htx}. It also searches for light BSM Higgs bosons through $e^+e^- \to HZ$ and $e^+e^- \to AH$, with $m_H + m_A > 209~\mathrm{GeV}$ required by the $AH$ production~\cite{ALEPH:2006tnd}.

The LHC has searched for exotic Higgs bosons through various decay channels, including $A/H \to \mu\mu$~\cite{CMS:2019mij,ATLAS:2019odt}, $A/H \to bb$~\cite{CMS:2018hir, CMS:2025frm, ATLAS:2019tpq}, $A/H \to \tau\tau$~\cite{CMS:2018rmh, CMS:2019hvr, ATLAS:2020zms}, $A/H \to \gamma\gamma$~\cite{ATLAS:2014jdv, ATLAS:2024bjr, CMS:2018cyk, CMS:2024nht}, $A/H \to tt$~\cite{CMS:2019pzc}, $H \to ZZ$~\cite{CMS:2018amk, ATLAS:2017tlw}, $H \to WW$~\cite{ATLAS:2017uhp, CMS:2019bnu}, $H \to hh$~\cite{CMS:2017yfv, CMS:2018ipl, ATLAS:2015sxd, ATLAS:2024ish}, $A \to hZ \to bb\ell\ell$~\cite{CMS:2015flt, CMS:2019qcx, ATLAS:2015kpj, ATLAS:2022enb}, $A \to \tau\tau\ell\ell$~\cite{ATLAS:2015kpj, CMS:2015uzk, CMS:2025bvl}, and $A/H \to HZ/AZ \to bb\ell\ell$~\cite{ATLAS:2018oht, CMS:2019ogx, ATLAS:2020gxx}, $A/H \to HZ/AZ \to \tau\tau\ell\ell$~\cite{CMS:2016xnc}. These searches have excluded part of the parameter space in the Type-I 2HDM. 
For the degenerate case with $m_A = m_H$, the channels $H/A \to \tau\tau$ and $\gamma\gamma$ exclude the low-mass region $m_{A,H} < 2m_t$ for $\tan\beta < 3$. 
For the non-degenerate case, the channels $H/A \to \tau\tau$ and $\gamma\gamma$ exclude the $\tan\beta < 1$ region. For $m_A - m_H = m_{H^\pm} - m_H = 200~\mathrm{GeV}$, the $A \to HZ$ channel excludes the region $m_{A,H} < 2m_t$ when $\tan\beta < 5$. Top quarks related searches, $4t$ and $A/H \to tt$, exclude $m_H < 800~\mathrm{GeV}$ for $\tan\beta < 0.3$ and $m_H < 650~\mathrm{GeV}$ for $\tan\beta < 1.1$. 
At low masses $m_{A,H} < m_h/2$, the BSM Higgs bosons can be produced via the SM Higgs decay $h \to AA,\,HH$. We will introduce these exotic decays later at Sec.~\ref{sec:case} in detail.
A complete recasting of the LHC direct searches on neutral and charged Higgses in the 2HDM can be found in Refs.~\cite{Kling:2020hmi, Su:2020pjw, Li:2024kpd}.
%
\subsubsection{Flavour constrains}
Precise flavor observables, such as the branching ratios of $B$ meson decays $B \to X_s \gamma$ and $B_{s,d} \to \mu^+ \mu^-$, impose strong constraints on the mass of the charged Higgs and the value of $\tan\beta$. However, the interpretation of these constraints depends on the specific model. In the Type-II and Type-F 2HDMs, $\text{BR}(B \to X_s \gamma)$ sets a lower bound of $m_{H^\pm} \gtrsim 800~\mathrm{GeV}$~\cite{Atkinson:2021eox, Misiak:2020vlo}. In contrast, in the Type-I 2HDM, flavor constraints primarily affect the low $\tan\beta$ region. The strongest constraint comes from $\text{BR}(B_d \to \mu^+ \mu^-)$, which excludes the region $\tan\beta < 3$ for $m_{H^\pm} = 100~\mathrm{GeV}$. As the charged Higgs mass increases, the constraints weaken, for instance, at $m_{H^\pm} = 800~\mathrm{GeV}$, only $\tan\beta < 1.2$ is excluded.
%
\subsubsection{Other light scalar searches}
\label{sec:light_cons}
There are also existing experimental constraints directly on a light (pseudo)scalar. Here, we summarize several representative examples, which will be used to further constrain the mass and interactions of the light scalar complementarily when we discuss the reach of FASER in \autoref{sec:faser}.

The CHARM fixed-target beam-dump experiment at the CERN SPS, using a $400~\mathrm{GeV}$ proton beam on a copper target, has been used to probe axion-like particles and set limits on light scalars~\cite{CHARM:1985anb, Winkler:2018qyg, Gorbunov:2021ccu}. Supernova (SN) observations, in particular from SN1987A, impose strong constraints through nucleon bremsstrahlung $NN \to NN A$, since excessive scalar emission would shorten the neutrino burst duration~\cite{Turner:1987by, Ellis:1987pk, Krnjaic:2015mbs, Batell:2019nwo, Dev:2020eam, Balaji:2022noj}. Meson decays also provide stringent bounds. $B$ meson decays, including $B \to K^*\phi(\mu^+\mu^-)$~\cite{LHCb:2015nkv} and $B^+ \to K^+\chi(\mu^+\mu^-)$~\cite{LHCb:2016awg}, as well as the measured branching ratio $\text{Br}(B \to X_s \nu \bar{\nu}) = 6.4 \times 10^{-4}$~\cite{BaBar:2013npw}, restrict (pseudo)scalar production below the $B$ threshold. Similarly, kaon decays such as $K^+ \to \pi^+ X(\nu\bar{\nu})$ at NA62~\cite{NA62:2021zjw}, $K^+ \to \pi^+\chi(e^+ e^-)$ at MicroBooNE~\cite{MicroBooNE:2021sov}, and $K^+ \to \pi^+ X$ at E949~\cite{BNL-E949:2009dza} provide constraints based on the light scalar decay lifetime. $D$ meson decay constraints are usually very weak, with relevant limits reported by the PDG~\cite{ParticleDataGroup:2022pth} and LHCb~\cite{LHCb:2020car}. Finally, LEP searches for $e^+e^- \to Z^*\phi$ at OPAL, ALEPH, and L3~\cite{L3:1996ome, ALEPH:1993sjl, OPAL:2007qwz}, considering both invisible and prompt decays of $\phi$, also set strong bounds on light scalar particles~\cite{Clarke:2013aya, Winkler:2018qyg}. 

\subsubsection{Higgs invisible decay and exotic decay}
\label{sec:case}

To explore light A and H cases, Higgs invisible decay and exotic decay may get involved. The existence of a particle with a mass below $m_h/2$ can contribute to extra Higgs decay channels like $h \to HH / AA$. The explicit form of the branching fraction can be expressed as
\begin{equation}
\begin{aligned}
\text{Br}(h \to HH / AA) &= \frac{\Gamma(h \to HH / AA)}{\Gamma_h} \approx \frac{1}{\Gamma_h^{\text{SM}}} \frac{g_{hHH / hAA}^2}{8 \pi m_h^2} \left( 1 - \frac{4m_{H/A}^2}{m_h^2} \right)^{1/2} \\
&\approx 4700 \cdot \left( \frac{g_{hHH / hAA}}{v} \right)^2,
\end{aligned}
\end{equation}
where the tri-linear couplings in the 2HDM are given as,
\begin{equation}
\begin{aligned}
g_{hHH} &= \frac{s_{\beta - \alpha}}{2v} \left[ \left( m_H^2 - 3 \lambda v^2 - m_h^2 \right) \left( 2 t_{2 \beta}^{-1} s_{\beta - \alpha} c_{\beta - \alpha} - c_{\beta - \alpha}^2 + s_{\beta - \alpha}^2 \right) + \left( \lambda v^2 - m_H^2 \right) \right], \\
g_{hAA} &= \frac{1}{2v} \left[ \left( 2 m_H^2 - 2 \lambda v^2 - 2 m_A^2 - m_h^2 \right) s_{\beta - \alpha} + 2 \left( m_H^2 - \lambda v^2 - m_h^2 \right) t_{2 \beta}^{-1} c_{\beta - \alpha} \right].
\label{eq:invisible couplings}
\end{aligned}
\end{equation}

If $A/H$ is sufficiently long-lived, it can pass through the main detectors at the LHC without leaving a signal. Therefore the exotic decay channels of $h \to HH / AA$ contribute to the Higgs invisible decay width, which is constrained $\text{Br}(h \to HH / AA) < 0.107$ by the current Higgs measurements~\cite{ATLAS:2023tkt, CMS:2023sdw}. This constraint can be easily satisfied by suppressed values of $g_{hHH}$ or $g_{hAA}$. Specifically, under the approximations of large $\tan \beta$, and small $\lambda v^2$ and $m_{H/A}$, to have a negligible Higgs invisible decay contribution, the following conditions can be imposed,
\begin{itemize}
\item Light $H$:
\begin{equation}
\cos(\beta - \alpha) = \frac{1}{2 \left( \frac{1}{\tan^2 \beta} + 1 \right)}  \sqrt{2 + \frac{4}{\tan^2 \beta} +  \frac{2}{\tan^4 \beta} + 2 \left( \frac{\cos \beta}{\tan \beta} - \sin \beta \right)} \approx \frac{1}{\tan \beta},  
\label{eq:lightH_invisible}
\end{equation}  
\item Light $A$:
\begin{equation}
\begin{aligned}
\cos(\beta - \alpha) &= \frac{1}{\tan\beta} \frac{2m_H^2 - m_h^2}{m_H^2 - m_h^2} 
\frac{1}{\sqrt{1 - \frac{2}{\tan^2\beta} + \left(\frac{2m_H^2 - m_h^2}{m_H^2 - m_h^2}\right)^2 
\frac{1}{\tan^2\beta} + \frac{1}{\tan^4\beta}}} \\
&\approx \frac{1}{\tan \beta} \frac{2 m_H^2 - m_h^2}{m_H^2 - m_h^2}. 
\end{aligned}
\label{eq:lightA_invisible}
\end{equation}
\end{itemize}
For the light $H$, $g_{hHH}$ is determined solely by $\tan\beta$ and $\cos(\beta-\alpha)$. In contrast, for the light $A$ with the same approximation, $g_{hAA}$ depends not only on $\tan\beta$ and $\cos(\beta-\alpha)$ but also on the value of $m_H$. 

Besides the light particle (either $H$ or $A$), there is also an extra neutral scalar ($A$ if $H$ is the light particle or vice versa) which could further contribute to Higgs exotic decays if kinematically allowed. Hence when $m_{A/H} < m_{h/2}$ for a light $H/A$, the exotic decay searches $h \to AA/HH \to ffff$ and $4\gamma$ should also be considered. 
We mainly refer to current LHC searches for such channels, including $h \to HH/AA \to bbbb$~\cite{ATLAS:2018pvw, CMS:2024zfv}, $h \to HH/AA \to bb\tau\tau$~\cite{CMS:2018zvv}, $h \to HH/AA \to \mu\mu\mu\mu$~\cite{CMS:2018jid}, $h \to HH/AA \to bb\mu\mu$~\cite{ATLAS:2021hbr, CMS:2018nsh}, $h \to HH/AA \to \tau\tau\tau\tau$~\cite{CMS:2019spf}, $h \to HH/AA \to \tau\tau\mu\mu$~\cite{CMS:2018qvj} and $h \to HH/AA \to \gamma\gamma\gamma\gamma$~\cite{CMS:2022fyt}. These searches typically focus on particles with masses above $4$–$5~\mathrm{GeV}$. According to the mass hierarchy, we divide the following discussion on the constraints from Higgs invisible and exotic decays into 5 cases. 

\begin{itemize}
\item \textbf{Case 0: Both $m_A$ and $m_H$ are light}
\label{itm:case0}
\end{itemize}
We first consider the case where both $H$ and $A$ are very light with $m_A, m_H < 5~\mathrm{GeV}$. However, this region of parameter space has already been excluded by EWPO, as discussed at the end of Sec.\ref{sec:ewpo}, and the corresponding results are shown in \autoref{fig:EWPO_ex}. Therefore, this scenario will not be considered in the following analysis.

\begin{itemize}
\item \textbf{Case 1: Light $H$ with $m_A \in (5 \text{~GeV}, m_h/2)$}
\end{itemize}
\begin{figure}[htbp]
    \centering
    \includegraphics[width=0.49\linewidth]{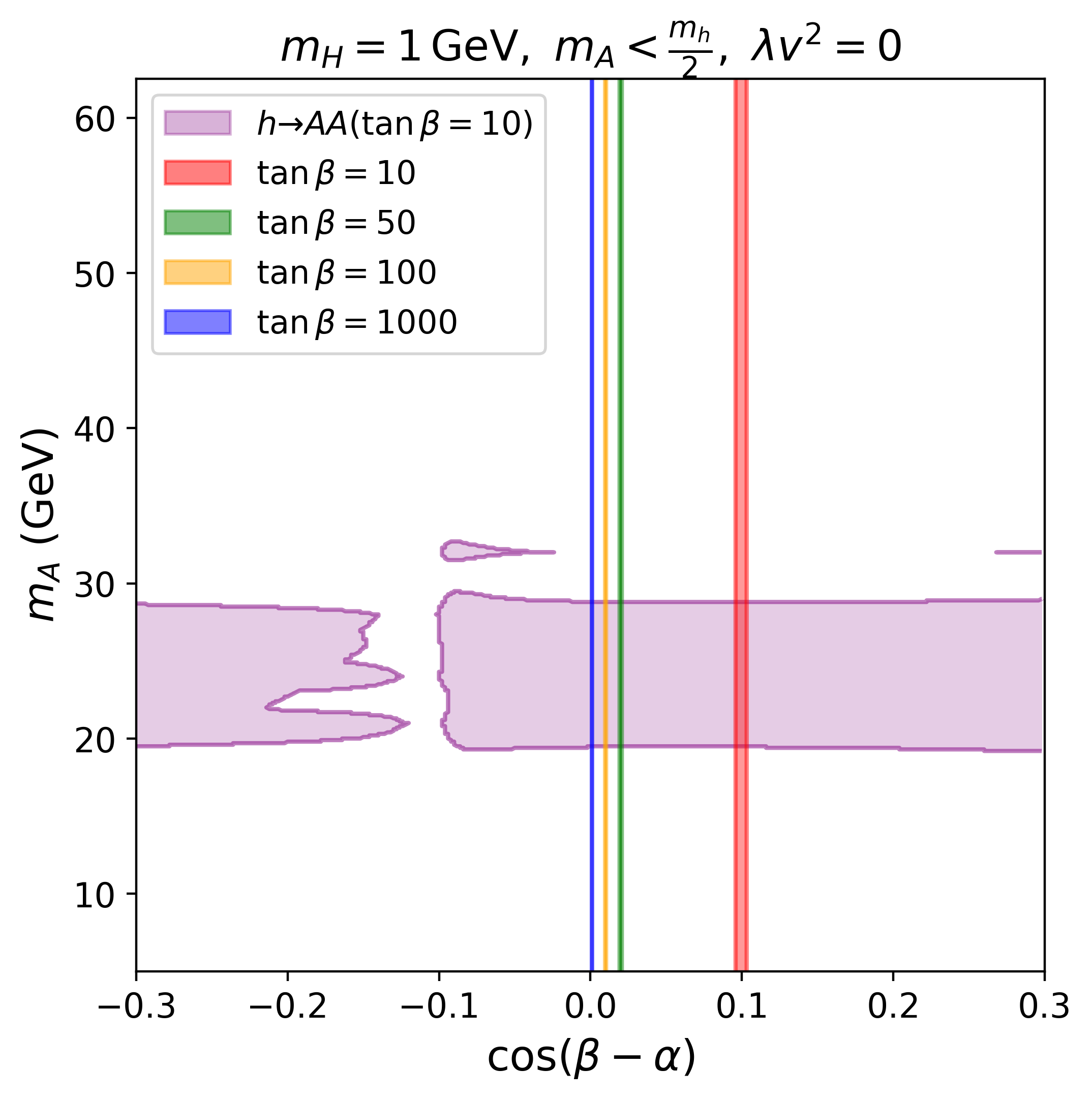}    \includegraphics[width=0.49\linewidth]{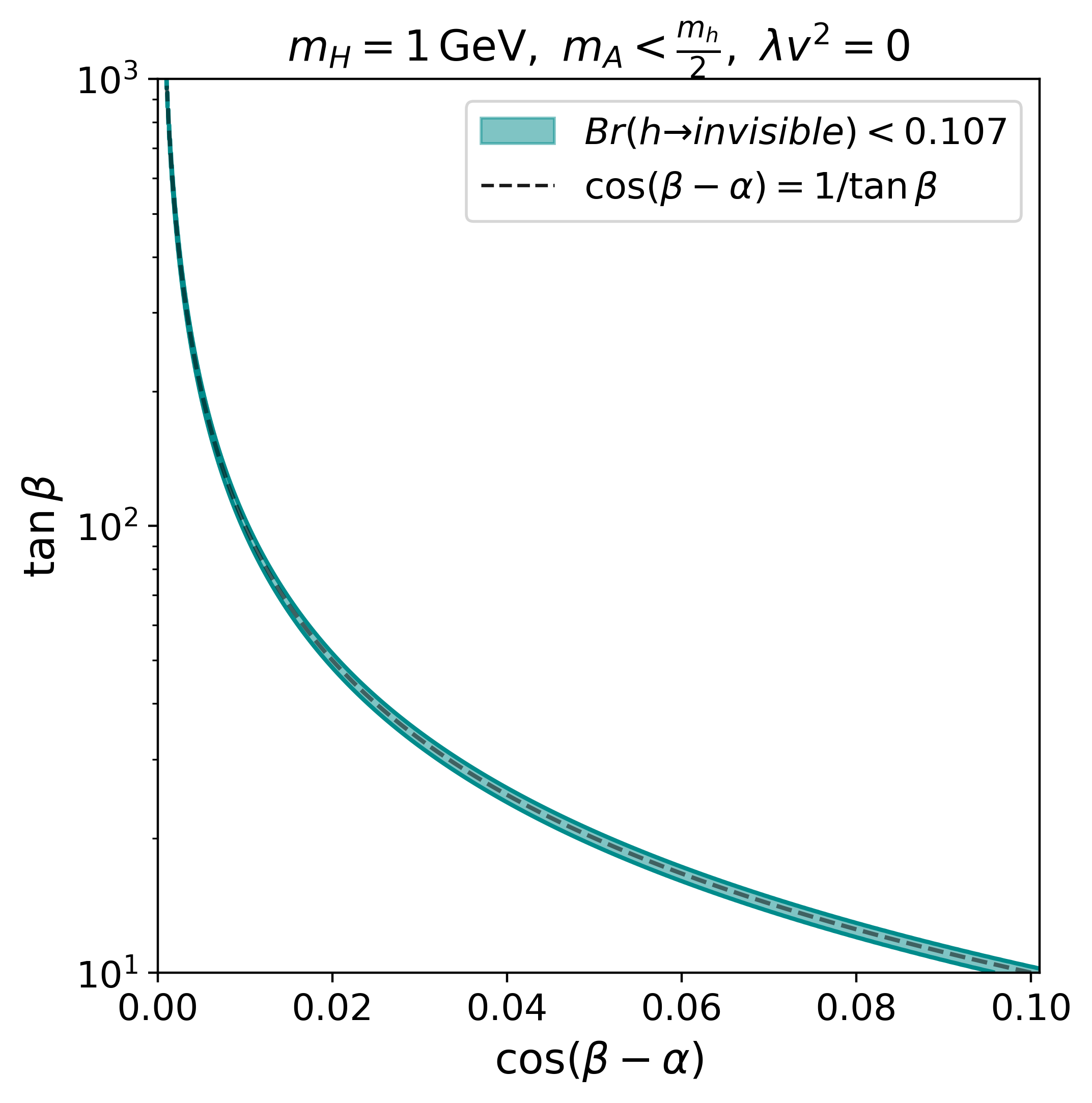}
    \caption{(Left Panel): Allowed region in the $\cos(\beta - \alpha)$ vs. $m_{A}$ plane, under the constraint of invisible Higgs decay $\mathrm{Br}(h \to \text{invisible}) < 0.107$, for $m_H = 1~\mathrm{GeV}$ and $m_A < \frac{m_h}{2}$, with $\tan\beta$ = 10 (red), 50 (green), 100 (orange), 1000 (blue). The purple region is excluded by direct searches for the exotic decay $h \to AA$, with $m_H = 1~\mathrm{GeV}$, $m_{H^\pm} = 80~\mathrm{GeV}$ and $\tan\beta$ = 10. (Right Panel): The allowed region in the $\cos(\beta - \alpha)$ vs.\ $\tan\beta$ plane, under the constraint of invisible Higgs decay $\mathrm{Br}(h \to \text{invisible}) < 0.107$, for $m_H = 1~\mathrm{GeV}$ and $m_A < \frac{m_h}{2}$, with $\lambda v^2 = 0$. The dashed black line corresponds to $\cos(\beta - \alpha) = 1 / \tan\beta$. Here, we have $\lambda v^2 = 0$, and  the allowed region works for $m_A > \frac{m_h}{2}$.}
    \label{fig:llp_H_ie}
\end{figure}
We now consider the case where the SM-like Higgs decays into both $HH$ and $AA$, for the light $H$. In this case, we take into account constraints from the invisible Higgs decay, $\mathrm{Br}(h \to \text{invisible}) = \mathrm{Br}(h \to HH)$, and from direct searches for the exotic decays $h \to AA \to ffff$. In the left panel of~\autoref{fig:llp_H_ie}, we show the allowed parameter space in the $\cos(\beta - \alpha)$ vs. $m_A$ plane, under the constraint $\mathrm{Br}(h \to \text{invisible}) < 0.107$, for $m_H = 1~\mathrm{GeV}$ and $m_A \in (5 \text{~GeV}, m_h/2)$, with $\tan\beta = 10$ (red), 50 (green), 100 (orange), and 1000 (blue). The purple region indicates the exclusion region from direct searches for $h \to AA \to ffff$ in the Type-I 2HDM, where $m_H = 1~\mathrm{GeV}$, $m_{H^\pm} = 80~\mathrm{GeV}$, and $\tan\beta = 10$. Here, we fix $\lambda v^2 = 0$. We find that for any value of $ \tan\beta $, the region $ 5~\mathrm{GeV} < m_A < \frac{m_h}{2} $ is always allowed by the invisible Higgs decay constraint. For the exotic decay constraint with $ \tan\beta = 10 $, the region $ 20~\mathrm{GeV} < m_A < 30~\mathrm{GeV} $ is excluded. This constraint vanishes near $\cos(\beta - \alpha) \simeq -0.1$, because the $h \to AA$ coupling is suppressed. Due to the dependence of the process $AA \to bb\mu\mu$ on $\cos(\beta - \alpha)$, an exclusion still appears around $m_A \simeq 31~\mathrm{GeV}$. Therefore, under the invisible Higgs decay constraint with $ \tan\beta = 10 $, the allowed parameter space is $ 5~\mathrm{GeV} < m_A < 20~\mathrm{GeV} $ and $ 30~\mathrm{GeV} < m_A < \frac{m_h}{2} $. As $ \tan\beta $ increases, the exotic decay constraint quickly weakens, because the couplings of $A$ to SM particles share the same $1/\tan\beta$ dependence. Around $\tan \beta \simeq 50$, the constraints vanish. Thus, at large $ \tan\beta $, there exists viable parameter space satisfying both the invisible and exotic decay constraints, within the range $ 5~\mathrm{GeV} < m_A < \frac{m_h}{2} $. Meanwhile, the allowed values of $ \cos(\beta - \alpha) $ tend to zero as $ \tan\beta $ increases.

To further determine the specific parameter space, in the right panel of~\autoref{fig:llp_H_ie}, we show the allowed region in the $\cos(\beta - \alpha)$ vs. $\tan\beta$ plane, under the constraint $\mathrm{Br}(h \to \text{invisible}) < 0.107$, for $m_H = 1~\mathrm{GeV}$, $m_A < \frac{m_h}{2}$, and $\lambda v^2 = 0$. The dashed black line represents the relation $\cos(\beta - \alpha) = 1/\tan\beta$, as given in Eq.~\eqref{eq:lightH_invisible}. The viable parameter space is tightly concentrated near this line, independent of the non-SM Higgs mass. Considering the constraint from direct searches for exotic decays, the allowed parameter space  with $\lambda v^2 = 0$ is:
\begin{equation}
\begin{aligned}
\text{For } \tan \beta = 10:\quad
& \cos(\beta - \alpha) \simeq \frac{1}{\tan \beta},\quad
m_A \in (5,\,20) \cup \left(30,\,\frac{m_h}{2}\right)~\mathrm{GeV}, \\
\text{For } \tan \beta \gtrsim 50:\quad
& \cos(\beta - \alpha) \simeq \frac{1}{\tan \beta},\quad
m_A \in \left(5,\,\frac{m_h}{2}\right)~\mathrm{GeV},
\end{aligned}
\end{equation}

\begin{itemize}
\item \textbf{Case 2: Light $H$ with $m_A > m_h/2$}
\end{itemize}

%
We next focus the case of light $H$, where the SM-like Higgs decays only into $HH$. In this case, we only consider the constraint from the invisible Higgs decay, $\mathrm{Br}(h \to \text{invisible}) = \mathrm{Br}(h \to HH)$, which depends on $\cos(\beta - \alpha)$ and $\tan\beta$, with no constraint on $m_A$.  The allowed region is highly constrained and located near this line, as determined by Eq.~\eqref{eq:lightH_invisible}. Thus the region plot is same to the right plot of \autoref{fig:llp_H_ie}. The difference is that $m_A$ is no longer affected by $\tan\beta$ due to the absence of decays to light $AA$.

At large $\tan\beta$, $\cos(\beta - \alpha)$ naturally approaches zero. 
For \(\lambda v^2 = 0\), the parameter space satisfying the invisible Higgs decay constraint is approximately given by:
\begin{equation}
\text{For } \tan \beta \gtrsim 10:\quad
\cos(\beta - \alpha) \simeq \frac{1}{\tan \beta}, \quad
m_A \in \left(\frac{m_h}{2},\,600\right)~\mathrm{GeV}.
\end{equation}
Here we have $\tan \beta \gtrsim 10$ from other light scalar searches.

\begin{itemize}
\item \textbf{Case 3: Light $A$ with $m_H \in (5 \text{~GeV}, m_h/2)$}
\end{itemize}

\begin{figure}[htbp]
    \centering
    \includegraphics[width=0.49\linewidth]{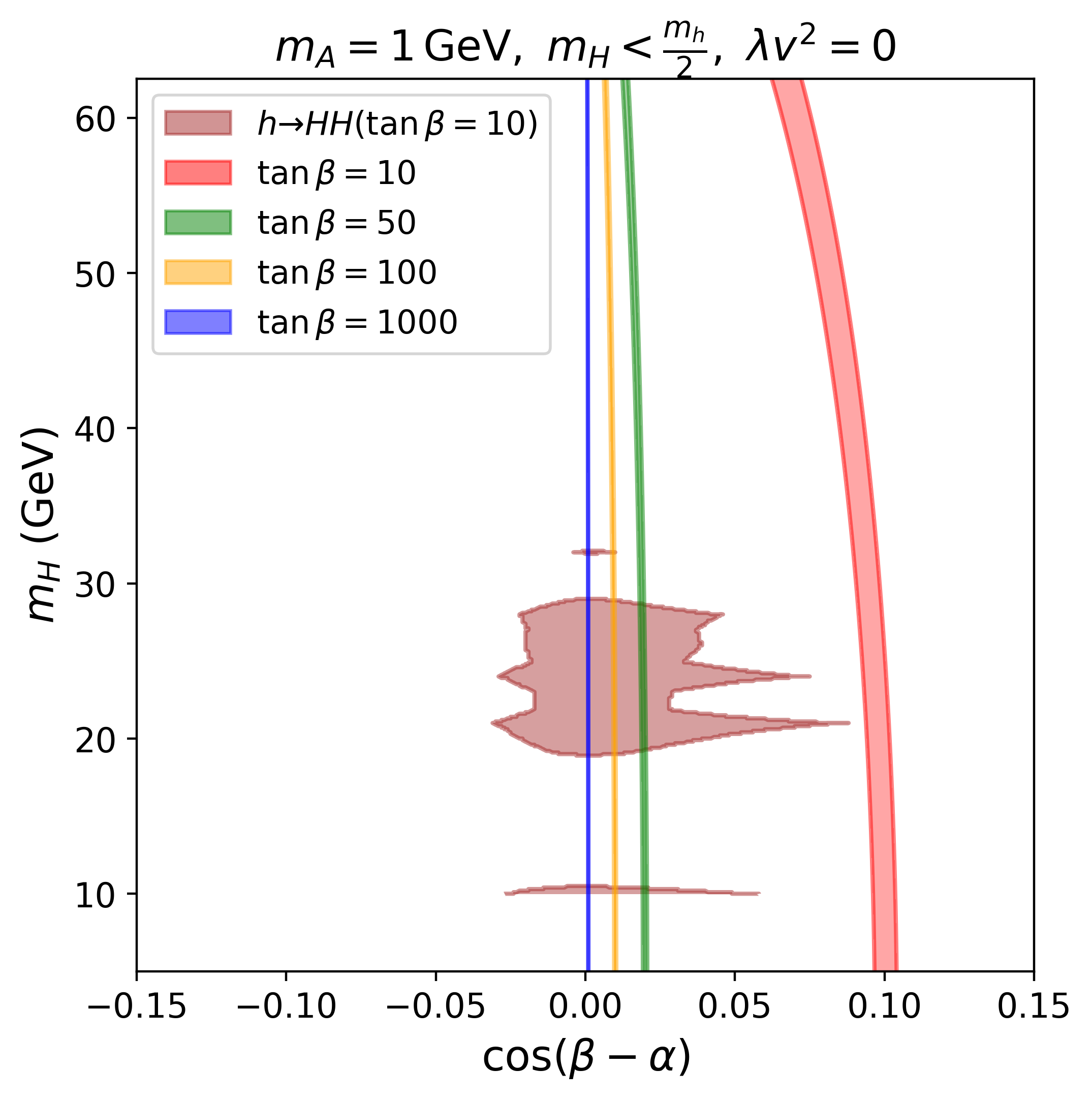}
    \includegraphics[width=0.49\linewidth]{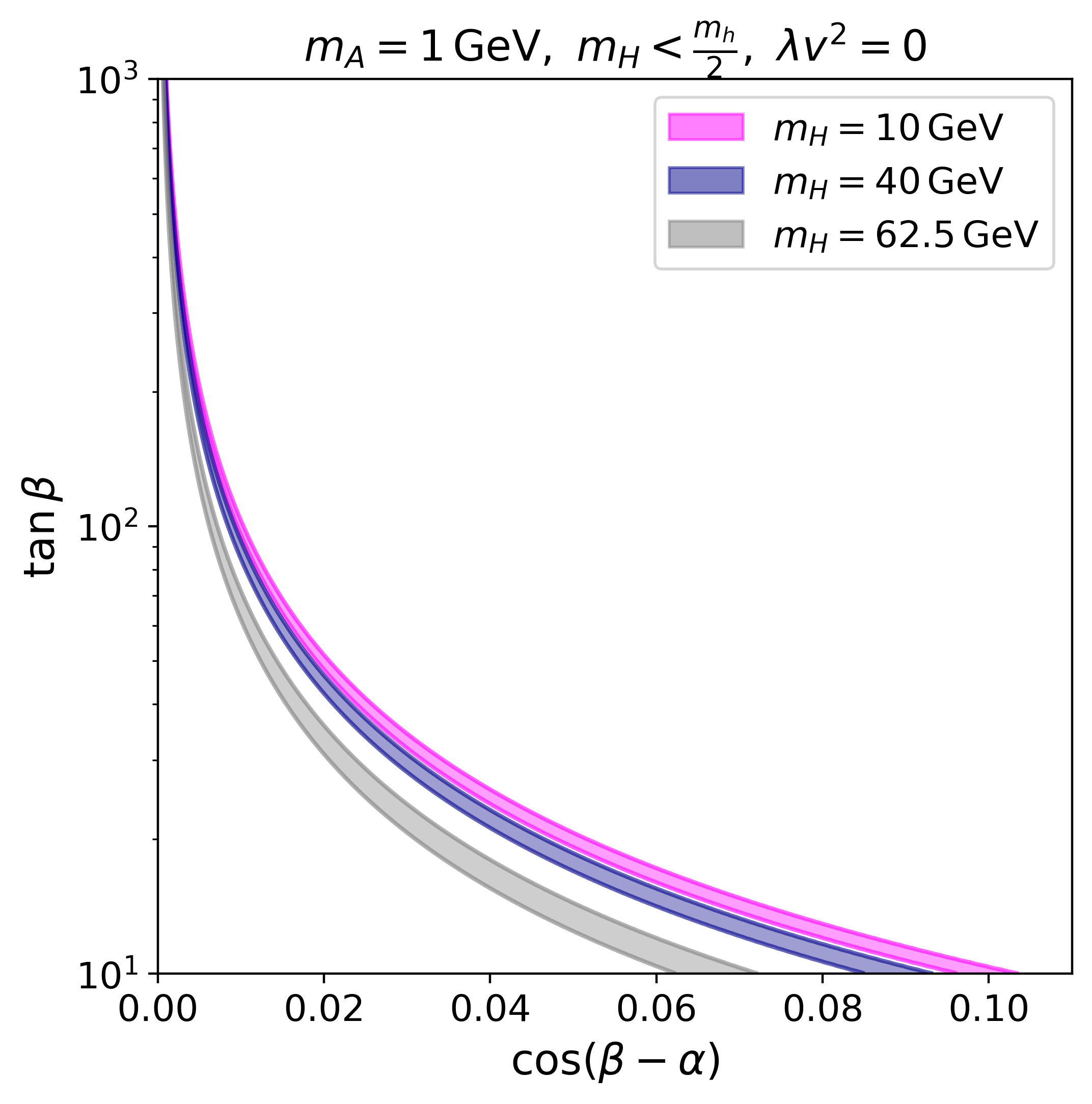}
    \caption{(Left Panel): Allowed region in the $\cos(\beta - \alpha)$ vs. $m_{H}$ plane, under the constraint of invisible Higgs decay $\mathrm{Br}(h \to \text{invisible}) < 0.107$, for $m_A = 1~\mathrm{GeV}$ and $m_H \in (5 \text{~GeV}, m_h/2)$, with $\tan\beta$ = 10 (red), 50 (green), 100 (orange), 1000 (blue). The brown region is excluded by direct searches for the exotic decay $h \to HH$ in the Type-I 2HDM, with $m_A = 1~\mathrm{GeV}$, $m_{H^\pm} = 80~\mathrm{GeV}$ and $\tan\beta$ = 10. (Right Panel): The allowed regions in the $\cos(\beta - \alpha)$ vs.\ $\tan\beta$ plane, under the invisible Higgs decay $\mathrm{Br}(h \to \text{invisible}) < 0.107$, for $m_A = 1~\mathrm{GeV}$, and $m_H = 10$ (magenta), 40 (dark blue), and 62.5 (grey)$~\mathrm{GeV}$.}
    \label{fig:llp_A_ie}
\end{figure}
We now consider the case of the light $A$, where the SM-like Higgs can decay into both $HH$ and $AA$. In this case, we not only consider constraints from the invisible Higgs decay, where $\mathrm{Br}(h \to \text{invisible}) = \mathrm{Br}(h \to AA)$, but also from direct searches for exotic decays $h \to HH \to ffff$. In the left panel of~\autoref{fig:llp_A_ie}, we show the allowed parameter space in the $\cos(\beta - \alpha)$ vs. $m_H$ plane, under the constraint $\mathrm{Br}(h \to \text{invisible}) < 0.107$, for $m_A = 1~\mathrm{GeV}$ and $m_H \in (5 \text{~GeV}, m_h/2)$ , with $\tan\beta$ = 10 (red), 50 (green), 100 (orange), and 1000 (blue). The brown region presents the exclusion region from direct searches for $h \to HH \to ffff$, for $m_A = 1~\mathrm{GeV}$, $m_{H^\pm} = 80~\mathrm{GeV}$, $\tan\beta = 10$. We fix $\lambda v^2 = 0$. We find that, for different values of $\tan\beta$, there always exists viable parameter space satisfying the invisible Higgs decay constraint, for any $m_H$ within the range $5~\mathrm{GeV} < m_H < \frac{m_h}{2}$, satisfying the condition given in Eq.~\eqref{eq:lightA_invisible}. At $\tan\beta = 10$, $m_H$ has a significant impact on $\cos(\beta - \alpha)$, and the allowed values of $\cos(\beta - \alpha)$ decrease as $m_H$ increases. In contrast, at large $\tan\beta$, the allowed values of $\cos(\beta - \alpha)$ tend to zero, and the influence of $m_H$ becomes negligible. For the exotic decay constraint with $\tan\beta = 10$, the channels $HH \to \tau \tau \tau \tau$ and $HH \to bb\mu\mu$, which depend on $\cos(\beta - \alpha)$, exclude different regions near $\cos(\beta - \alpha) \sim 0$. Specifically, $HH \to \tau \tau \tau \tau$ excludes the region at $m_H = 10~\mathrm{GeV}$, while $HH \to bb\mu\mu$ excludes the regions $19~\mathrm{GeV} \le m_H \le 29~\mathrm{GeV}$ and $m_H \simeq 32~\mathrm{GeV}$. However, these exclusions do not affect the parameter space allowed by the invisible Higgs decay constraint with same $\tan\beta$. As $\tan\beta$ increases, the couplings of $H$ to SM particles are suppressed, leading to a rapid weakening of the exotic decay constraint, which continues to have no impact on the invisible decay allowed region.

In the right panel of~\autoref{fig:llp_A_ie}, we show the allowed regions in the $\cos(\beta - \alpha)$ vs. $\tan\beta$ plane, under the invisible Higgs decay constraint $\mathrm{Br}(h \to \text{invisible}) < 0.107$, for $m_A = 1~\mathrm{GeV}$ and $\lambda v^2 = 0$, with $m_H = 10$ (magenta), 40 (dark blue), and 62.5 (grey)$~\mathrm{GeV}$. As $m_H$ increases, the allowed region shifts toward smaller values of $\cos(\beta - \alpha)$, and for large $\tan\beta$, $\cos(\beta - \alpha)$ approaches zero. Moreover, when $m_H = 10~\mathrm{GeV}$, the allowed region satisfies the relation $\cos(\beta - \alpha) = \frac{1}{\tan\beta}$. We summarize the allowed parameter space in this case, under the condition of $\lambda v^2 = 0$, as follows:
\begin{equation}
\text{For } \tan \beta \gtrsim 10: \quad \cos(\beta - \alpha) \simeq \frac{1}{\tan \beta} \frac{2 m_H^2 - m_h^2}{m_H^2 - m_h^2}, \quad 
m_H \in \left(5,\,\frac{m_h}{2}\right)~\mathrm{GeV}.
\end{equation}
In particular, for $5~\mathrm{GeV} < m_H < 25~\mathrm{GeV}$, a special benchmark scenario is given by
\begin{equation}
\text{For } \tan \beta \gtrsim 10: \quad \cos(\beta - \alpha) \simeq \frac{1}{\tan \beta}, \quad 
m_H \in (5,\,25)~\mathrm{GeV}.
\end{equation}

\begin{itemize}
\item \textbf{Case 4: Light $A$ with $m_H > m_h/2$}
\end{itemize}
\begin{figure}[htbp]
    \centering
    \includegraphics[width=0.49\linewidth]{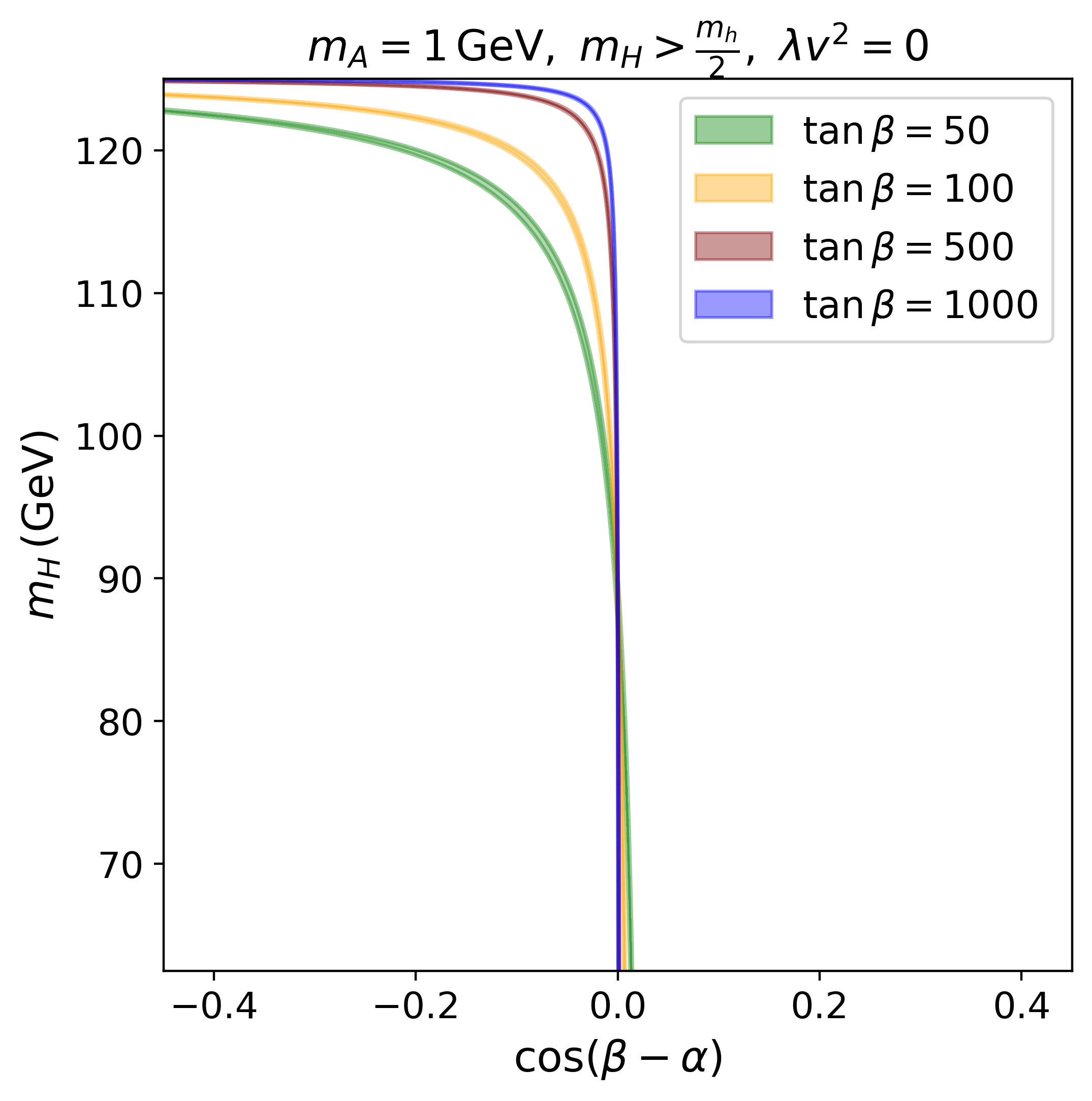}
    \includegraphics[width=0.49\linewidth]{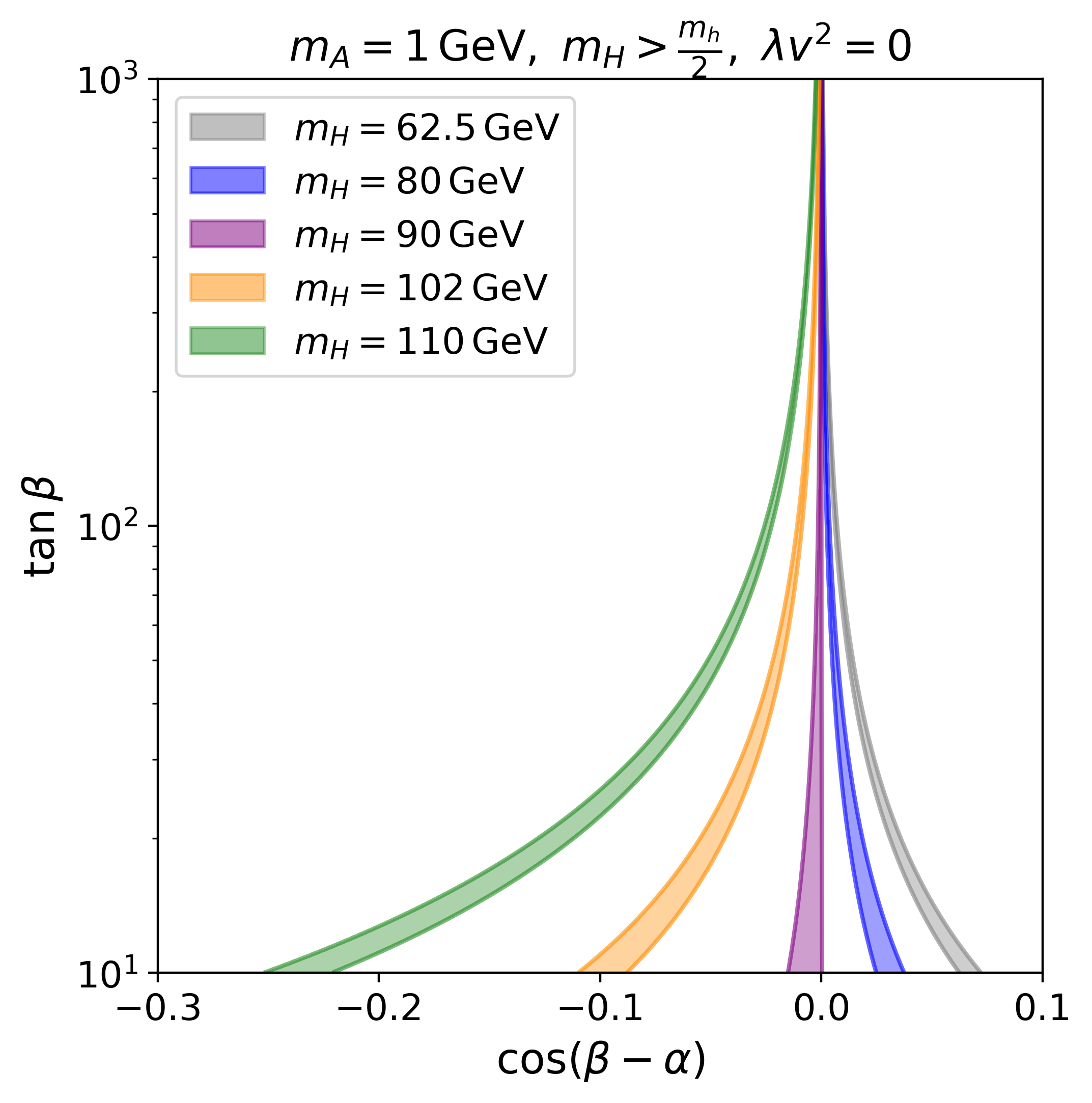}
    \caption{(Left Panel): The allowed regions in the $\cos(\beta - \alpha)$ vs. $m_H$ plane, under the constraint $\mathrm{Br}(h \to \text{invisible}) < 0.107$, for $m_A = 1~\mathrm{GeV}$, $m_H > m_h / 2$, with $\tan\beta = 50$ (green), 100 (orange), 500 (maroon), 1000 (blue). (Right Panel): The allowed regions in the $\cos(\beta - \alpha)$ vs.\ $\tan\beta$ plane, under the invisible Higgs decay $\mathrm{Br}(h \to \text{invisible}) < 0.107$, for $m_A = 1~\mathrm{GeV}$, and $m_H$ = 62.5 (grey), 80 (blue), 90 (purple), 102 (orange), 110 (green)$~\mathrm{GeV}$. Here, $\lambda v^2 = 0$.}
    \label{fig:llp_A_p}
\end{figure}

We finally consider the case of light $A$, where the SM-like Higgs decays only into $AA$. In this case, only consider the constraint from the invisible Higgs decay, $\mathrm{Br}(h \to \text{invisible}) = \mathrm{Br}(h \to AA)$, which depends on the parameters $\cos(\beta - \alpha)$, $\tan\beta$, and $m_H$. In the left panel of~\autoref{fig:llp_A_p}, we show the allowed regions in the $\cos(\beta - \alpha)$ vs.\ $m_H$ plane, under the invisible Higgs decay constraint, for $m_A = 1~\mathrm{GeV}$, $m_H > \frac{m_h}{2}$, and $\lambda v^2 = 0$, with $\tan\beta = 50$ (green), 100 (orange), 500 (magenta), 1000 (blue). We find that for various values of $\tan\beta$, there exists viable parameter space, for any $m_H$ in the range $\frac{m_h}{2} < m_H < m_h$, and as $\tan\beta$ increases, the allowed regions approach $\cos(\beta - \alpha) = 0$, consistent with the theoretical requirement for accommodating light long-lived particles.

To explore the parameter space for different values of $m_H$, in the right panel of~\autoref{fig:llp_A_p}, we show the allowed regions in the $\cos(\beta - \alpha)$ vs.\ $\tan\beta$ plane, under the constraint from the invisible Higgs decay branching fraction, for $m_A = 1~\mathrm{GeV}$, $\lambda v^2 = 0$, and $m_H$ = 62.5 (grey), 80 (blue), 90 (purple), 102 (orange), 110 (green)$~\mathrm{GeV}$. We find that as $m_H$ increases, the allowed regions shift toward smaller values of $\cos(\beta - \alpha)$, and for large $\tan\beta$, $\cos(\beta - \alpha)$ approaches zero, consistent with the theoretical requirements for accommodating light long-lived particles. Meanwhile, different values of $m_H$ can realize either the alignment limit $\cos(\beta - \alpha) = 0$ or the special relation between $\cos(\beta - \alpha)$ and $\tan\beta$. As shown in Eq.~\eqref{eq:lightA_invisible}, the alignment limit $\cos(\beta - \alpha) = 0$ can be realized around $m_H = \frac{m_h}{\sqrt{2}} \approx 88~\mathrm{GeV}$, while the relation $\cos(\beta - \alpha) = -\frac{1}{\tan\beta}$ can be achieved around $m_H = \sqrt{\frac{2}{3}}\,m_h \approx 102~\mathrm{GeV}$. The allowed parameter space in this case, for $\lambda v^2 = 0$, can be summarized as follows:
\begin{equation}
\text{For } \tan \beta \gtrsim 10: \quad \cos(\beta - \alpha) \simeq \frac{1}{\tan \beta} \frac{2 m_H^2 - m_h^2}{m_H^2 - m_h^2}, \quad 
m_H \in \left(\frac{m_h}{2},\,m_h\right).
\end{equation}
In particular, for a light $A$, we have two special regions:
    \begin{equation}
    \text{For } \tan \beta \gtrsim 10: \quad \cos(\beta - \alpha) \simeq 0,\quad
    m_H \in (86.9,\,90)~\mathrm{GeV},
    \end{equation}
    \begin{equation}
    \text{For } \tan \beta \gtrsim 10: \quad \cos(\beta - \alpha) \simeq -\frac{1}{\tan\beta},\quad
    m_H \in (101.2,\,103)~\mathrm{GeV}.
    \end{equation}

\subsubsection{Electroweak precision measurements}
\label{sec:ewpo}

\begin{table}[tb]
\centering
\resizebox{\textwidth}{!}{
  \begin{tabular}{|l|c|r|r|r|c|r|r|r|c|r|r|r|c|r|r|r|}
   \hline
    & \multicolumn{4}{c|}{2018~\cite{Chen:2018shg}} & \multicolumn{4}{c|}{2020~\cite{stupdg:2020}} & \multicolumn{4}{c|}{2022~\cite{stupdg:2022}} & \multicolumn{4}{c|}{2024~\cite{stupdg:2024}} \\
   \hline
   \multirow{2}{*}{} & \multirow{2}{*}{$\sigma$} & \multicolumn{3}{c|}{correlation} & \multirow{2}{*}{$\sigma$} & \multicolumn{3}{c|}{correlation} & \multirow{2}{*}{$\sigma$} & \multicolumn{3}{c|}{correlation} & \multirow{2}{*}{$\sigma$} & \multicolumn{3}{c|}{correlation} \\
   \cline{3-5}\cline{7-9}\cline{11-13}\cline{15-17}
   & & $S$ & $T$ & $U$ & & $S$ & $T$ & $U$ & & $S$ & $T$ & $U$ & & $S$ & $T$ & $U$ \\
   \hline
   $S$ & $0.04 \pm 0.11$ & $1$ & $0.92$ & $-0.68$ & $-0.01 \pm 0.10$ & $1$ & $0.92$ & $-0.80$ & $-0.02 \pm 0.10$ & $1$ & $0.92$ & $-0.80$ & $-0.04 \pm 0.10$ & $1$ & $0.93$ & $-0.70$ \\
   \hline
   $T$ & $0.09 \pm 0.14$ & $-$ & $1$ & $-0.87$ & $0.03 \pm 0.12$ & $-$ & $1$ & $-0.93$ & $0.03 \pm 0.12$ & $-$ & $1$ & $-0.93$ & $0.01 \pm 0.12$ & $-$ & $1$ & $-0.87$ \\
   \hline
   $U$ & $-0.02 \pm 0.11$ & $-$ & $-$ & $1$ & $0.02 \pm 0.11$ & $-$ & $-$ & $1$ & $0.01 \pm 0.11$ & $-$ & $-$ & $1$ & $-0.01 \pm 0.09$ & $-$ & $-$ & $1$ \\
   \hline
  \end{tabular}
}
\caption{S, T, U ranges and correlation matrices $\rho_{ij}$ from 2018 to 2024 $Z$-pole precision measurements.}
\label{tab:STU}
\end{table}
In Table~\ref{tab:STU}, we summarize the Z-pole precision measurements of the oblique parameters $S$, $T$, and $U$, and the correlations among them. Based on both the 2018 and 2024 fits, we scan the 2HDM parameter space as defined in Eq.~\eqref{eq:scan_H} and Eq.~\eqref{eq:scan_A}, analyzing the impacts of each parameter on the 95\%\ C.L. global fit. Note that the 2018 fit is based on the LEP-I results~\cite{ALEPH:2005ab}, which are stronger than the PDG fit~\cite{stupdg:2018}; the 2020 and 2022 fits are very similar to the 2024 fit.

We perform a $\chi^2$ fit to the oblique parameters $S$, $T$, and $U$ using a profile-likelihood method based on $Z$-pole precision measurements. The $\chi^2$ function is defined as
\begin{align}
\chi^{2} \equiv \sum_{ij} (X_i - \hat{X}_i) (\sigma^2)^{-1}_{ij} (X_j - \hat{X}_j),
\label{eq:chi}
\end{align}
where $X_i = (\Delta S, \Delta T, \Delta U)_{\mathrm{2HDM}}$ are the 2HDM predictions, and $\hat{X}_i = (\Delta S, \Delta T, \Delta U)$ the best-fit values. The $\sigma_{ij}$ are the error elements, constructed from $\sigma_i$ and $\rho_{ij}$, as provided in Table~\ref{tab:STU}. The 95\% C.L. region corresponds to $\Delta \chi^2 \equiv \chi^2 - \chi^2_{\text{min}} < 5.99$.

\begin{figure}[htb]
  \centering
\includegraphics[width=0.42 \linewidth]{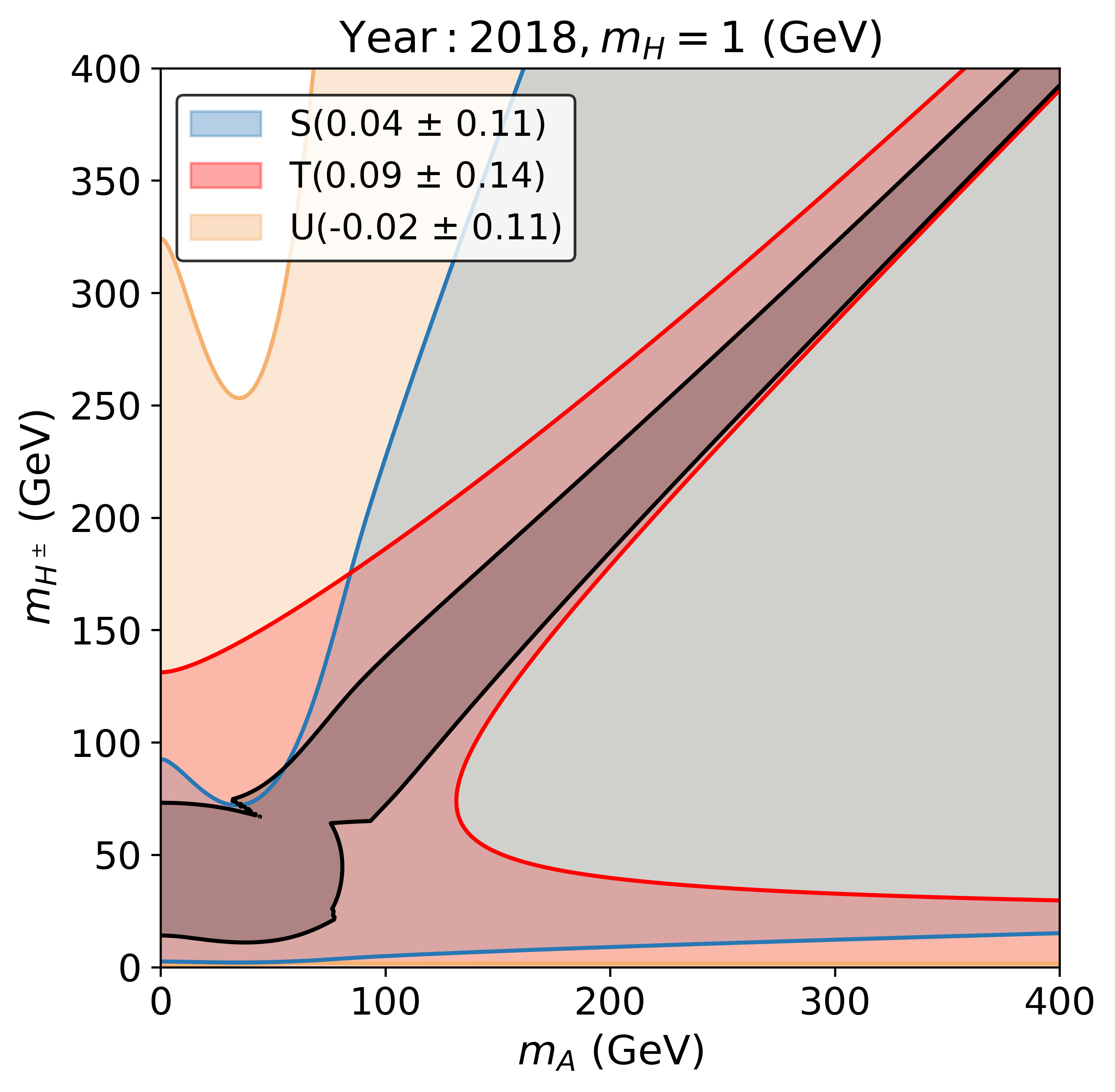}
\includegraphics[width=0.42 \linewidth]{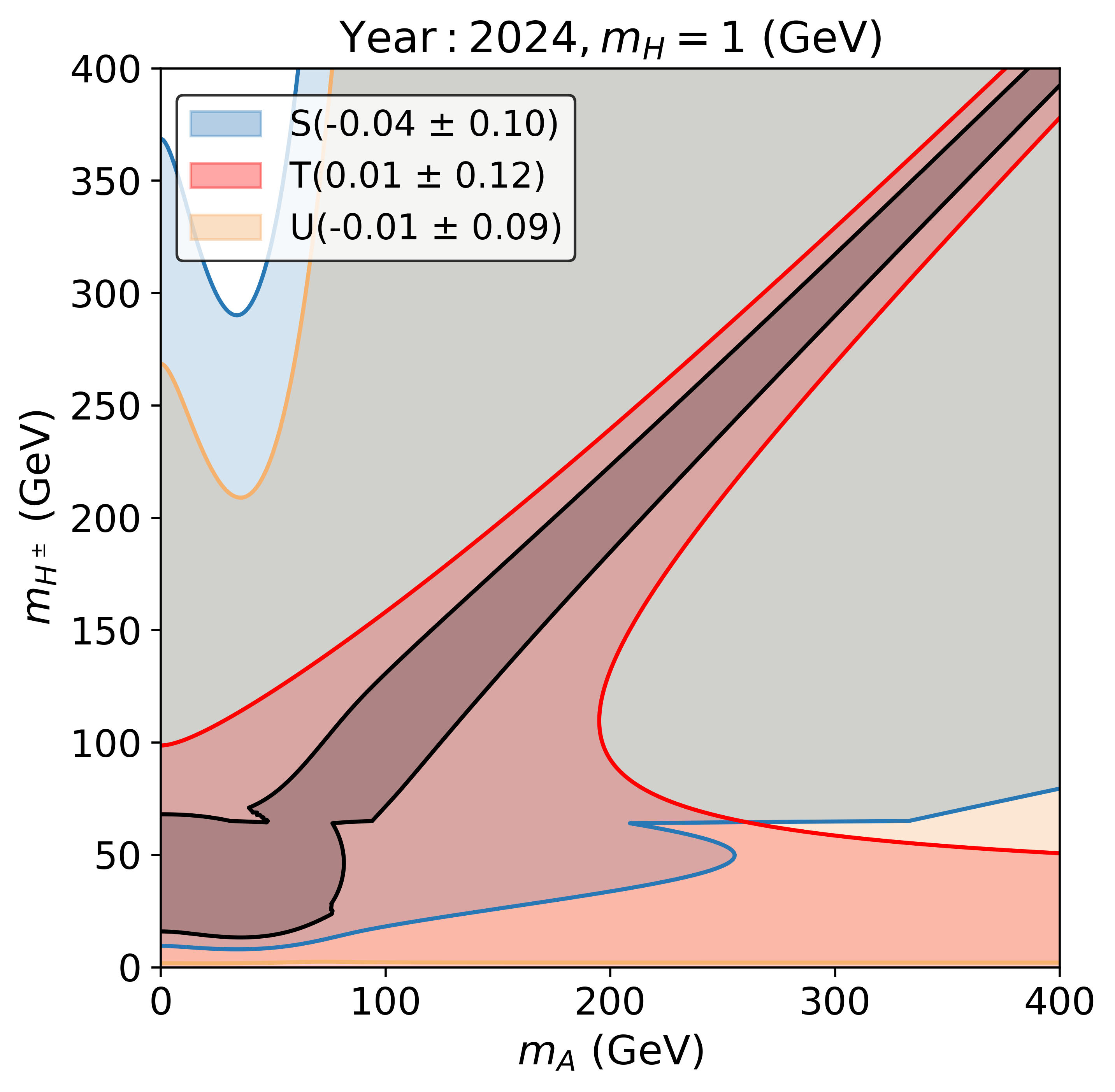}
\includegraphics[width=0.42 \linewidth]{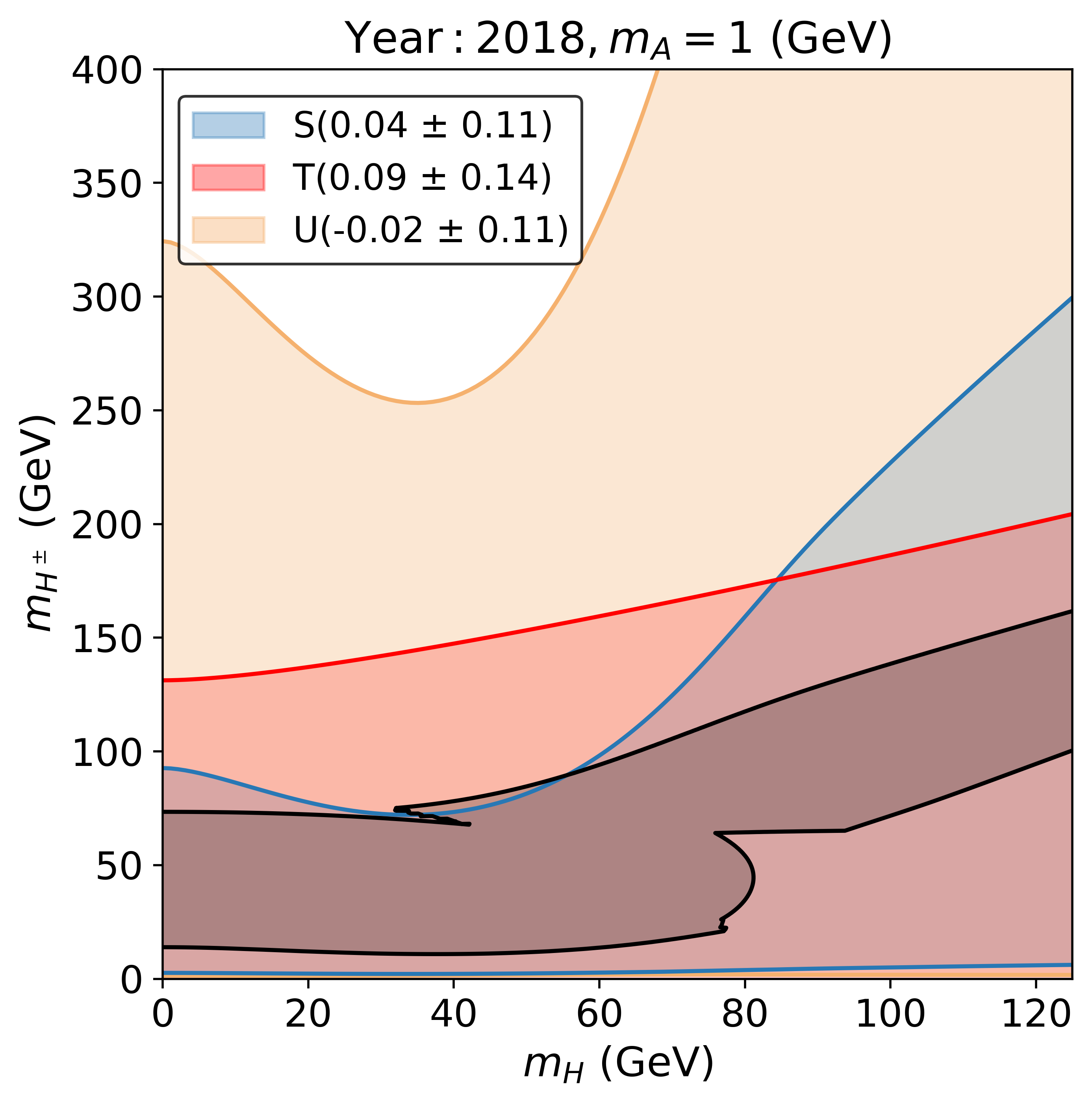}
\includegraphics[width=0.42 \linewidth]{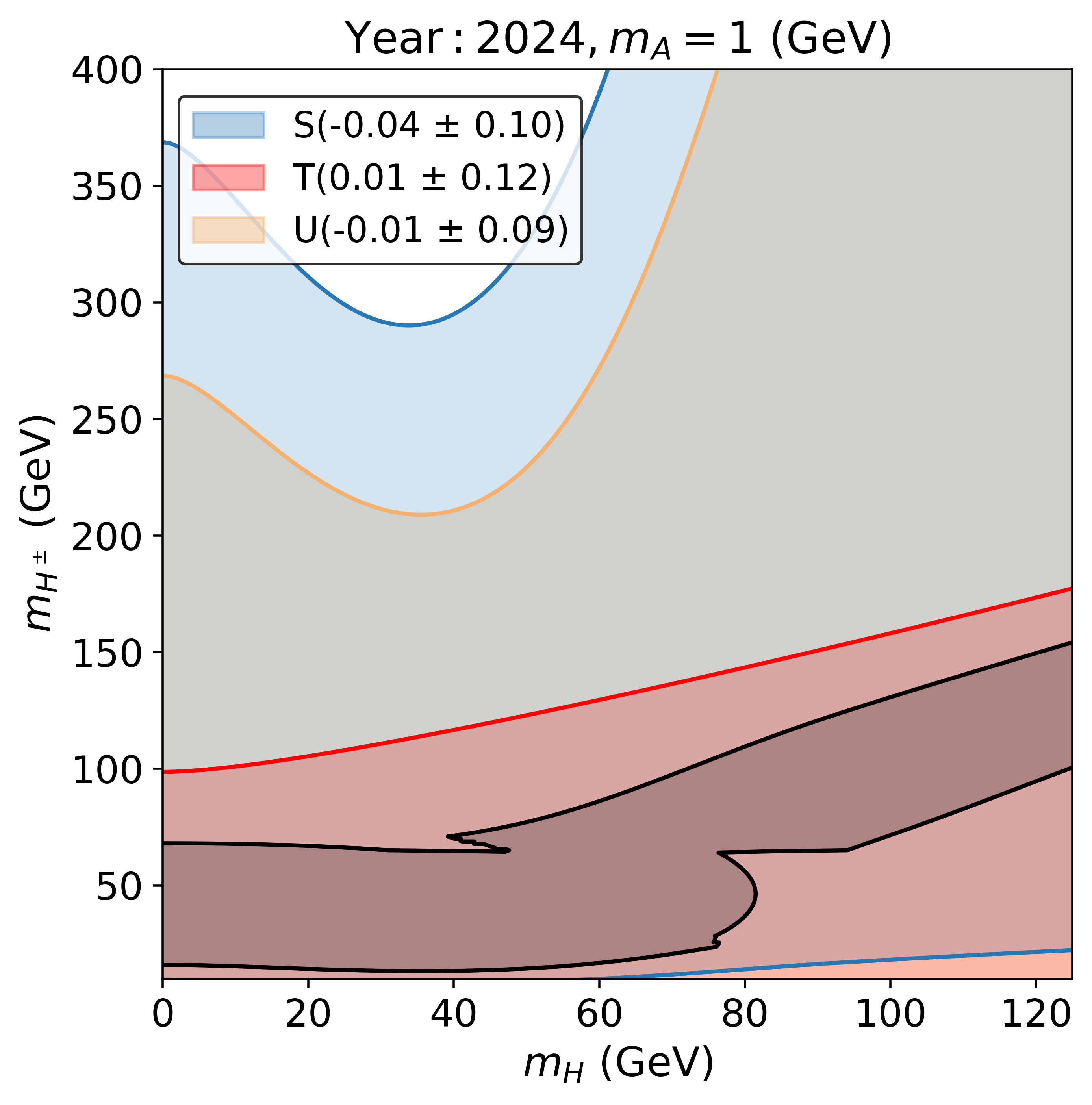}
\caption{Allowed regions of Type-I 2HDM in the $m_{A/H}$ vs. $m_{H^\pm}$ plane from the oblique parameters $S$ (blue), $T$ (red), $U$ (orange), and the 95\%\ C.L. global fit (black), for the 2018 (left) and the 2024 (right) Z-pole precision measurements. The upper panels corresponds to $m_H = 1~\mathrm{GeV}$ and $\cos(\beta-\alpha) = 1/\tan\beta$, and the lower panels to $m_A = 1~\mathrm{GeV}$ and $\cos(\beta-\alpha) = -1/\tan\beta$. Here we have $\tan\beta = 1000$ and $\lambda v^2 = 0$. 
}
\label{fig:cut_stu}
\end{figure}
In \autoref{fig:cut_stu}, we show the allowed region of Type-I 2HDM in the $m_{A/H}$ vs. $m_{H^\pm}$, under the constraints of oblique parameter $S$ (blue), $T$ (red), and $U$ (orange) and the 95\%\ C.L. global fit (black), based on Z-pole precision measurements from 2018 (left) and 2024 (right). The upper panels are for $m_H = 1~\mathrm{GeV}$ and $\cos(\beta-\alpha) = 1/\tan\beta$, while the lower panels are for $m_A = 1~\mathrm{GeV}$ and $\cos(\beta-\alpha) = -1/\tan\beta$. Here we have $\tan\beta = 1000$ and $\lambda v^2 = 0$. 

For the case of $m_H = 1~\mathrm{GeV}$ (upper panels), the constraints on $m_{H^\pm}$ from oblique parameters vary between the 2018 and 2024 data. For the 2018 data, the upper limit of the global fit is mainly determined by the $S$ parameter when $m_A < 85 ~\mathrm{GeV}$, followed by the $T$ parameter. For $m_A > 85 ~\mathrm{GeV}$, the $T$ parameter dominates, while the influence of $S$ becomes negligible. The $S$ and $T$ parameters set the upper limit on $m_{H^\pm}$ in different regions. The lower limit is determined jointly by $S$, $T$, and $U$ when $m_A < 130 ~\mathrm{GeV}$, with a slightly larger contribution from the $S$ parameter. For larger $m_A$, the $T$ parameter imposes stronger constraints on $m_{H^\pm}$. For the 2024 data, the upper limit of the global fit is primarily determined by the $T$ parameter, setting the upper limit on $m_{H^\pm}$, while the contributions from $S$ and $U$ are negligible. For the lower limit, the $S$ parameter dominates when $m_A < 195 ~\mathrm{GeV}$, providing stronger constraints than $T$ and $U$, and playing a larger role than in the 2018 data. At larger $m_A$, the $T$ parameter again becomes dominant, setting the lower limit on $m_{H^\pm}$.

For the case of $m_A = 1 ~\mathrm{GeV}$ (lower panels), shows similar behavior to the $m_H = 1 ~\mathrm{GeV}$ case. For the 2018 data, the upper limit of the global fit is mainly determined by the $S$ parameter when $m_H < 85 ~\mathrm{GeV}$, followed by the $T$ parameter. For $m_H > 85 ~\mathrm{GeV}$, the $T$ parameter becomes dominant. The $S$ and $T$ parameters set the upper limit on $m_{H^\pm}$ in different regions. The lower limit of the global fit is determined jointly by $S$, $T$, and $U$. For the 2024 data, the upper limit of the global fit is primarily determined by the $T$ parameter, setting the upper limit on $m_{H^\pm}$, while the contributions from $S$ and $U$ are negligible. For the lower limit, the $S$ parameter dominates and determines the lower limit on $m_{H^\pm}$, with a stronger impact compared to the 2018 data.

Overall, the $S$ parameter provides stronger constraints on $m_{H^\pm}$ than $T$ and $U$ in the low-mass region at $m_{H/A}=1~\mathrm{GeV}$. For 2018 data, the $S$ parameter mainly set the upper limit on $m_{H^\pm}$, while for 2024 it primarily determined the lower limit. Such kind of differences mainly come from that the center values of oblique parameters vary between year 2018 and 2024. 

This also implies that EWPO impose strong constraints on the mass spectrum of the 2HDM, requiring the charged Higgs mass $ m_{H^\pm}$ to be around the neutral Higgs bosons of either $m_H$ or $m_A$~\cite{Kling:2016opi, Haber:2015pua}. Based on these experimental limits, the allowed parameter space for weakly coupled light scalars is further constrained:
\begin{align}
m_H \sim 0 &: \quad m_{H^\pm} \sim m_A, \quad m_{A/H^\pm} \lesssim 600~\mathrm{GeV} \label{eq:ewpo_H}\\
m_A \sim 0 &: \quad m_{H^\pm} \sim m_H, \quad m_{H^\pm} \lesssim 600~\mathrm{GeV}, \quad m_H \lesssim m_h, \label{eq:ewpo_A}
\end{align}
with $\lambda v^2 = 0$ and $\cos(\beta - \alpha) \simeq0$.

\begin{figure}[htbp]
    \centering
    \includegraphics[width=0.49\linewidth]{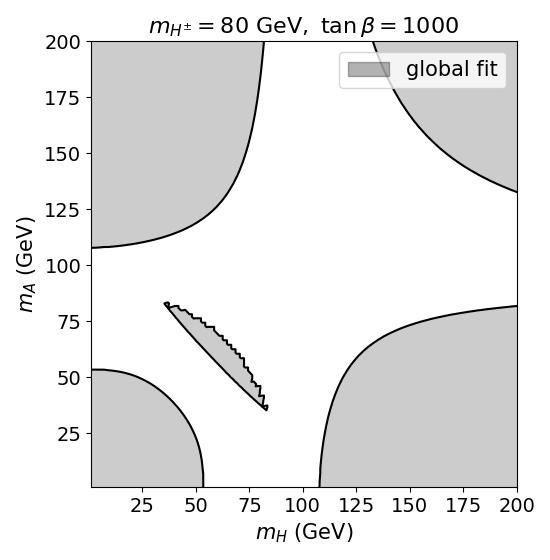}
    \caption{Excluded region at 95\% C.L.\ from global fit based on the 2024 $S$, $T$, and $U$ parameters in the $m_H$ vs. $m_A$ plane, for $m_{H^\pm} = 80~\mathrm{GeV}$ and $\tan\beta = 1000$ in the Type-I 2HDM, valid for $\cos(\beta - \alpha) \sim 0$. The region with $m_A (m_H) < 54~\mathrm{GeV}$ is excluded for the light $H$ ($A$) case.}
    \label{fig:EWPO_ex}
\end{figure}

Another space is the Case-0 discussed in Sec.\ref{sec:case}, where we have both light $m_A$ and $m_H$.
We need $m_{H^\pm} \sim m_H$ or $m_A$, and at same time the LEP charged Higgs search~\cite{ALEPH:2013htx, Arbey:2017gmh} excluded $m_{H^\pm} < 80~\mathrm{GeV}$. Therefore, for light (pseudo)scalar particles in the Type-I 2HDM with low mass region, $m_A, m_H < m_h/2$, we have our results at~\autoref{fig:EWPO_ex}

In~\autoref{fig:EWPO_ex}, we show the excluded region at 95\%\ C.L. from the global fit in the $m_H$ vs. $m_A$ plane, for $m_{H^\pm} = 80~\mathrm{GeV}$ and $\tan\beta = 1000$ in the Type-I 2HDM. This result applies for $\cos(\beta - \alpha) \sim 0$. Keep in mind that we only consider the case where $m_A, m_H < m_h/2$ here. When $m_H$ is very light, the region with $m_A < 54~\mathrm{GeV}$ is excluded; similarly, when $A$ is very light, the region with $m_H < 54~\mathrm{GeV}$ is also excluded. The excluded region around $m_H/A \sim 75~\mathrm{GeV}$ arises from significant positive shifts in $S$ and significant negative shifts in $U$, with $T$ remaining small and close to zero. Consequently, the following mass ranges remain viable: in the light $H$ case with $m_A < m_h/2$, the allowed range is $54~\mathrm{GeV} < m_A < m_h/2$; in the light $A$ case with $m_H < m_h/2$, the allowed range is $54~\mathrm{GeV} < m_H < m_h/2$, where the relation $\cos(\beta - \alpha) = 1/\tan\beta$ can no longer be realized.
%
%
\section{Parameter Space for LLPs in Type-I 2HDM}
\label{sec:parameter}

We summarize the reasonable parameter space for both surviving from all the above constraints and accommodating the existence of at least one light particle in the Type-I 2HDM as follows:
\begin{itemize} 
\item \textbf{Light $H$:} 
  \begin{equation}
  \cos(\beta - \alpha) \simeq \frac{1}{\tan \beta}, \quad m_A \in (54, 600)~\mathrm{GeV}, \quad m_{H^\pm} \sim m_A, \quad \lambda v^2 = 0.
  \label{eq:parameter_H}
  \end{equation}

\item \textbf{Light $A$:} 
\begin{equation}
  \cos(\beta - \alpha) \simeq \dfrac{1}{\tan\beta} \dfrac{2m_H^2 - m_h^2}{m_H^2 - m_h^2}, \quad m_H \in (54, m_h)~\mathrm{GeV}, \quad m_{H^\pm} \sim m_H, \quad \lambda v^2 = 0.
  \label{eq:parameter_A}
  \end{equation}
Specifically, for light A, two representative benchmark cases can be identified depending on the range of $m_H$: 
\begin{itemize} 
\item[-] For $86.9~\mathrm{GeV} \lesssim m_H \lesssim 90~\mathrm{GeV}$:
  \begin{equation}
  \cos(\beta - \alpha) = 0, \quad m_H \in (86.9, 90)~\mathrm{GeV}, \quad m_{H^\pm} \sim m_H, \quad \lambda v^2 = 0,
  \label{eq:benchmark_A1}
  \end{equation}
\item[-] For $101.2~\mathrm{GeV} \lesssim m_H \lesssim 103~\mathrm{GeV}$:
  \begin{equation}
  \cos(\beta - \alpha) = -\frac{1}{\tan \beta}, \quad m_H \in (101.2, 103)~\mathrm{GeV}, \quad m_{H^\pm} \sim m_H, \quad \lambda v^2 = 0.
  \label{eq:benchmark_A2}
  \end{equation}
\end{itemize}
\end{itemize} 
In all cases, we take large $\tan\beta$ to accommodate a light long-lived particle, with $80~\mathrm{GeV} < m_{H^\pm} < 600~\mathrm{GeV}$. The detailed region depends on netural heavy scalar. Therefor for fixed $m_A$ or $m_H$, we have the corresponding allowed ranges of $m_{H^\pm}$ in the light $H$ and light $A$ scenarios listed in \autoref{sec:mc}.
\begin{figure}[htbp]
  \centering
\includegraphics[width=0.49 \linewidth]{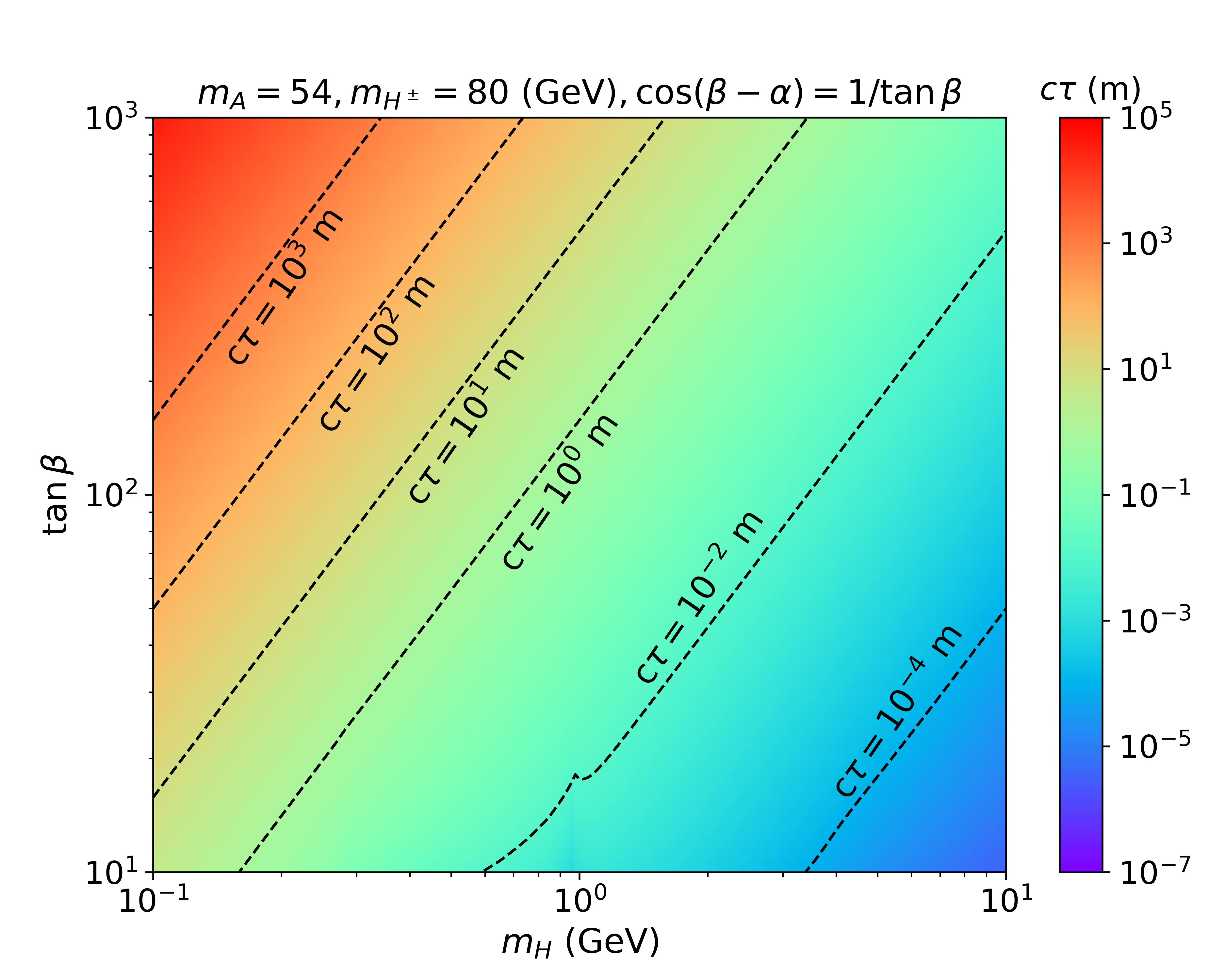}
\includegraphics[width=0.49 \linewidth]{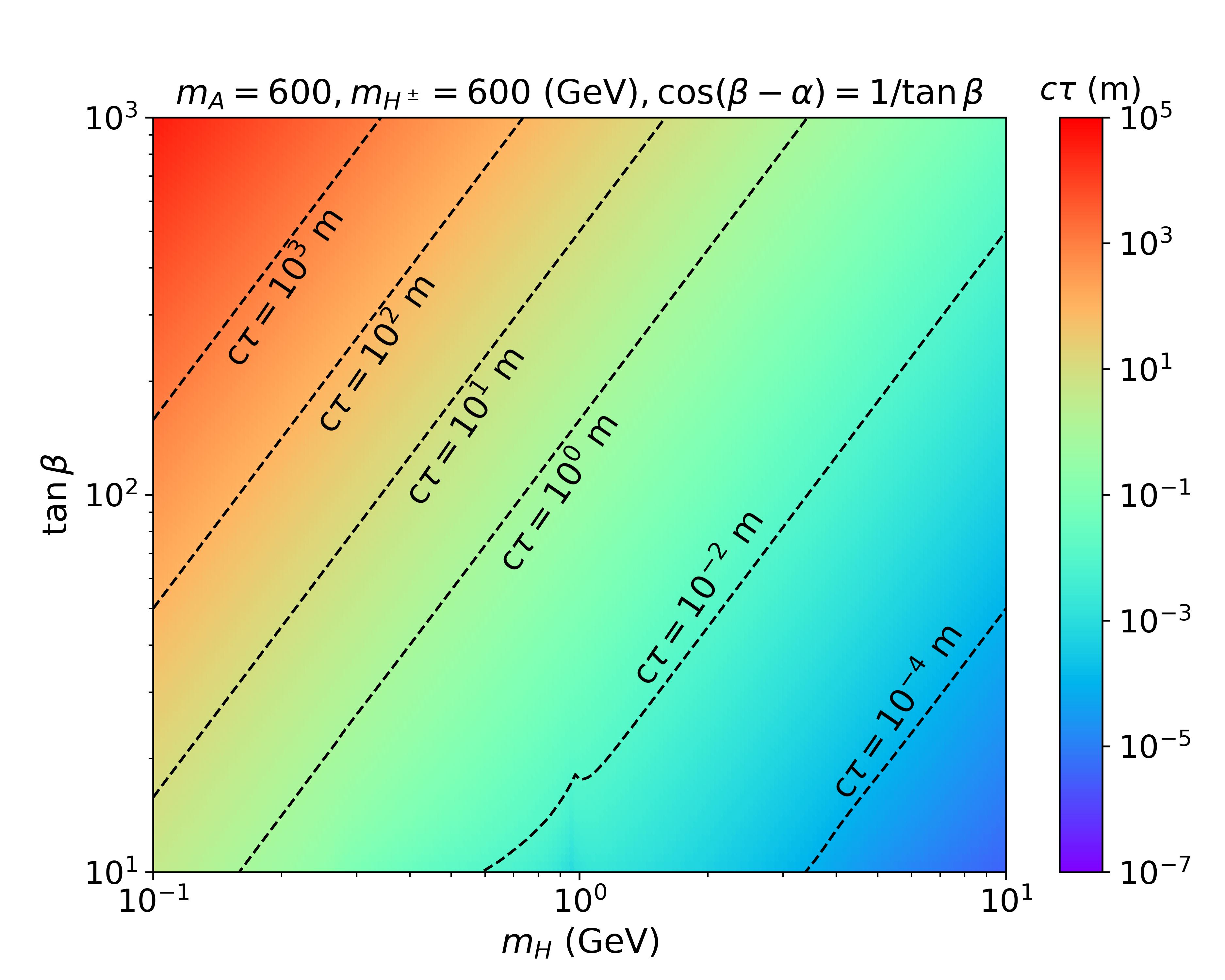}
\includegraphics[width=0.49 \linewidth]{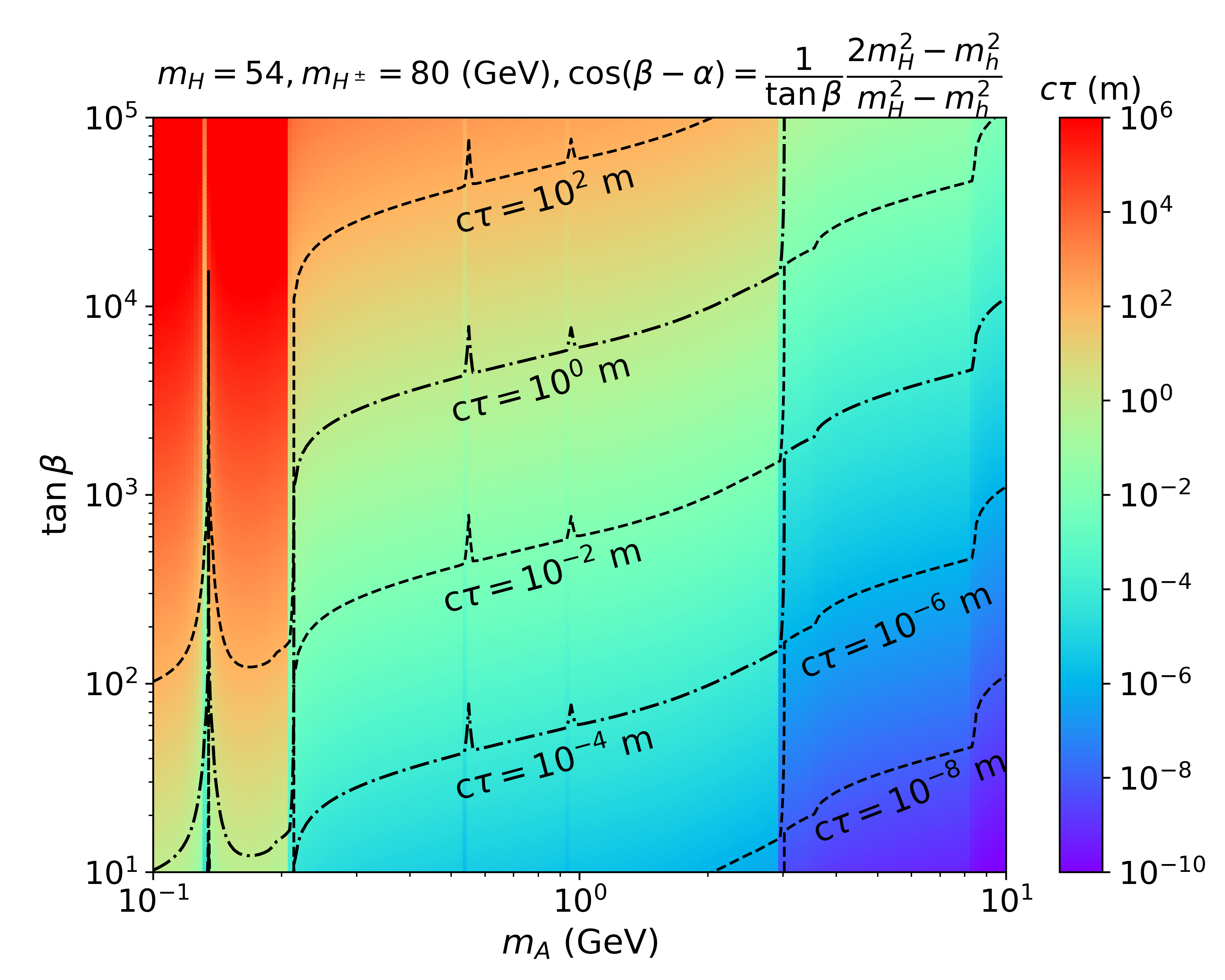}
\includegraphics[width=0.49 \linewidth]{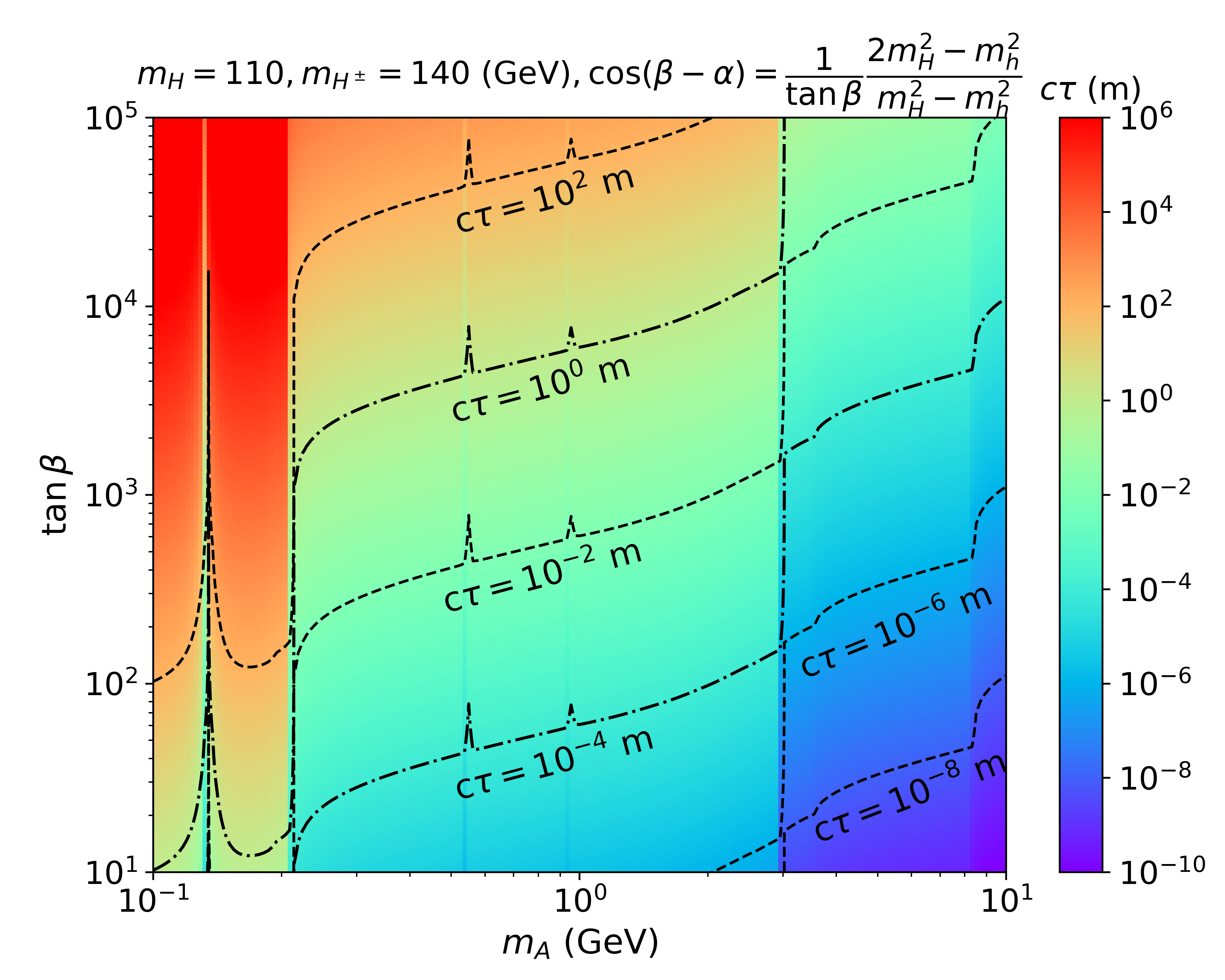}
\caption{(Upper Panel): Decay length $c\tau$ in the $m_H$ vs. $\tan\beta$ plane in the Type-I 2HDM for a light $H$, with $m_A = 54~\mathrm{GeV}$ and $m_{H^\pm} = 80~\mathrm{GeV}$ (left) and $m_H = 600~\mathrm{GeV}$ and $m_{H^\pm} = 600~\mathrm{GeV}$ (right), $\cos(\beta-\alpha) = 1/\tan\beta$. The colors are values of $c\tau$, varying from $10^{-7}$ to $10^5~\mathrm{m}$. We have 6 dashed black lines for $c\tau = 10^{-4}, 10^{-2}, 1, 10, 10^2, 10^3~\mathrm{m}$. (Lower Panel): Decay length $c\tau$ in the $m_A$ vs. $\tan\beta$ plane in the Type-I 2HDM for a light $A$, with $m_H = 54~\mathrm{GeV}$ and $m_{H^\pm} = 80~\mathrm{GeV}$ (left) and $m_H = 110~\mathrm{GeV}$ and $m_{H^\pm} = 140~\mathrm{GeV}$ (right), $\cos(\beta-\alpha) \simeq \dfrac{1}{\tan\beta}\dfrac{2m_H^2 - m_h^2}{m_H^2 - m_h^2}$. Colors indicate $c\tau$ from $10^{-10}$ to $10^6~\mathrm{m}$. We have 4 dashed black lines for $c\tau = 10^{-8}, 10^{-6}, 10^{-2},10^2~\mathrm{m}$, and 2 dash-dotted lines for $c\tau = 10^{-4}, 1~\mathrm{m}$. }
\label{fig:ctau}
\end{figure}

We further step into the LLP parameter space. To determine the lifetime of the light $H/A$, in \autoref{sec:length}, we analyze the total decay width $\Gamma$ and decay length $c\tau$ in the Type-I 2HDM. In both the light $H$ and light $A$ scenarios, $\Gamma$ and $c\tau$ are insensitive to $m_{A/H}$ or $m_{H^\pm}$. In the upper panels of \autoref{fig:ctau}, the decay length $c\tau$ is presented as a function of $m_H$ and $\tan\beta$ in the Type-I 2HDM for the light $H$ scenario. Here, we choose two representative sets of non-SM masses: $m_A = 54~\mathrm{GeV}$, $m_{H^\pm} = 80~\mathrm{GeV}$ (left), and $m_A = 600~\mathrm{GeV}$, $m_{H^\pm} = 600~\mathrm{GeV}$ (right), with $\cos(\beta-\alpha) = 1/\tan\beta$. The colors are values of $c\tau$, varying from $10^{-7}$ to $10^5~\mathrm{m}$. We have 6 dashed black lines for $c\tau = 10^{-4}, 10^{-2}, 1, 10, 10^2, 10^3~\mathrm{m}$. In both cases, the resulting $c\tau$ values are similar, as the loop contribution of $m_{H^\pm}$ to the $H \to \gamma\gamma$ decay width is small. The value of $c\tau$ increases with $\tan\beta$, reaching to several centimeters to tens of meters for $\tan\beta > 10$, several hundred meters for $\tan\beta > 50$, and up to several kilometers for $\tan\beta > 200$. It confirms that diphoton decays dominate in these regions.

In the lower panels of \autoref{fig:ctau}, the decay length $c\tau$ is shown as a function of $m_A$ and $\tan\beta$ in the Type-I 2HDM for the light $A$ scenario. We choose two benchmark cases: $m_H = 54~\mathrm{GeV}$, $m_{H^\pm} = 80~\mathrm{GeV}$ (left) and $m_H = 110~\mathrm{GeV}$, $m_{H^\pm} = 140~\mathrm{GeV}$ (right), with $\cos(\beta-\alpha) \simeq \dfrac{1}{\tan\beta}\dfrac{2m_H^2 - m_h^2}{m_H^2 - m_h^2}$. These two cases are nearly the boundary of the allowed parameter space. Colors indicate $c\tau$ from $10^{-10}$ to $10^6~\mathrm{m}$. We have 4 dashed black lines for $c\tau = 10^{-8}, 10^{-6}, 10^{-2},10^2~\mathrm{m}$, and 2 dash-dotted lines for $c\tau = 10^{-4}, 1~\mathrm{m}$. Since the coupling of $A$ to SM particles is independent of the non-SM Higgs masses, the resulting $c\tau$ is nearly identical in both panels. The value of $c\tau$ increases with $\tan\beta$, reaching to several meters to tens of meters for $\tan\beta > 10$, several hundred meters for $\tan\beta > 100$, and several kilometers for $\tan\beta > 320$. The peaks around $m_A \sim 1~\mathrm{GeV}$ or below originate from the $\pi^0$, $\eta$, and $\eta'$ resonances. The sudden increases of $\tan\beta$ at $m_A = 0.21, 3~\mathrm{GeV}$ are primarily due to the opening of the $\mu\bar{\mu}, c\bar{c}$ decay channels, respectively. After $3~\mathrm{GeV}$, decays into $c\bar{c}$ and gluon pairs become dominant, which requires larger values of $\tan\beta$ to maintain the same $c\tau$. It is worth noting that, unlike the light $H$ case in the upper panel, the $\tan\beta$ dependence of $c\tau$ here appears only as an overall shift. This is because all couplings of $A$ to SM particles exhibit the same $1/\tan\beta$ dependence. For both light H and A cases, $m_A$ ($m_H$) and $m_{H^\pm}$ do not change the LLP lifetimes inside of the allowed parameter space.

\section{$m_W$ in Type-I 2HDM}
\label{sec:mw}

As discussed, the mass splittings between BSM scalars are strongly constrained. At same time, these mass splittings also contribute to $m_W$ effectively in 2HDM framework. Thus we take them into account here, to have a detailed analysis. They may meet either the LHC measurements~\cite{CMS:2024lrd, ATLAS:2024erm} or $W$ boson mass anomaly from CDF-II~\cite{CDF:2022hxs}.
Here, we choose the light $H$ benchmark in Eq.~\eqref{eq:parameter_H} and the light $A$ benchmark in Eq.~\eqref{eq:benchmark_A1} and Eq.~\eqref{eq:benchmark_A2}.
\begin{figure}[htbp]
    \centering
    \includegraphics[width=0.32\linewidth]{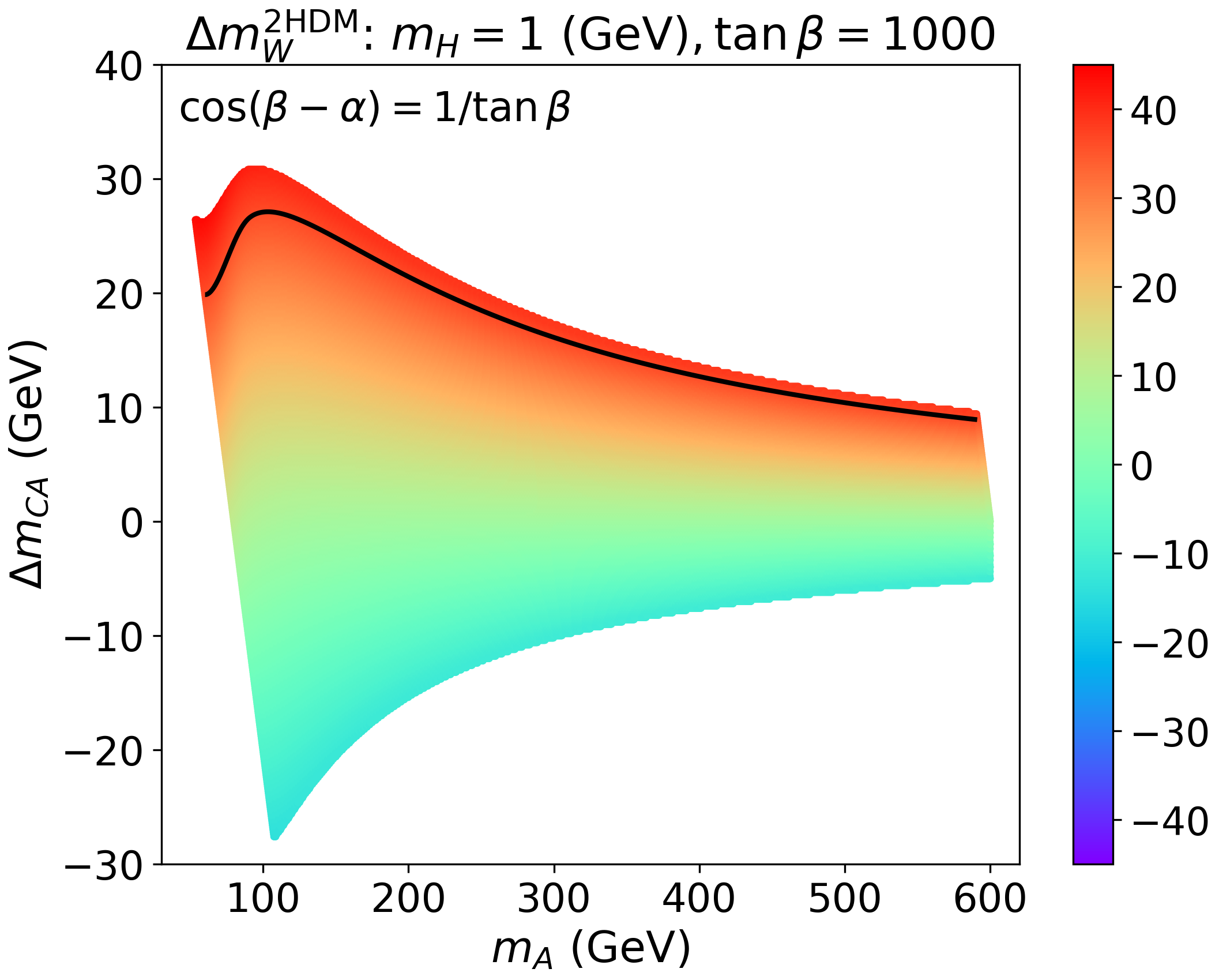}
    \includegraphics[width=0.32\linewidth]{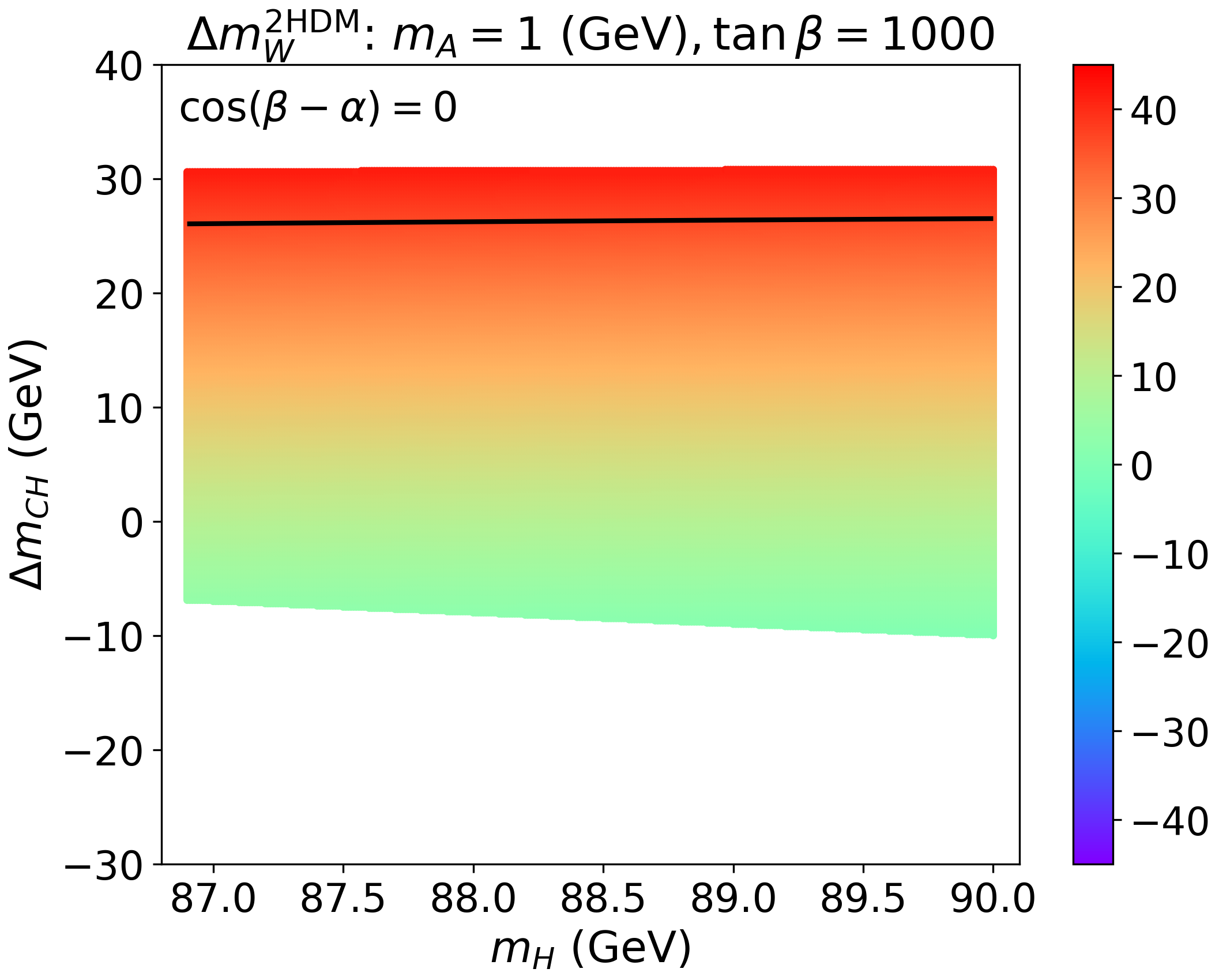}
    \includegraphics[width=0.32\linewidth]{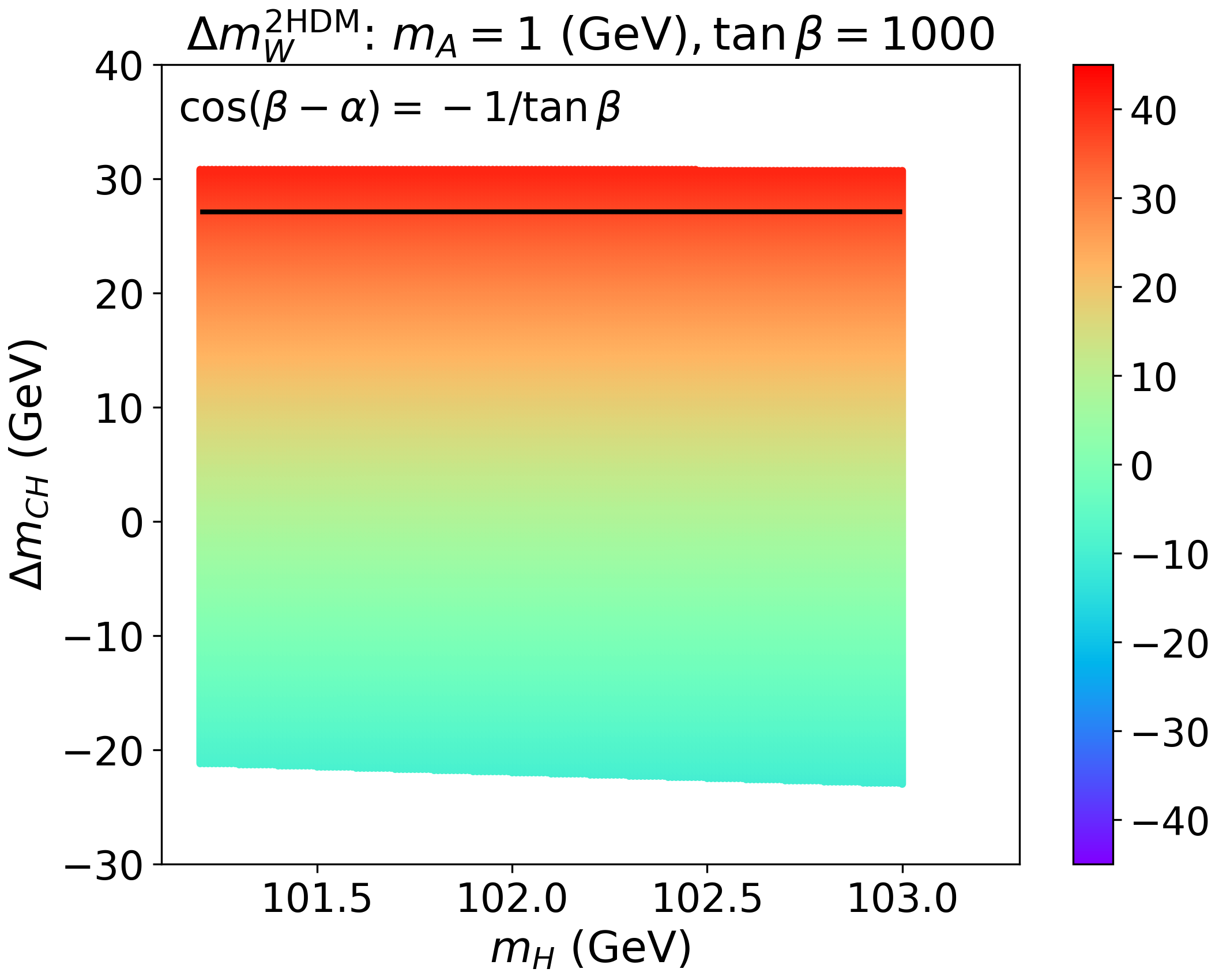}
    \caption{The 95\% C.L. allowed region in the Type-I 2HDM, with the black line above presenting the region where the $m_W^{\mathrm{2HDM}}$ meets the experimental measurement at CDF-II. (Left Panel): $\Delta m_W^{\mathrm{2HDM}}$ in the $m_A$ vs. $\Delta m_{CA}$ plane with $m_H = 1~\mathrm{GeV}$, $\cos(\beta - \alpha) = 1 / \tan \beta$, and $\tan\beta = 1000$. The colors indicate values of $\Delta m_W^{\mathrm{2HDM}}$, ranging from $-45~\mathrm{MeV}$ to $45~\mathrm{MeV}$. (Middle Panel): $\Delta m_W^{\mathrm{2HDM}}$ in the $m_H$ vs. $\Delta m_{CH}$ plane with $m_A = 1~\mathrm{GeV}$, $\cos(\beta - \alpha) = 0$, and $\tan\beta = 1000$. (Right Panel): $\Delta m_W^{\mathrm{2HDM}}$ in the $m_H$ vs. $\Delta m_{CH}$ plane with $m_A = 1~\mathrm{GeV}$, $\cos(\beta - \alpha) = -1 / \tan \beta$, and $\tan\beta = 1000$. In all three panels, $\tan \beta = 1000$ and $\lambda v^2 = 0$. The colors are same to the left panel.}
\label{fig:mass_split}
\end{figure}
In the 2HDM, $m_W$ associated corrections can be expressed by~\cite{Sirlin:2012mh}:
\begin{equation}
    m_W^{\text{2HDM}} = m_W^{\text{SM}} 
    \left[
    1 + \frac{\alpha c_W^2}{2(c_W^2 - s_W^2)} T(1 + \delta \rho^{\text{2HDM}}) 
    + \frac{\alpha}{8s_W^2} U 
    - \frac{\alpha}{4(c_W^2 - s_W^2)} S 
    \right]
\label{eq:m_W}
\end{equation}
to order $\mathcal{O}(\alpha^2)$. We define the deviation $\Delta m_W^{\text{2HDM}}\equiv m_W^{\text{2HDM}} - m_W^{\text{SM}} $.
To determine the impact of the mass splitting between $m_{A/H}$ and $m_{H^\pm}$ on $m_W$, we have:
\begin{equation}
\Delta m_{CA} = m_{H^\pm} - m_A, \quad \Delta m_{CH} = m_{H^\pm} - m_H.
\end{equation} 
We have the 95\% C.L. allowed region shown in \autoref{fig:mass_split} in the Type-I 2HDM, under various constraints discussed above. Generally the region above the black line presents the parameter space where the $m_W^{\mathrm{2HDM}}$ meets the experimental measurement at CDF-II~\cite{CDF:2022hxs}, and the lower region is consistent with the LHC~\cite{CMS:2024lrd, ATLAS:2024erm}.
In the left panel of \autoref{fig:mass_split}, we show $\Delta m_W^{\mathrm{2HDM}}$ in the $m_A$ vs. $\Delta m_{CA}$ plane for the light $H$ benchmark, Eq.~\eqref{eq:parameter_H}, with $m_H = 1~\mathrm{GeV}$, $\cos(\beta - \alpha) = 1 / \tan \beta$ and $\tan\beta=1000$. The colors show values of $\Delta m_W^{\mathrm{2HDM}}$, ranging from $-45~\mathrm{MeV}$ to $45~\mathrm{MeV}$. 
At large $m_A$, the allowed range of $\Delta m_{CA}$ becomes narrower because the mass splitting between $m_H$ and $m_{H^\pm}$ is large. When $m_{H^\pm} > m_A$, $\Delta m_W^{\mathrm{2HDM}}$ shows positive shift, and $\Delta m_W^{\mathrm{2HDM}}$ increases with $\Delta m_{CA}$. A region providing the theoretical correction meeting the new experimental measurement at CDF-II~\cite{CDF:2022hxs} first appears at $\Delta m_{CA} \approx 9~\mathrm{GeV}$. This region moves to larger $\Delta m_{CA}$ as $m_A$ decreases, which implies that a smaller $m_A$ requires a larger mass splitting between $m_H$ and $m_{H^\pm}$ to meet the CDF-II experimental measurement. 

In the middle panel, we show $\Delta m_W^{\mathrm{2HDM}}$ in the $m_H$ vs. $\Delta m_{CH}$ plane for the light $A$ benchmark, Eq.~\eqref{eq:benchmark_A1}, with $m_A = 1~\mathrm{GeV}$, $\cos(\beta - \alpha) = 0$, and $\tan\beta = 1000$. We can see that $\Delta m_W^{\mathrm{2HDM}}$ increases with $\Delta m_{CH}$ and exhibits a positive shift when $m_{H^\pm} > m_H$. The region that satisfies the CDF-II measurement~\cite{CDF:2022hxs} appears at $\Delta m_{CH} \gtrsim 26~\mathrm{GeV}$. The right panel corresponds to another light $A$ benchmark scenario, as defined in Eq.~\eqref{eq:benchmark_A2}, where $m_A = 1~\mathrm{GeV}$, $\cos(\beta - \alpha) = -1/\tan\beta$, and $\tan\beta = 1000$. The behavior of $\Delta m_W^{\mathrm{2HDM}}$ is similar and increases with $\Delta m_{CH}$, showing a positive shift for $m_{H^\pm} > m_H$. In this case, the region compatible with the CDF-II measurement~\cite{CDF:2022hxs} appears at $\Delta m_{CH} \gtrsim 27~\mathrm{GeV}$. Since the mass range of $m_H$ is narrow in both light $A$ scenarios, $\Delta m_W^{\mathrm{2HDM}}$ is insensitive to $m_H$, and the allowed range of $\Delta m_{CH}$ shows little dependence on $m_H$.

Combining with other constraints that lead to Eq.~\eqref{eq:parameter_H}, Eq.~\eqref{eq:benchmark_A1} and Eq.~\eqref{eq:benchmark_A2}, we identify four benchmark scenarios in the Type-I 2HDM that can simultaneously accommodate a light long-lived particle and provide the theoretical correction consistent meeting the CDF-II measurement of $m_W$~\cite{CDF:2022hxs},

For light $H$:
\begin{equation}
\cos(\beta - \alpha) = \frac{1}{\tan \beta}, \quad m_A = 200~\mathrm{GeV}, \quad  m_{H^\pm} = 222.3 \pm 0.9~\mathrm{GeV}, \quad \lambda v^2 = 0,
\label{eq:benchmark_Hdw_1}
\end{equation}
\begin{equation}
\cos(\beta - \alpha) = \frac{1}{\tan \beta}, \quad m_A = 400~\mathrm{GeV}, \quad  m_{H^\pm} = 413.1 \pm 0.4~\mathrm{GeV}, \quad \lambda v^2 = 0,
\label{eq:benchmark_Hdw_2}
\end{equation}

For light $A$:
\begin{equation}
\cos(\beta - \alpha) = 0, \quad m_H = 90~\mathrm{GeV}, \quad m_{H^\pm} = 118.7 \pm 2.2~\mathrm{GeV}, \quad \lambda v^2 = 0,
 \label{eq:benchmark_Adw_1}
\end{equation}
\begin{equation}
\cos(\beta - \alpha) = -\frac{1}{\tan \beta}, \quad m_H = 102~\mathrm{GeV}, \quad m_{H^\pm} = 130.9 \pm 1.8~\mathrm{GeV}, \quad \lambda v^2 = 0.
 \label{eq:benchmark_Adw_2}
\end{equation}
Here, $m_{H^\pm}$ is for the upper and lower limit of corresponding values, and we choose $\tan\beta = 1000$ to accommodate a long-lived particle.

\section{Results for FASER and FASER~2}
\label{sec:faser}
A light CP-even Higgs $H$ can be produced through the decays of $\pi$, $K$, $\eta$, $D$ and $B$ mesons, as well as the radiative decay of bottomonium $\Upsilon$. While a light CP-odd scalar $A$ can be produced in any processes via its mixing with $\pi^0$, $\eta$, or $\eta^\prime$ meson states, weak decays of mesons, such as $K \to \pi A$ and $B \to X_s A$, and radiative decays of bottomonium $\Upsilon$ and charmonium $J/\psi$, which only provide subdominant contributions. A detailed discussion of these production mechanisms and the corresponding rates can be found in Ref.~\cite{Kling:2022uzy}. 
\paragraph{FASER and FASER~2.}
%
FASER is an experiment at the LHC designed to search for light weakly-interacting particles produced at the ATLAS interaction point (IP). Its detector has a cylindrical configuration with a radius of 10 cm and a length of 1.5 m and is installed in tunnel TI12, approximately 480~m away from the ATLAS IP along the beam axis, and is located in the very forward direction with an acceptance angle $\theta \lesssim 2\times 10^{-4}$~\cite{Feng:2017uoz, FASER:2022hcn, Feng:2022inv, Adhikary:2024nlv}. During the Run 3 of the LHC, it has collected data with an integrated luminosity of $195.9~\mathrm{fb}^{-1}$ by the end of year 2024 and its first results has been reported in Ref.~\cite{FASER:2023tle} for a dark photon with an integrated luminosity of $27~\mathrm{fb}^{-1}$ and Ref.~\cite{FASER:2024bbl} for a LLP (mainly axion-like particle) with a luminosity of $57.7~\mathrm{fb}^{-1}$. At the HL-LHC, an upgraded detector FASER~2 has been proposed with the total integrated luminosities up to $3~\mathrm{ab}^{-1}$ reached~\cite{Feng:2022inv}. It may located either 480 m downstream of the ATLAS IP or at the Forward Physics Facility (FPF) 620 m downstream. For a concrete study of the FASER~2 reach, we consider a setup of FASER~2: a cylindrical detector with a length of 10 m and a radius of 1 m at 480 m away from the IP. We have checked that changing to other considered options does not modify our results significantly.

\begin{figure}[htbp]
  \centering
 \includegraphics[width=0.49 \linewidth]{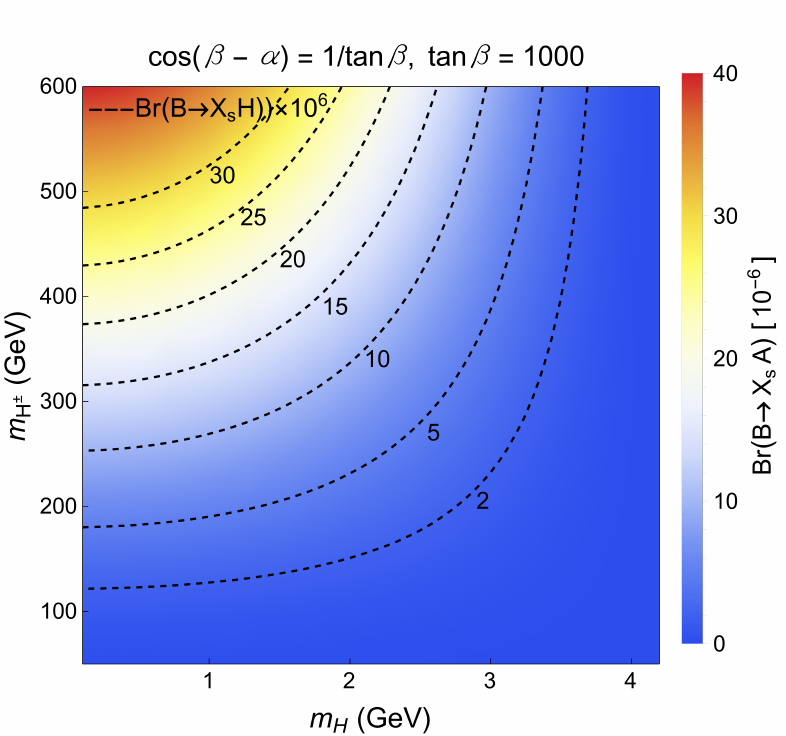}
 \includegraphics[width=0.49 \linewidth]{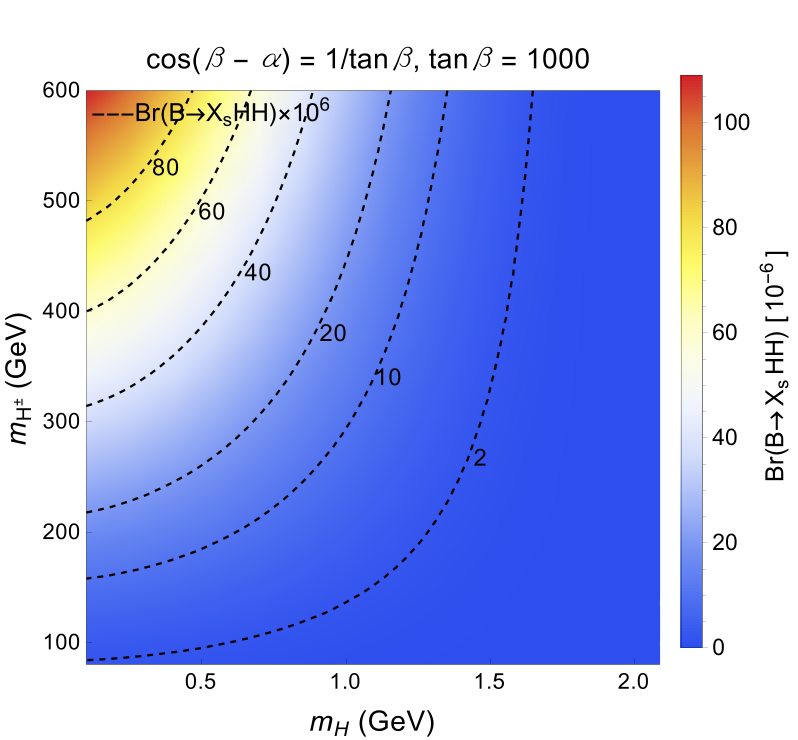}
 \includegraphics[width=0.49 \linewidth]{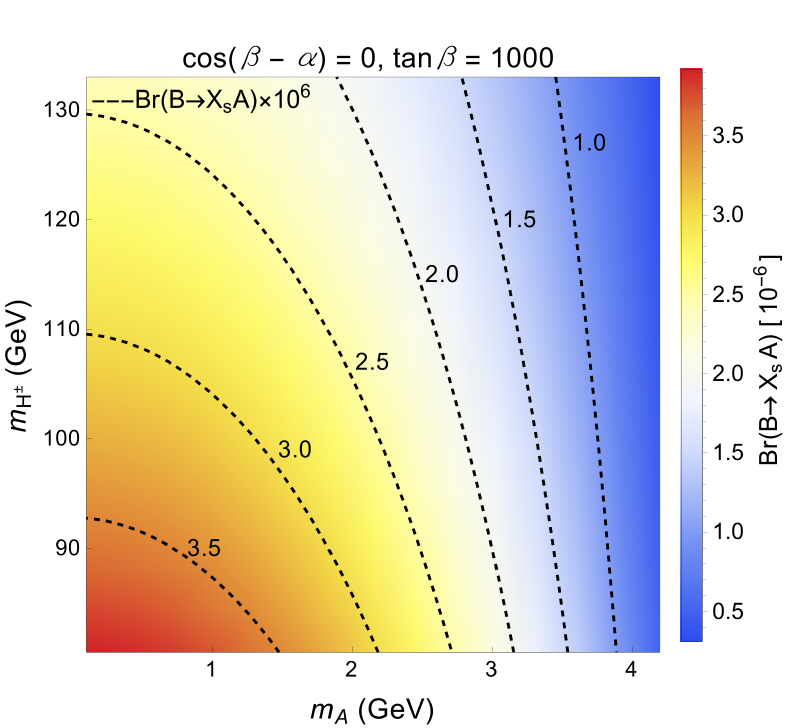}
 \includegraphics[width=0.49 \linewidth]{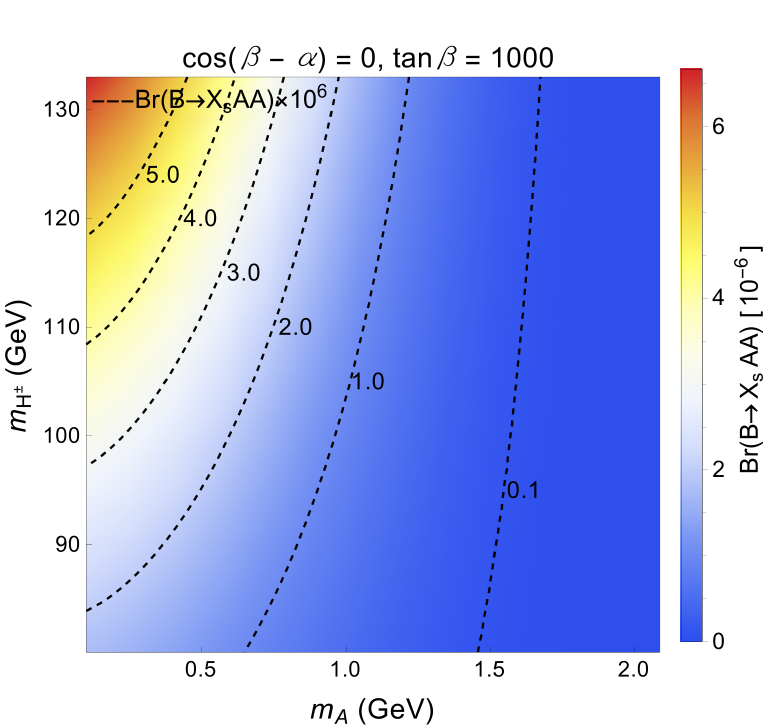}
\caption{(Upper Panel): The branching fractions for scalar production from $B$ meson two-body decay (left) and scalar pair production from $B$ meson three-body decay (right) in the $m_H$ vs.\ $m_{H^\pm}$ plane, with $\cos(\beta - \alpha) = 1 / \tan \beta$ and $\tan \beta = 1000$. The colors represent the values of branching fraction, and the dashed black lines show the corresponding values. (Lower Panel): The branching fractions for pseudoscalar production from $B$ meson two-body decay (left) and pseudoscalar pair production from $B$ meson three-body decay (right) in the $m_A$ vs.\ $m_{H^\pm}$ plane, with $\cos(\beta - \alpha) = 0$ and $\tan \beta = 1000$. The colors represent the values of branching fraction, and the dashed black lines show the corresponding values.}
\label{fig:faser_mc}
\end{figure}

The decays of $B$ mesons to light CP-even scalar $H$ and CP-odd scalar $A$ are mainly governed by the flavor changing effective interactions between $b$ and $s$ quarks. To study the impact of $m_{H^\pm}$ on the production of $H/A$, in \autoref{fig:faser_mc}, we show the branching fractions for scalar production from $B$ meson decay in the $m_{H/A}$ vs.\ $m_{H^\pm}$ plane, for two-body decays (left) and three-body decays (right). The upper panels correspond to the production of CP-even scalar $H$, with $\cos(\beta - \alpha) = 1 / \tan \beta$ and $\tan \beta = 1000$. The lower panels correspond to the production of CP-odd pseudoscalar $A$, with $\cos(\beta - \alpha) = 0$ and $\tan \beta = 1000$. The colors indicate values of branching fractions, and the dashed black lines represent the corresponding values.

For the case of CP-even scalar $H$ production from $B$ meson decays (upper panels), the two-body decay $B \to X_s H$ is kinematically allowed up to $m_H \sim 4.2~\mathrm{GeV}$, while the three-body decay $B \to X_s H H$ is only allowed for $m_H \lesssim 2~\mathrm{GeV}$ due to phase space suppression. At low $m_H$, both branching fractions $\text{Br}(B \to X_s H)$ and $\text{Br}(B \to X_s H H)$ increase with increasing $m_{H^\pm}$, and then decrease significantly as $m_H$ increases.

For the case of CP-odd scalar $A$ production from $B$ meson decays (lower panels), the two-body decay $B \to X_s A$ is kinematically allowed up to $m_A \sim 4.2~\mathrm{GeV}$, while the three-body decay $B \to X_s A A$ is strongly suppressed by phase space and only allowed up to $m_A \sim 2~\mathrm{GeV}$. At low $m_A$, the branching fraction $\text{Br}(B \to X_s A)$ decreases with increasing $m_{H^\pm}$, while $\text{Br}(B \to X_s A A)$ increases with $m_{H^\pm}$. Both decay modes decrease in branching ratio as $m_A$ increases. By the way, the branching fractions $\text{Br}(B \to X_s A)$ and $\text{Br}(B \to X_s AA)$ are almost insensitive to $\cos(\beta - \alpha) \simeq 0$, since these processes are suppressed by $\cos(\beta - \alpha)$ or by $1/\tan\beta$ in the Type-I 2HDM. As a result, for $\cos(\beta - \alpha) = -1/\tan\beta$ with $\tan\beta = 1000$, the branching fractions show similar behavior.

\begin{figure}[htbp]
  \centering
 \includegraphics[width=0.49 \linewidth]{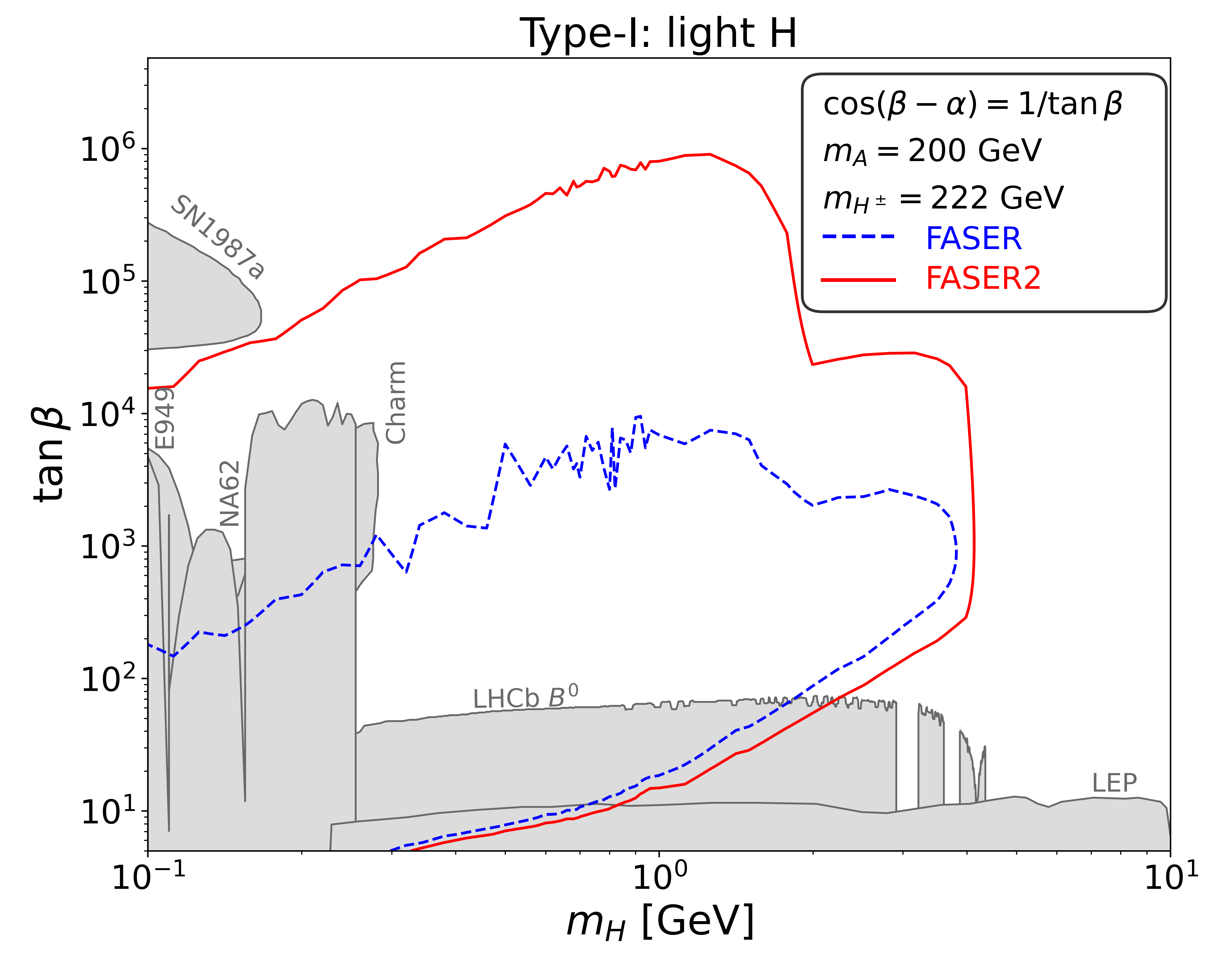}
 \includegraphics[width=0.49 \linewidth]{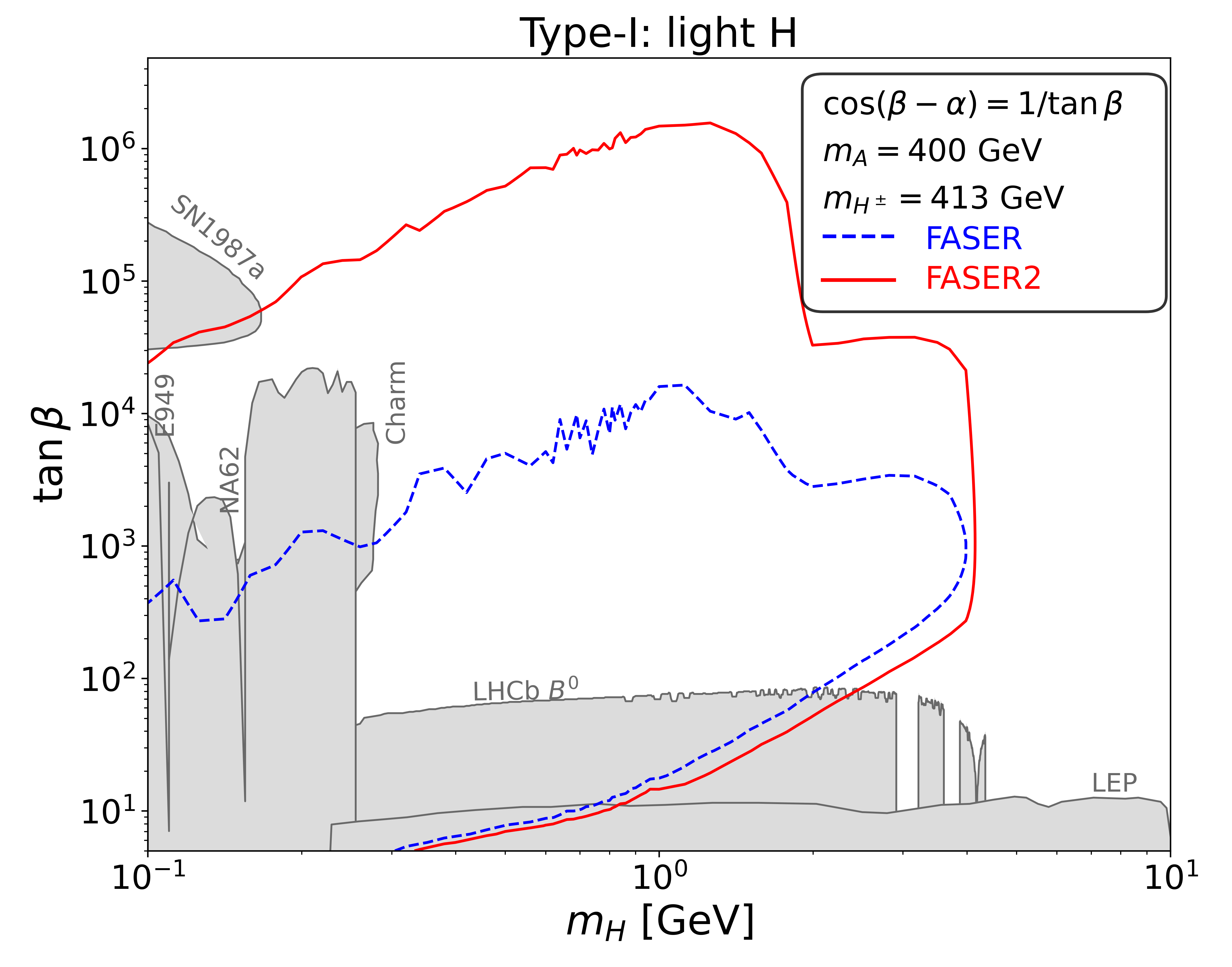}
 \includegraphics[width=0.49 \linewidth]{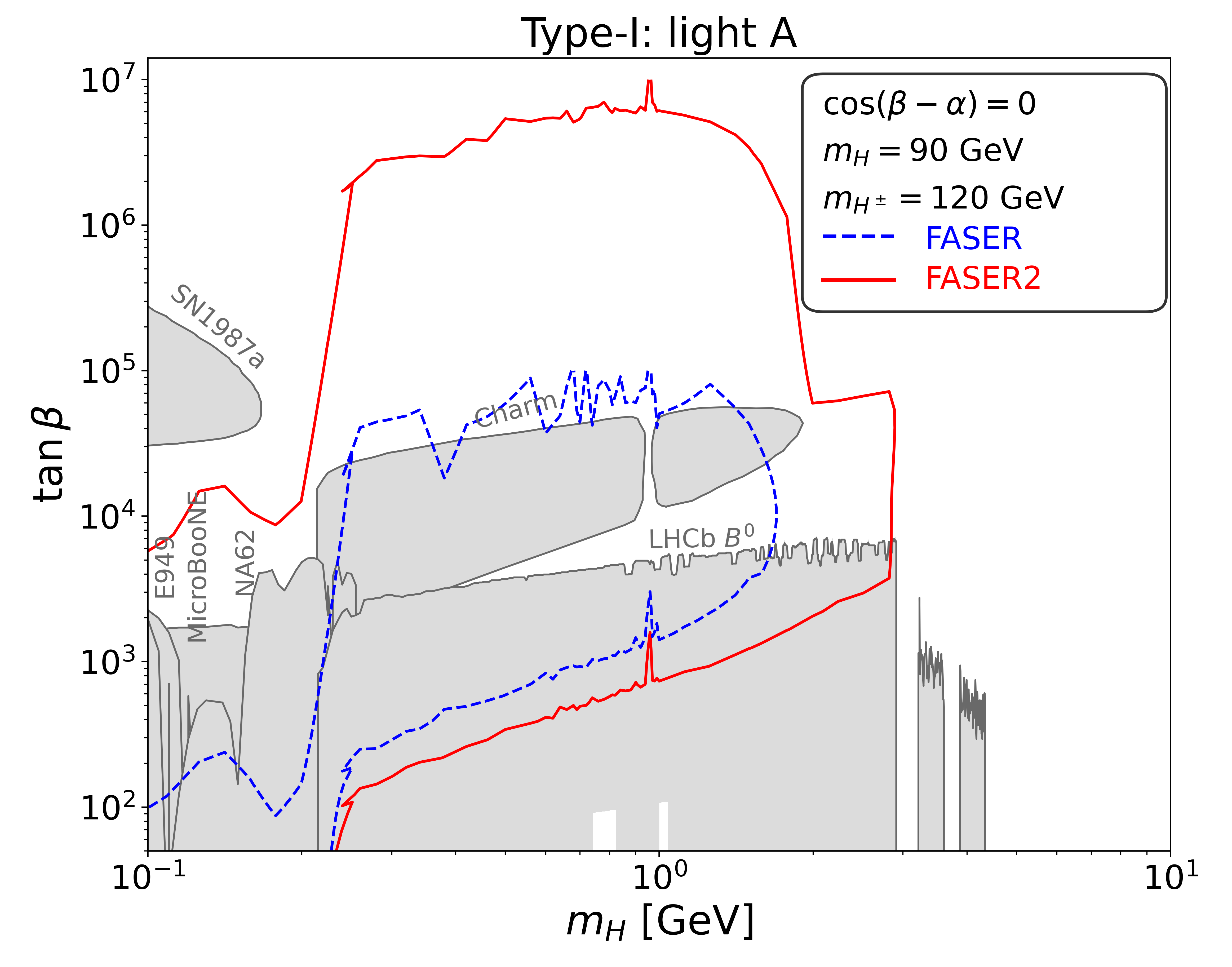}
 \includegraphics[width=0.49 \linewidth]{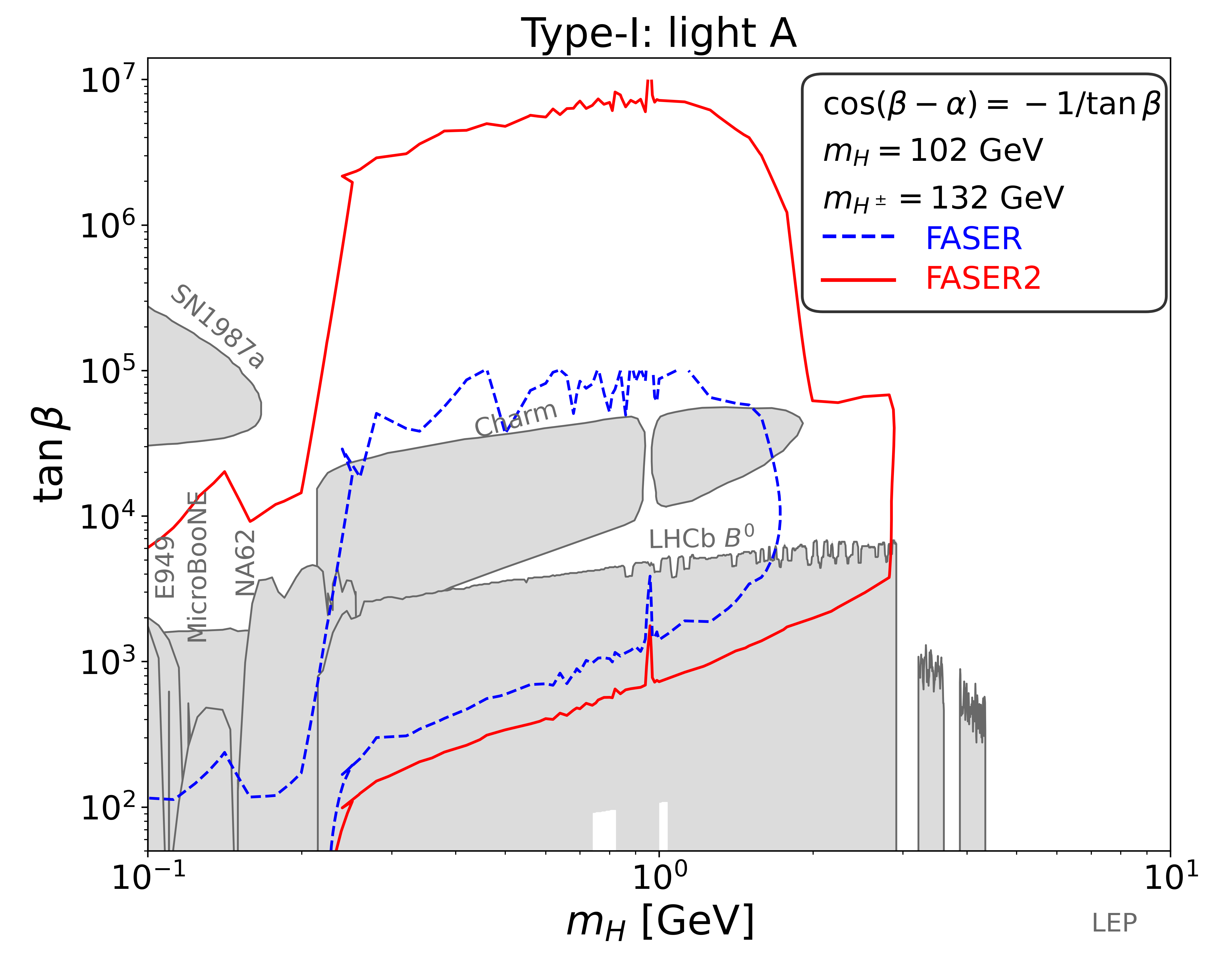}
\caption{(Upper panel): FASER (blue dashed curve) and FASER~2 (red solid curve) reach for the light CP-even Higgs $H$ in the $m_{H}$ vs.\ $\tan\beta$ plane, under different benchmark scenarios. (Lower panel): FASER (blue dashed curve) and FASER~2 (red solid curve) reach for the light CP-odd Higgs $A$ in the $m_{A}$ vs.\ $\tan\beta$ plane, under different benchmark scenarios. Various current experimental constraints are shown in grey region. 
}
\label{fig:foresee}
\end{figure}
We obtain the reaches at FASER/FASER~2 for the benchmark scenarios defined in Eqs.~(\ref{eq:benchmark_Hdw_1}), (\ref{eq:benchmark_Hdw_2}), (\ref{eq:benchmark_Adw_1}), and (\ref{eq:benchmark_Adw_2}). 
In the upper panels of \autoref{fig:foresee}, we show the potential reaches at FASER (blue dashed curve) and FASER~2 (red solid curve) in the $m_H$ vs. $\tan\beta$ plane for a light $H$. The left panel corresponds to a parameter choice of $\cos(\beta - \alpha) = 1/\tan\beta$, $m_A = 200~\mathrm{GeV}$, and $m_{H^\pm} = 222~\mathrm{GeV}$, while the right panel corresponds to $\cos(\beta - \alpha) = 1/\tan\beta$, $m_A = 400~\mathrm{GeV}$, and $m_{H^\pm} = 413~\mathrm{GeV}$. 
In the lower panels of \autoref{fig:foresee}, we show the reach in the $m_A$ vs. $\tan\beta$ plane for a light $A$, corresponds to the benchmark point with $\cos(\beta - \alpha) = 0$, $m_H = 90~\mathrm{GeV}$, and $m_{H^\pm} = 120~\mathrm{GeV}$. In the lower panels of \autoref{fig:foresee}, we show the reach in the $m_A$ vs. $\tan\beta$ plane for the light $A$. The left panel corresponds to the benchmark scenario with $\cos(\beta - \alpha) = 0$, $m_H = 90~\mathrm{GeV}$, and $m_{H^\pm} = 120~\mathrm{GeV}$; the right panel corresponds to $\cos(\beta - \alpha) = -1/\tan\beta$, $m_H = 102~\mathrm{GeV}$, and $m_{H^\pm} = 132~\mathrm{GeV}$. Various current experimental constraints are shown in grey region.

For the two cases of the light $H$ (upper panels), the beam dump and fixed-target experiments (CHARM, NA62 and E949) and SN1987A mainly exclude a size region for the mass of $H$ below 250 MeV. Constraints from MicroBooNE, Standard Model Higgs coupling measurements, Higgs invisible decay channels, and flavor physics bounds do not constrain the CP-even scalar scenario in the $\tanb>5$ parameter region. LEP and LHCb only exclude low $\tanb$ region $\tanb<5$ and $\tanb<50$ respectively, while FASER and FASER~2 can probe larger values of $\tanb$ up to around $10^4$ and $10^6$ separately. The reach of the scalar mass at FASER and FASER~2 can reach up to $m_B$ due to the possible production process $B \to X_s H$. When $m_H\gtrsim m_B/2$, the total production rate of $H$ decreases due to the double scalar production ($B \to X_s HH$) is forbidden kinematically, leading to significant reductions of the reaches. Notably, FASER~2 improves the reach in $\tan\beta$ by about two orders of magnitude compared to FASER in the large $\tan\beta$ region because of 20 times higher luminosity and larger detector volume. The reaches shown in the right panel are slightly greater than those in the left panel, since the branching ratio of $B$-meson decays into $H$ increases about a factor of 3 as $m_{H^\pm}$ increases from $222~\mathrm{GeV}$ to $413~\mathrm{GeV}$. However, this effect might be marginal given other uncertainties in spectra of mesons and theoretical calculations of the mesons decay rates, which indicated more work needs to be done in this direction.

For the case of the light $A$ (lower panels), similar to the upper panels, the grey shaded regions show current experimental constraints. The CHARM experiment provides stronger exclusion up to $m_A \sim 2~\mathrm{GeV}$, compared to the light $H$ case of $m_H \sim 300~\mathrm{MeV}$. Since we consider $\tan\beta$ values starting from 50, the LEP bounds in the low-$\tan\beta$ region are not shown. Compared to the light $H$ scenario, the reach extends to larger $\tan\beta$ values, with FASER and FASER~2 reaching up to $10^5$ and $10^7$, respectively. This improvement comes from the different $\tan\beta$ dependencies of the Yukawa couplings $\xi_H^f$ and $\xi_A^f$. Moreover, in the light $A$ scenario, the reach in $m_A$ is much more sensitive to the geometry of the detector, especially its radius. The left and right panels show nearly identical reaches, since the narrow possible values of $m_{H^\pm}$ and $\cos(\beta - \alpha)$ have a barely noticeable impact when we vary them. 

\section{Summary}
\label{sec:Conclude}
In this work, we analyzed the parameter space of the Type-I 2HDM. We first worked out the region that accommodates a light scalar $H/A$ under various constraints, and then figured out the region of light long-lived scalar $H/A$. We emphasized that the Type-I 2HDM has an advantage over the other three types in the large $\tan\beta$ regime, and discussed the impact of different $S$, $T$, and $U$ parameters on the global fit. The allowed region are sensitive to mass splittings of BSM scalars, which affects $m_W^{\text{2HDM}}$ much. We pointed out that the allowed region can explain the CDF-II measurement and latest LHC results of $m_W$ with a LLP candidate. In the benchmark scenarios we chose the former one. Further we displayed the reaches of FASER and FASER~2 in these regions. Our main findings included:
\begin{itemize}
    \item[1.]     %
    We summarized and compared the global fit results of the oblique parameters $S$, $T$, and $U$ obtained from $Z$-pole precision measurements between 2018 and 2024. When there is a light BSM scalar, $S$ can impose the dominant constraint on $m_{H^\pm}$, which is different to cases where only heavy BSM Higgs are involved. While $S$ with the 2018 data provides an dominant upper limit on $m_{H^\pm}$, the 2024 data dominates at a lower limit. This comes from the varied center values of $S$ parameter. 
    \item[2.] Under theoretical and experimental constraints, we presented the parameter space in the Type-I 2HDM that accommodates light long-lived scalars $H/A$.  Under the approximation $\lambda v^2 = 0$ and large $\tan\beta$, the allowed regions are:

    \item[] For light $H$: $\cos(\beta - \alpha) \simeq \frac{1}{\tan \beta}, \quad m_A \in (54, 600)~\mathrm{GeV}, \quad m_{H^\pm} \sim m_A$.

    \item[] For light $A$: $\cos(\beta - \alpha) \simeq \dfrac{1}{\tan\beta} \dfrac{2m_H^2 - m_h^2}{m_H^2 - m_h^2}, \quad m_H \in (54, m_h)~\mathrm{GeV}, \quad m_{H^\pm} \sim m_H$.

    For the light $A$ scenario, we summarized two representative benchmark scenarios, as shown in Eq.~\eqref{eq:benchmark_A1},~\eqref{eq:benchmark_A2}.Based on \autoref{fig:ctau}, we can get the regime with long-lived scalars.

    \item[3.] Based on benchmark scenarios accommodating long-lived scalars $H/A$, we explained the CDF-II precision measurement of $m_W$. We further presented the potential reach of FASER and FASER~2 in these regions, which enables probing the parameter space at large $\tan\beta$. For the light $H$ scenario, we selected two benchmark points and found that a heavier $m_{H^\pm}$ enhances the branching fraction of $B$ meson decays to $H$, enabling FASER and FASER~2 to cover larger $\tan\beta$ range. For light H (A) case, $m_A$ ($m_H$) and $m_{H^\pm}$ do not change the LLP lifetimes inside of the allowed parameter space.
\end{itemize}

Therefore, the Type-I 2HDM provides a well-motivated model for future searches of long-lived particles in the large $\tan\beta$ region, with promising experimental prospects.

\acknowledgments
We would like to thank Shufang Su for useful discussions.  XQ and WS are supported by the Natural Science Foundation of China (NSFC) under grant numbers 12305115, Shenzhen Science and Technology Program (Grant No. 202206193000001, 20220816094256002), Guangdong Provincial Key Laboratory of Gamma-Gamma Collider and Its Comprehensive Applications(2024KSYS001), and Guangdong Provincial Key Laboratory of Advanced Particle Detection Technology(2024B1212010005). HS is supported by IBS under the project code, IBS-R018-D1. 

\appendix
\section{Decay length in different types of 2HDM}
\label{sec:type}
%
\begin{figure}[htbp]
  \centering
\includegraphics[width=0.49 \linewidth]{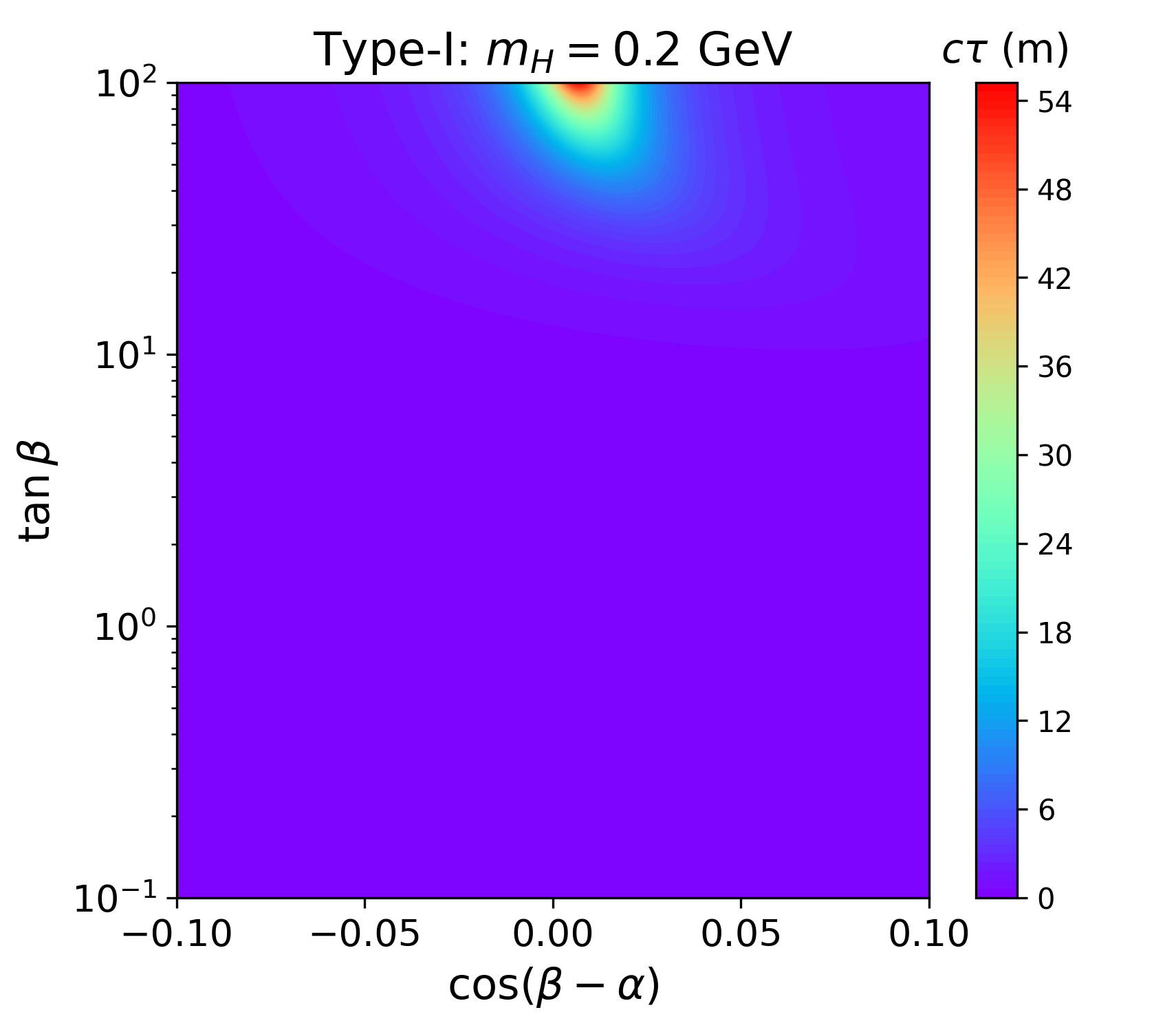}
\includegraphics[width=0.49 \linewidth]{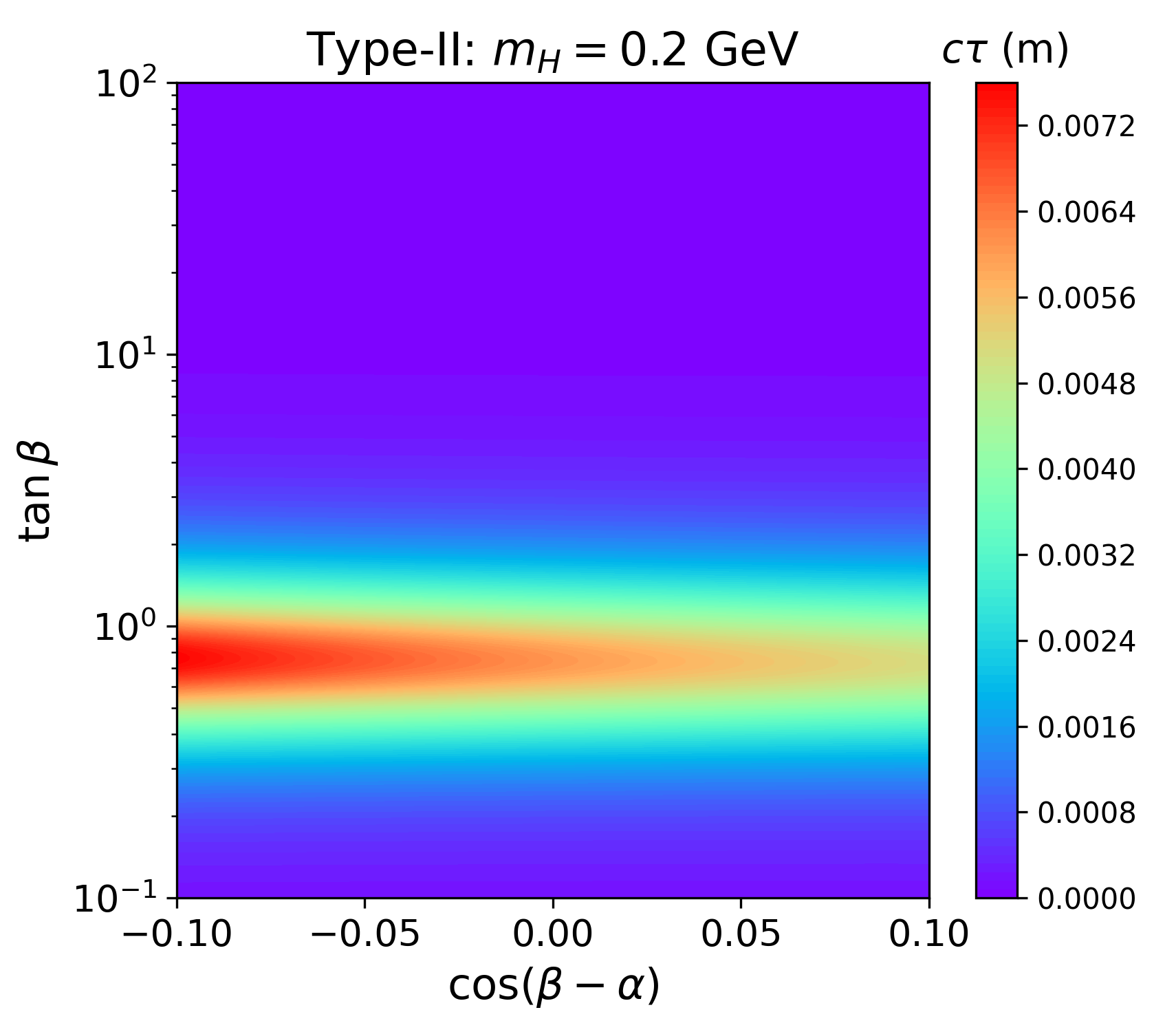}
\includegraphics[width=0.49 \linewidth]{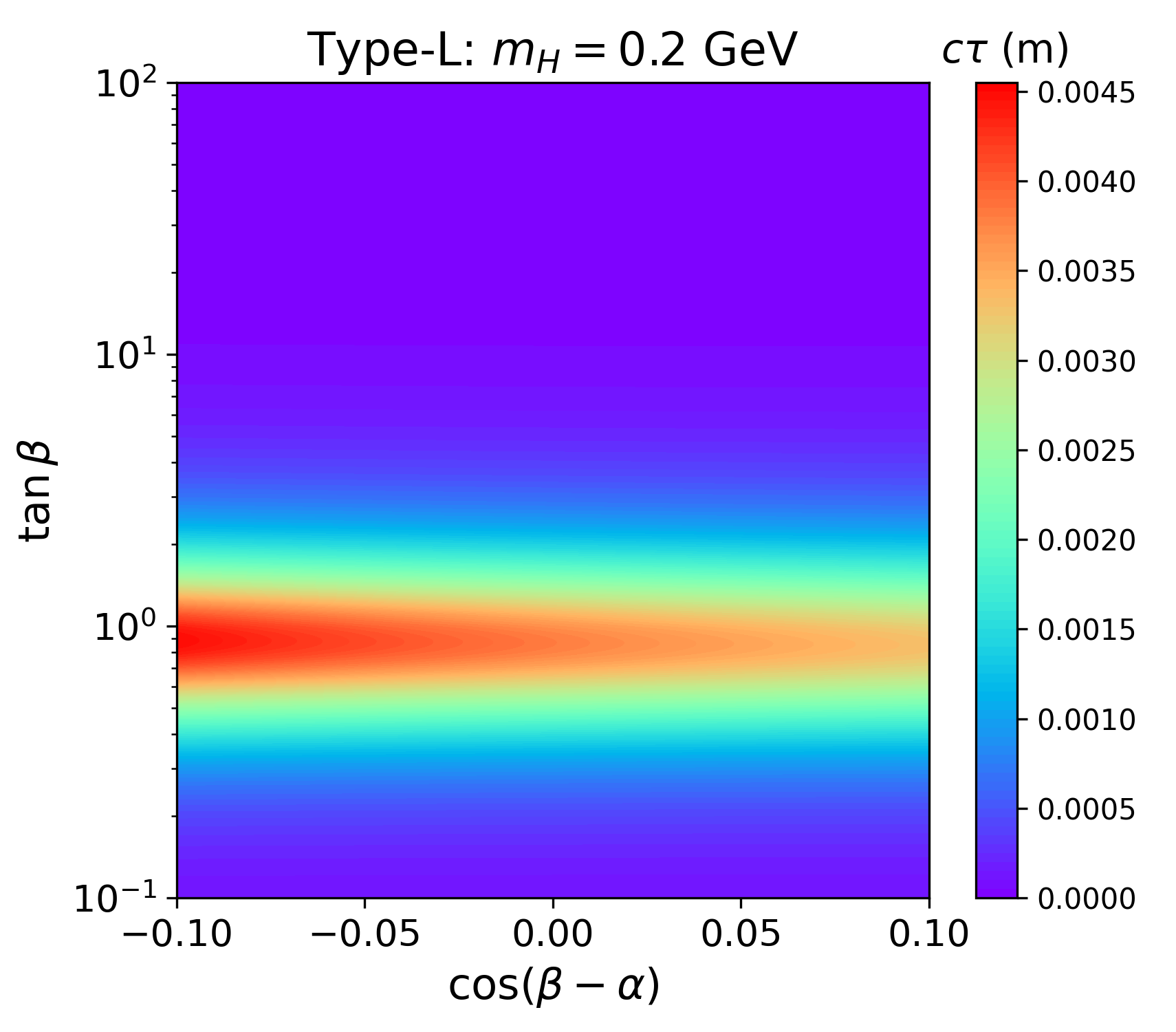}
\includegraphics[width=0.49 \linewidth]{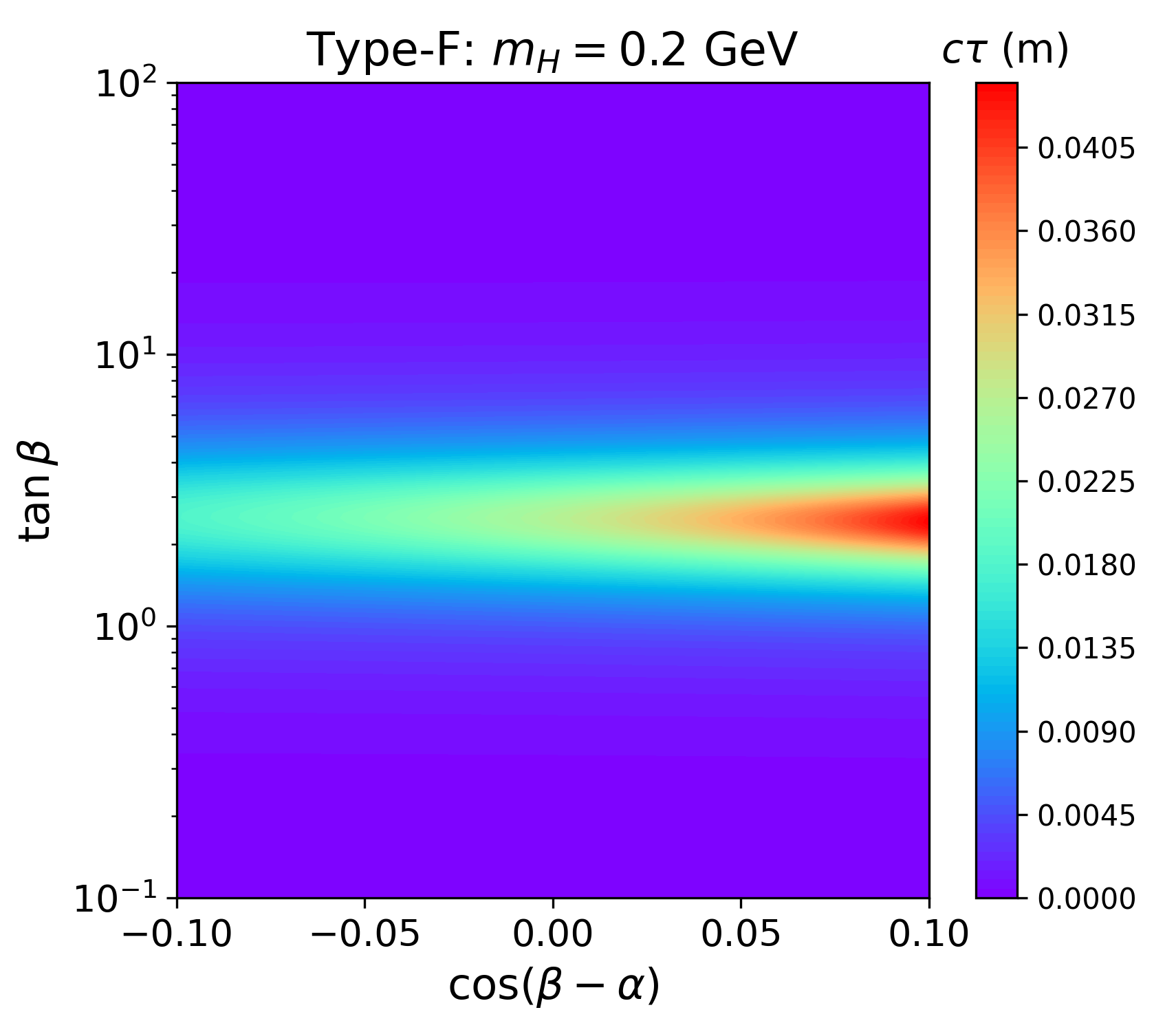}
\caption{Decay length $c\tau$(m) of the light $H$ in the $\cos(\beta-\alpha)$ vs. $\tan\beta$ plane for $m_H = 0.2~\mathrm{GeV}$ in the Type-I, II, L, and F 2HDMs. The colors indicate the corresponding $c\tau$ values, and the ranges (maximal values) are quite different for different types. } 
\label{fig:H_type}
\end{figure}
\begin{figure}[htbp]
  \centering
\includegraphics[width=0.49 \linewidth]{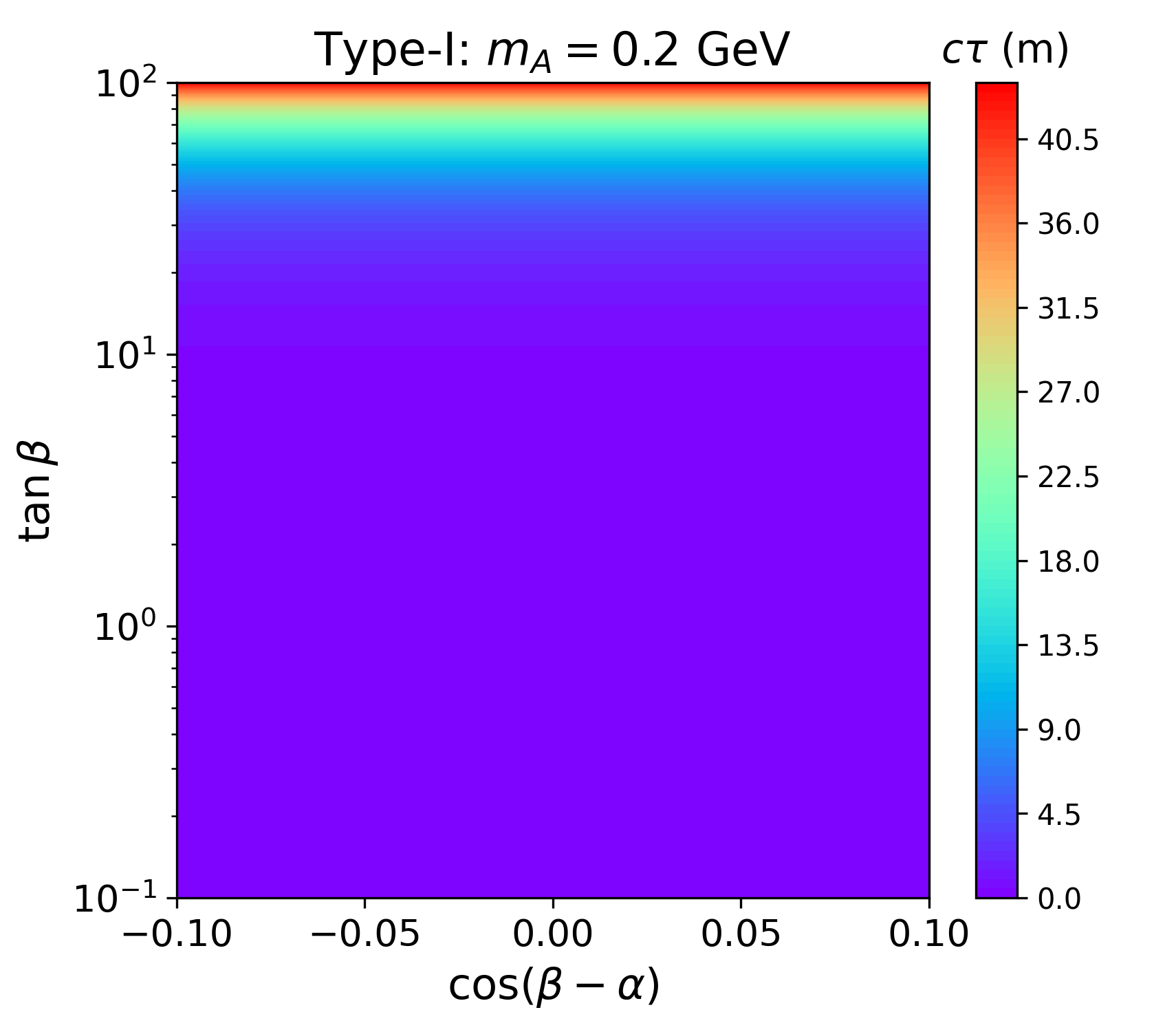}
\includegraphics[width=0.49 \linewidth]{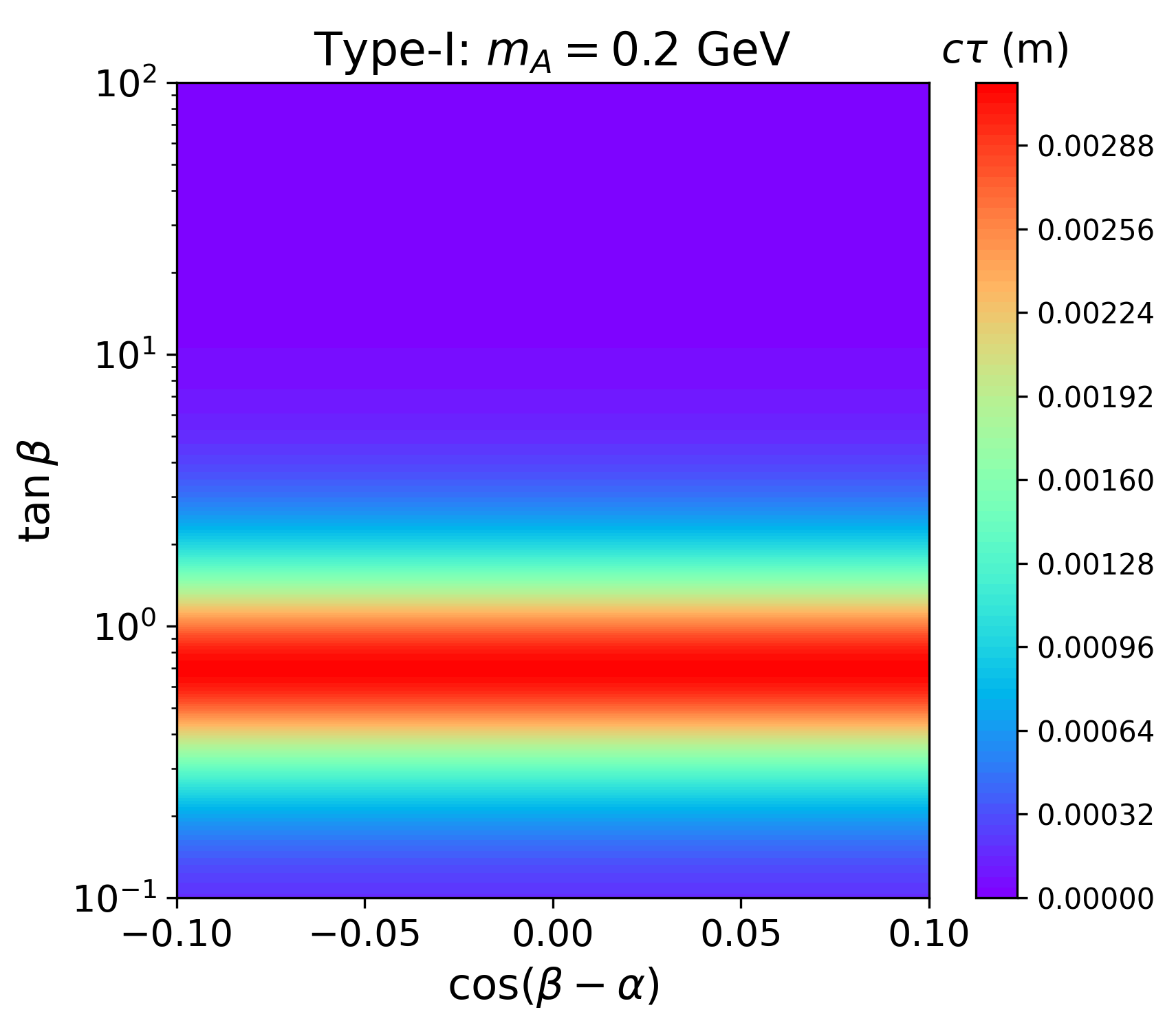}
\includegraphics[width=0.49 \linewidth]{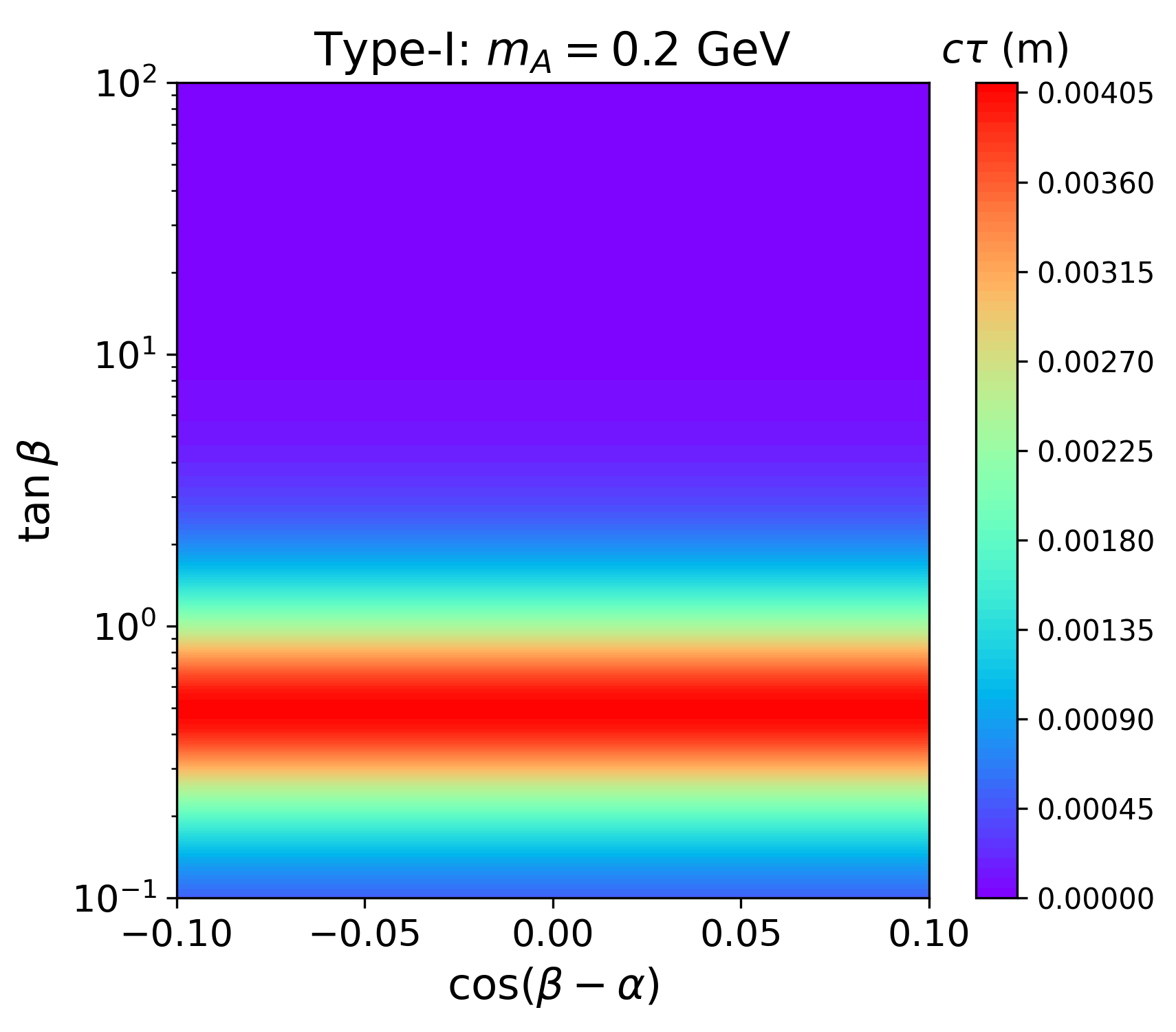}
\includegraphics[width=0.49 \linewidth]{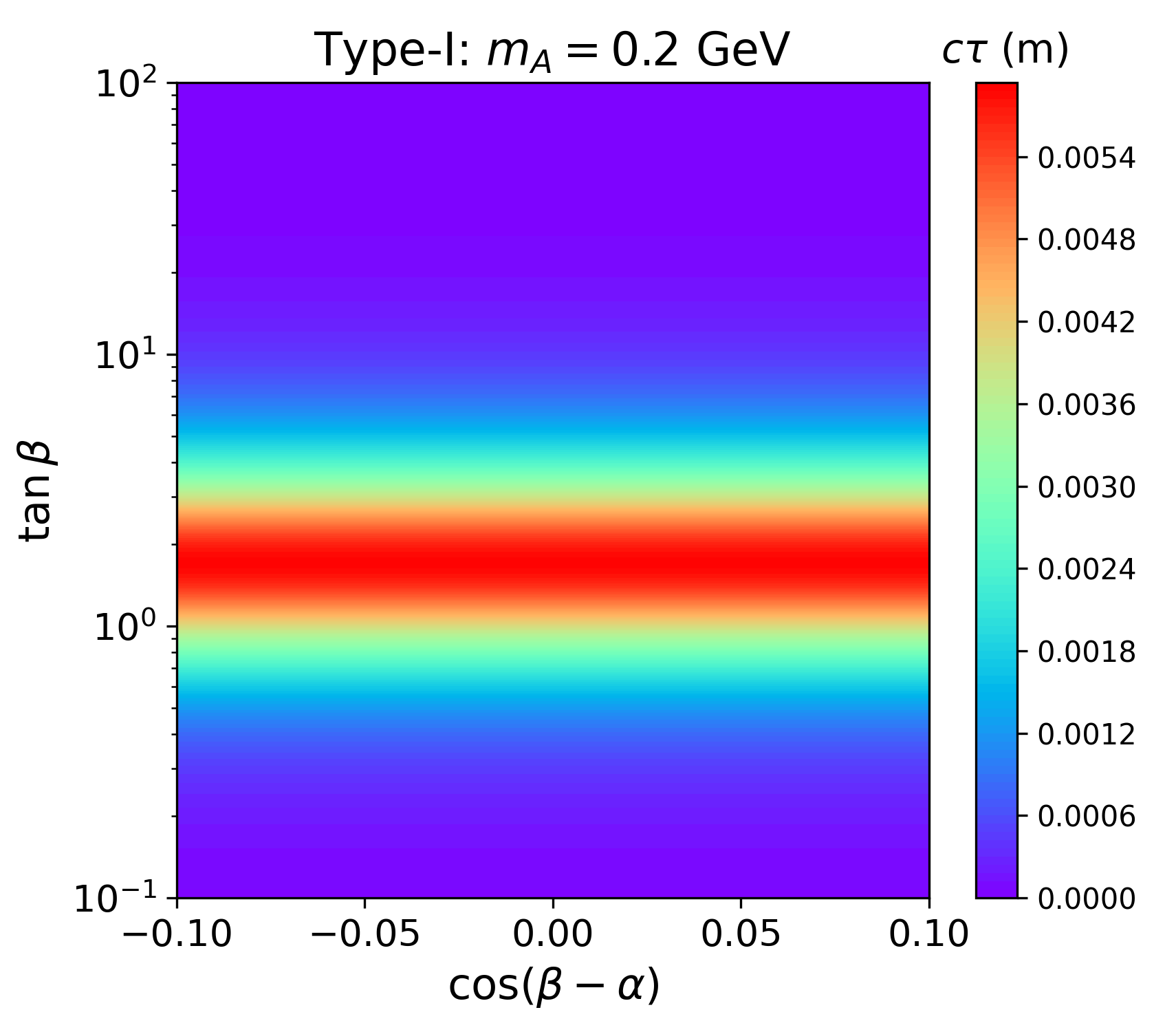}
\caption{Decay length $c\tau$(m) of the light $A$ in the $\cos(\beta-\alpha)$ vs. $\tan\beta$ plane for $m_A = 0.2~\mathrm{GeV}$ in the Type-I, II, L, and F 2HDMs. The colors indicate the corresponding $c\tau$ values.} 
\label{fig:A_type}
\end{figure}
In this section, we would like to illustrate the potential LLP candidate of different types of 2HDM.
In~\autoref{fig:H_type}, we show the decay length $c\tau$(m) of the light $H$ in the $\cos(\beta-\alpha)$ vs. $\tan\beta$ plane for $m_H = 0.2~\mathrm{GeV}$ in the Type-I, II, L, and F 2HDMs. The colors indicate the value of $c\tau$. Here we take $m_H = 0.2~\mathrm{GeV}$, where only the $e^+e^-$ and $\gamma\gamma$ decay channels \\re kinematically allowed, resulting in a relatively long lifetime and providing a suitable scenario for studying light long-lived scalars. If any type of 2HDM still cannot produce a sufficiently long $c\tau$ at this mass, it will not be able to do so at larger masses.
In the Type-I 2HDM, $c\tau$ increases with $\tan\beta$ around $\cos(\beta - \alpha) \sim 0$, with the maximum located at a small positive value of $\cos(\beta - \alpha)$, reaching up to $50~\mathrm{m}$ at $\tan\beta \sim 100$. Larger $\tan\beta$ is still allowed in Type-I, thus longer lifetime is allowed. In the Type-II and Type-L 2HDMs, $c\tau$ is only slightly large around $\tan\beta \sim 1$; in the Type-F 2HDM, $c\tau$ is slightly large around $\tan\beta \sim 2$. These regions are mostly located near $|\cos(\beta - \alpha)| \sim 0.1$. However, it should be noted that due to constraints from precision Higgs measurements~\cite{Kling:2020hmi}, the allowed values of $|\cos(\beta - \alpha)|$ are also very small when $\tan\beta$ is small, and such parameter regions may not be viable. Moreover, in all three of these types, the $c\tau$ values significantly smaller than in the Type-I 2HDM case. Therefore, these three types of 2HDM cannot accommodate a light long-lived particle $H$. 

In~\autoref{fig:A_type}, we show the decay length $c\tau$(m) of the light $A$ on the $\cos(\beta-\alpha)$ vs. $\tan\beta$ plane for $m_A = 0.2~\mathrm{GeV}$ in the Type-I, II, L, and F 2HDMs. The colors indicate the value of $c\tau$. Here, we choose $m_A = 0.2~\mathrm{GeV}$ for the same reason as for the light $H$. In the Type-I 2HDM, $c\tau$ increases with $\tan\beta$ and is independent of $\cos(\beta - \alpha)$, reaching up to $40~\mathrm{m}$ at $\tan\beta \sim 100$. In the Type-II and Type-L 2HDMs, $c\tau$ is slightly enhanced only for $\tan\beta < 1$, while in the Type-F 2HDM, it is slightly large around $\tan\beta \sim 2$. In these regions, $c\tau$ is also nearly independent of $\cos(\beta - \alpha)$. However, in all three of these types, the $c\tau$ values significantly smaller than in the Type-I 2HDM case. Therefore, these models cannot accommodate a light long-lived particle $A$.

\section{$m_{H^\pm}$ Ranges}
\label{sec:mc}
Table~\ref{tab:mc_ranges} summarizes the allowed $m_{H^\pm}$ ranges at 95\% C.L.\ 
The left part is for fixed values of $m_A$ in the light $H$ case, 
while the right part is for fixed values of $m_H$ in the light $A$ case.
\begin{table}[htbp]
\centering
\setlength{\tabcolsep}{11pt}       
\renewcommand{\arraystretch}{1.25}  
\begin{tabular}{|ccc|ccc|}
\cline{1-6}
\multicolumn{3}{|c|}{\textbf{Light $H$ (GeV)}} & \multicolumn{3}{c|}{\textbf{Light $A$ (GeV)}} \\
\cline{1-6}
$m_A$ & $(m_{H^\pm})_{\mathrm{min}}$ & $(m_{H^\pm})_{\mathrm{max}}$ & $m_H$ & $(m_{H^\pm})_{\mathrm{min}}$ & $(m_{H^\pm})_{\mathrm{max}}$ \\
\hline
54  & 80.0  & 80.5  & 54  & 80.0  & 80.5  \\
75  & 80.0  & 103.6 & 60  & 80.0  & 86.2  \\
100 & 80.0  & 130.8 & 63  & 80.0  & 89.5  \\
125 & 100.4 & 154.3 & 66  & 80.0  & 92.9  \\
150 & 129.4 & 177.2 & 69  & 80.0  & 96.4  \\
175 & 157.4 & 200.1 & 72  & 80.0  & 100.0 \\
200 & 184.6 & 223.2 & 75  & 80.0  & 103.6 \\
225 & 211.4 & 246.4 & 78  & 80.0  & 107.2 \\
250 & 237.8 & 269.9 & 81  & 80.0  & 110.8 \\
275 & 263.9 & 293.5 & 84  & 80.0  & 114.2 \\
300 & 289.8 & 317.3 & 87  & 80.0  & 117.6 \\
325 & 315.6 & 341.2 & 90  & 80.0  & 120.8 \\
350 & 341.3 & 365.2 & 93  & 80.0  & 123.9 \\
375 & 366.9 & 389.3 & 96  & 80.0  & 126.9 \\
400 & 392.4 & 413.5 & 99  & 80.0  & 129.8 \\
425 & 417.9 & 437.8 & 102 & 80.0  & 132.7 \\
450 & 443.3 & 462.2 & 105 & 80.0  & 135.6 \\
475 & 468.6 & 486.6 & 108 & 80.3  & 138.4 \\
500 & 494.0 & 511.1 & 111 & 83.8  & 141.3 \\
525 & 519.2 & 535.6 & 114 & 87.4  & 144.1 \\
550 & 544.5 & 560.1 & 117 & 91.0  & 146.9 \\
575 & 569.8 & 584.7 & 120 & 94.5  & 149.7 \\
600 & 595.0 & 600.0 & 125 & 100.4 & 154.3 \\
\cline{1-6}
\end{tabular}
\caption{Allowed $m_{H^\pm}$ ranges at 95\% C.L. for fixed $m_A$ (left) in the light $H$ case and fixed $m_H$ (right) in the light $A$ case.}
\label{tab:mc_ranges}
\end{table}

\section{Total decay width and decay length of light H/A}
\label{sec:length}
%
\begin{figure}[htbp]
  \centering
\includegraphics[width=0.49 \linewidth]{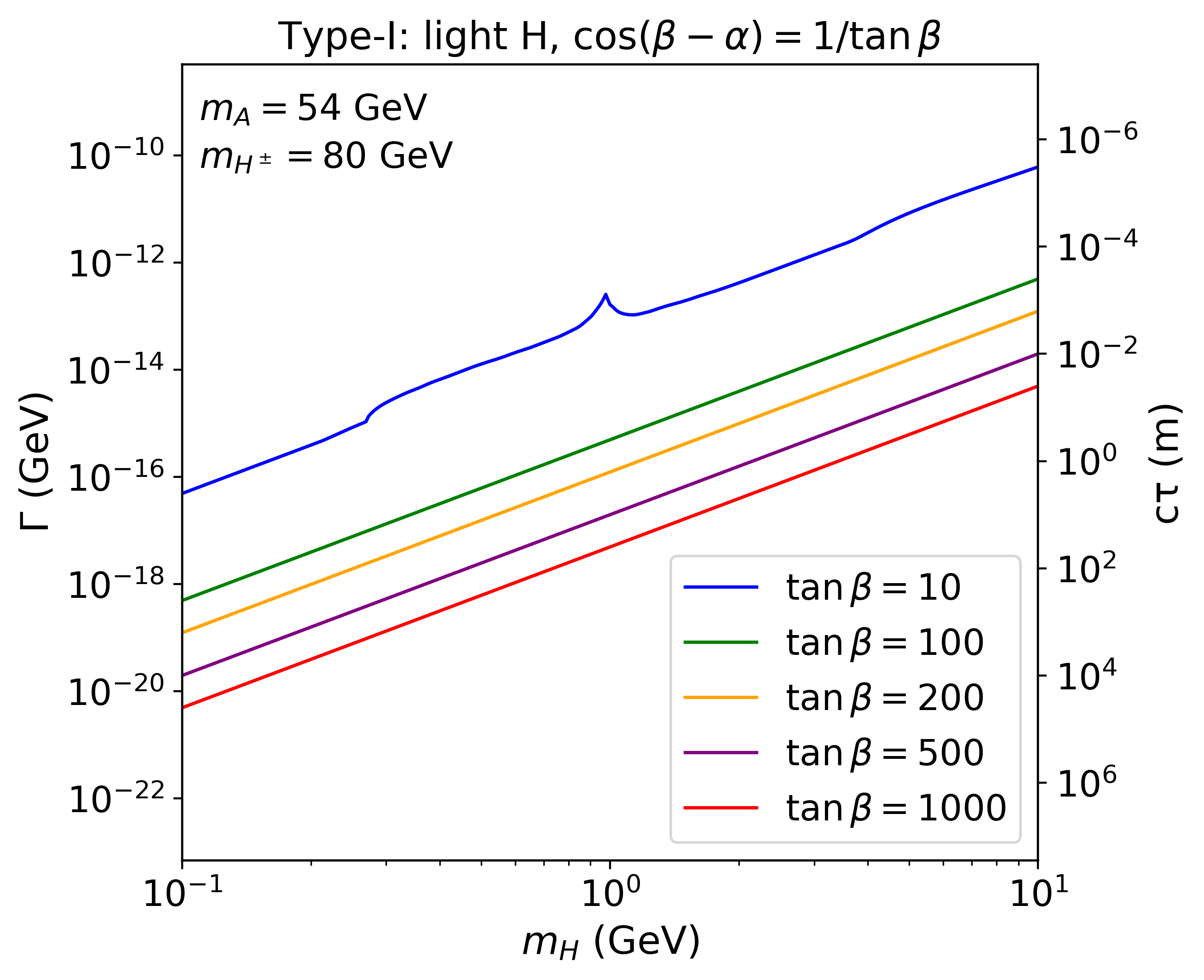}
\includegraphics[width=0.49 \linewidth]{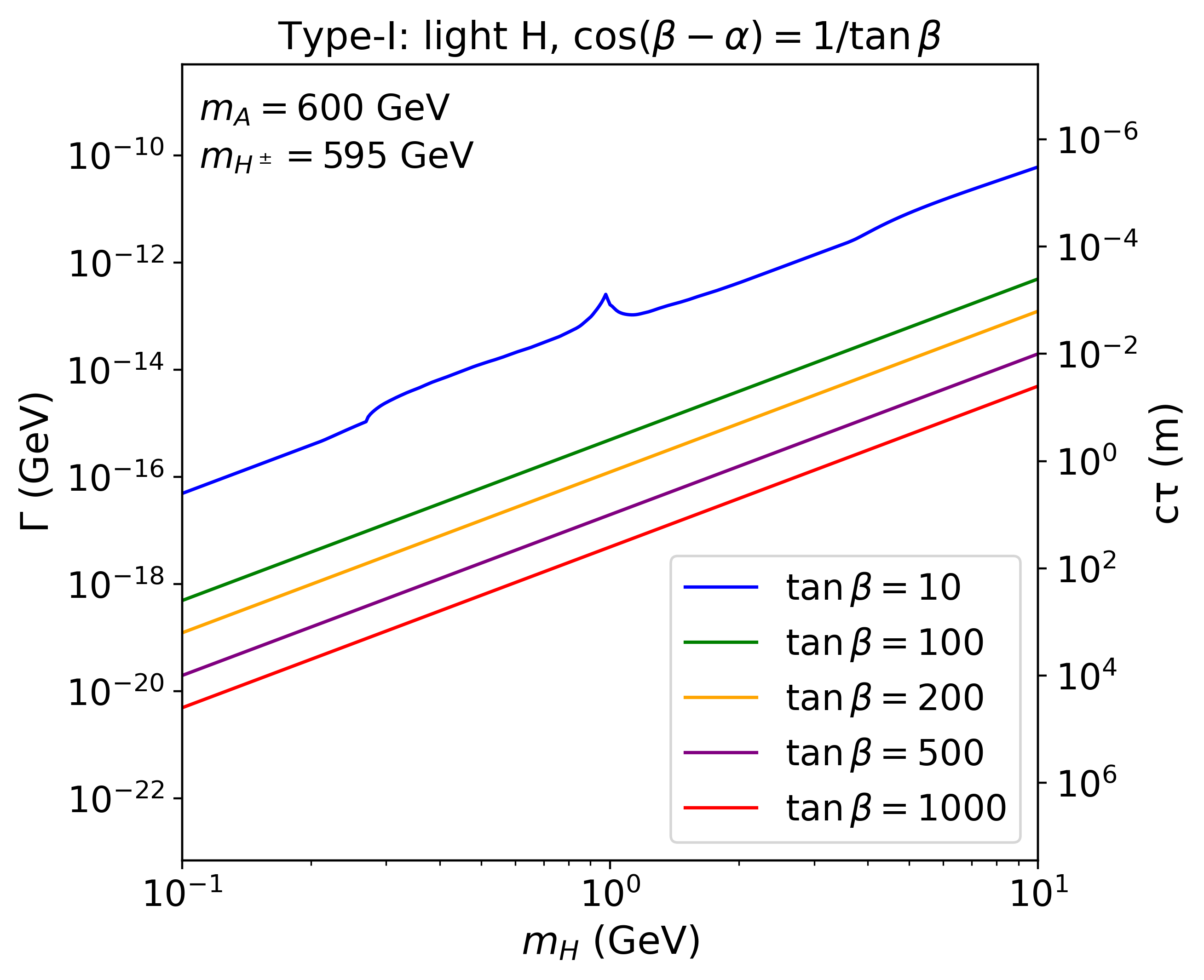}
\caption{The total decay width (left $y$-axis) and decay length $c\tau$ (right $y$-axis) of the light CP-even Higgs in the Type-I 2HDM for the light $H$ scenario. }
\label{fig:H_length}
\end{figure}
In \autoref{fig:H_length}, we show the total decay width and decay length of the light $H$ in the light $H$ scenario for different values of $m_{A/H^\pm}$. The left and right cases are nearly the lower and upper limits for corresponding parameters. It can be seen that $\Gamma$ and $c\tau$ are insensitive to $m_{A/H^\pm}$, which is also confirmed in \autoref{fig:ctau}. The $\Gamma$ and $c\tau$ become straight lines for very large $\tan\beta$ as a consequence of the dominant diphoton decay. Again here we confirm that $m_A$ and $m_{H^\pm}$ do not change the lifetimes inside of the allowed parameter space

\begin{figure}[htbp]
  \centering
\includegraphics[width=0.49 \linewidth]{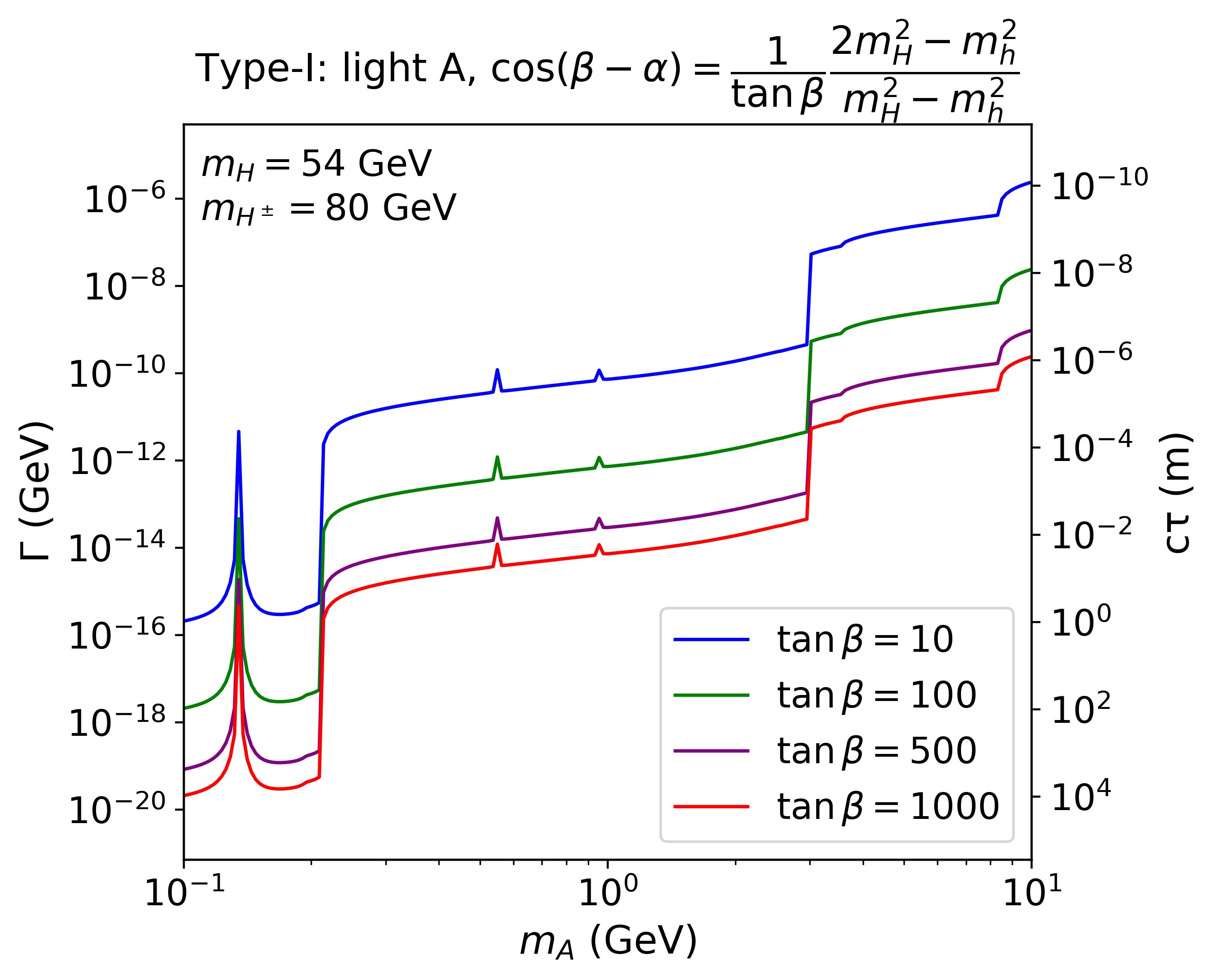}
\includegraphics[width=0.49 \linewidth]{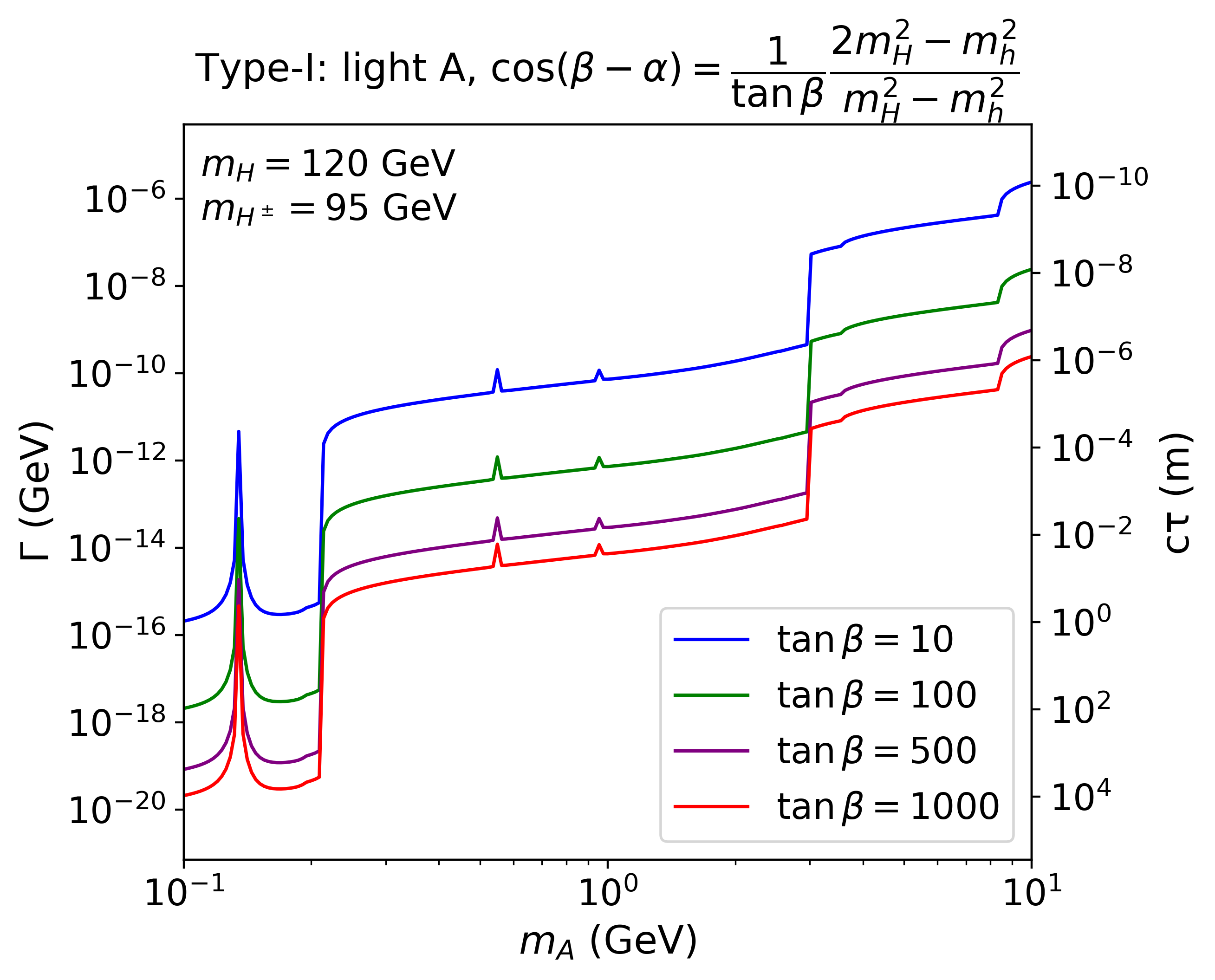}
\caption{The total decay width (left $y$-axis) and decay length $c\tau$ (right $y$-axis) of the light CP-odd Higgs in the Type-I 2HDM for the light $A$ scenario.}
\label{fig:A_length}
\end{figure}
In~\autoref{fig:A_length}, we show the total decay width and decay length of the light $A$ in the light $A$ scenario for different values of $m_{H/H^\pm}$. Similarly, the left and right cases are nearly the lower and upper limits for corresponding parameters. $\Gamma$ and $c\tau$ are insensitive to $m_{H/H^\pm}$. The peaks arise for the same reason discussed earlier. The $\tan\beta$ dependence of $\Gamma_A$ appears only as an overall shift with the same features, since all couplings of $A$ to SM particles scale have the same $1/\tan\beta$ dependence.



\bibliographystyle{JHEP}
\bibliography{ref_LLP_dw.bib}

\end{document}